\documentclass[a4paper]{article}
\usepackage{feynmp}
\begin{document}
\begin{fmffile}{knots_lanl_fig}
\def\today{\space\number\day\ \ifcase\month\or
January\or February\or March\or April\or May\or June\or July\or
August\or September\or October\or November\or December\fi
\ \number\year}
\fmfcmd{%
style_def fillvx expr p =
  fill (photon p)--cycle  withcolor .1[white,black];
enddef;
style_def fillwx expr p =
  fill p--cycle  withcolor .1[white,black];
enddef;
style_def unfillvx expr p =
  unfill (photon p)--cycle;
  draw currentpicture;
enddef;
style_def unfillwx expr p =
  unfill p--cycle;
  draw currentpicture;
enddef;}
\def\topfraction{0.9}
\def\bottomfraction{0.9}
\def\Break{\hfil\break}
\setlength{\unitlength}{1mm}
\overfullrule=0pt  
\def\mynote#1{{[{\it NOTE: #1}]}}
\def\fEQN#1#2{$$\hbox{\it #1\hfil}\EQN{#2}$$}
\def\Acknowledgements{{\bigskip\leftline
{{\bf Acknowledgments}}\medskip}}
\renewcommand{\baselinestretch}{1.0}
\small\normalsize

\begin{titlepage}


 \vspace{0cm}
 
\begin{center}
{\large\bf 
A dynamic programming algorithm for\\ 
RNA structure prediction including pseudoknots}

 \vspace{0.4cm}
 
{\bf Elena~Rivas
and Sean~R.~Eddy\footnote{To whom correspondence should be addressed.
Tel: +1 314 362 7666; Fax: +1 314 362 7855; Email: eddy@genetics.wustl.edu.}
}   

 \vspace{0.2cm}

{\sl Department of Genetics,\\ 
Washington University, St. Louis, MO 63130 USA}

\vspace{0.2cm}

\end{center}
\vspace{0.2cm}
\begin{abstract}
 
We describe a dynamic programming algorithm for predicting optimal RNA
secondary structure, including pseudoknots. The algorithm has a worst
case complexity of ${\cal O}(N^6)$ in time and ${\cal O}(N^4)$ in
storage. The description of the algorithm is complex, which led us to
adopt a useful graphical representation (Feynman diagrams) borrowed
from quantum field theory. We present an implementation of the
algorithm that generates the optimal minimum energy structure for a
single RNA sequence, using standard RNA folding thermodynamic
parameters augmented by a few parameters describing the thermodynamic
stability of pseudoknots. We demonstrate the properties of the
algorithm by using it to predict structures for several small
pseudoknotted and non-pseudoknotted RNAs. Although the time and memory
demands of the algorithm are steep, we believe this is the first
algorithm to be able to fold optimal (minimum energy) pseudoknotted
RNAs with the accepted RNA thermodynamic model.

\end{abstract}

\vspace{1.0 cm}
\begin{description}

\item [Running title] 
RNA pseudoknot prediction by dynamic programming.

\item [Keywords] 
RNA, secondary structure prediction, pseudoknots, dynamic programming, 
thermodynamic stability.

\end{description}

\end{titlepage}

\section{INTRODUCTION}

Many RNAs fold into structures that are important for regulatory,
catalytic, or structural roles in the cell. An RNA's structure is
dominated by base pairing interactions, most of which are Watson-Crick
pairs between complementary bases. The base paired structure of an RNA
is called its secondary structure. Because Watson-Crick pairs are such
a stereotyped and relatively simple interaction, accurate RNA
secondary structure prediction appears to be an achievable goal.

A rather reliable approach for RNA structure prediction is comparative
sequence analysis, in which covarying residues (e.g. compensatory
mutations) are identified in a multiple sequence alignment of RNAs
with similar structures but different sequences (Woese \& Pace, 1993).
Covarying residues, particularly pairs which covary to maintain
Watson-Crick complementarity, are indicative of conserved base pairing
interactions. The accepted secondary structures of most structural and
catalytic RNAs were generated by comparative sequence analysis.

If one has only a single RNA sequence (or a small family of RNAs with
little sequence diversity) comparative sequence analysis cannot be
applied. Here the best current approaches are energy minimization
algorithms (Schuster {\it et al.}, 1997). While not as accurate as
comparative sequence analysis, these algorithms have still proven to
be useful research tools. Thermodynamic parameters are available for
predicting the $\Delta$G of a given RNA structure (Freier {\it et
al.}, 1986; Serra {\it et al.}, 1995). The Zuker algorithm
(implemented in the programs MFOLD (Zuker, 1989a) and ViennaRNA
(Schuster {\it et al.}, 1994)) is an efficient dynamic programming
algorithm for identifying the globally minimal energy structure for a
sequence, as defined by such a thermodynamic model (Zuker \& Stiegler,
1981; Zuker \& Sankoff, 1984; Sankoff, 1985). The Zuker algorithm
requires $O(N^3)$ time and $O(N^2)$ space for a sequence of length
$N$, and so is reasonably efficient and practical even for large RNA
sequences. The Zuker dynamic programming algorithm was subsequently
extended to allow experimental constraints, and to sample suboptimal
folds (Zuker, 1989b).  McCaskill's variant of the Zuker algorithm
calculates probabilities (confidence estimates) for particular base
pairs (McCaskill, 1990).

One well-known limitation of the Zuker algorithm is that it is
incapable of predicting so-called {\em RNA pseudoknots}. This is the
problem that we address in this paper.

\begin{figure}[ht]
\begin{center}
\begin{picture}(55,40)

\put(10.0,2.0){$5^\prime$---}
\put(17.0,2.0){C}
\put(20.0,5.0){\shortstack{C$\bullet$G\\
                           U$\bullet$G\\
                           U$\bullet$A}}
\put(27.0,2.0){A\hspace{0.5mm}A\hspace{0.5mm}U\hspace{0.5mm}G\hspace{0.5mm}A
                \hspace{0.5mm}C\hspace{0.5mm}---$3^\prime$}
\put(17.0,16.0){A}
\put(19.0,20.0){U}
\put(24.0,20.0){C}
\put(27.0,16.0){A}
\bezier{15}(34.5,5.0)(35.0,16.0)(30.0,17.0)
\bezier{20}(37.4,5.0)(38.0,18.5)(28.0,20.0)
\bezier{25}(40.3,5.0)(40.0,28.0)(21.0,23.0)
\put(60.0,8.0){$\longrightarrow$}

\end{picture}
\begin{picture}(55,40)

\put(15.0,16.0){$5^\prime$--- C}
\put(26.0,10.0){\shortstack{U U C\\$\bullet$ \hspace{0.5mm}
$\bullet$ \hspace{0.5mm}$\bullet$\\A G G}}
\put(42.0,20.0){A}
\put(38.0,4.0){\shortstack{A C U\\$\bullet$ \hspace{0.5mm}
$\bullet$ \hspace{0.5mm}$\bullet$\\U G A}}
\put(50.0,4.0){C ---$3^\prime$}
\put(27.0,4.0){A}
\put(33.0,2.0){A}
\qbezier(36.0,19.0)(37.0,20.0)(41.0,21.0)
\qbezier(45.0,21.0)(50.0,17.0)(49.0,12.0)

\qbezier(25.0,11.0)(24.0,7.0)(27.0,5.0)
\qbezier(30.0,4.0)(31.0,3.5)(33.0,3.0)
\qbezier(35.5,3.0)(36.0,3.0)(38.0,4.0)

\end{picture}
\end{center}
\caption{\label{fig:pk} A simple pseudoknot. In a pseudoknot,
nucleotides inside a hairpin loop pair with nucleotides outside the
stem-loop.}
\end{figure}
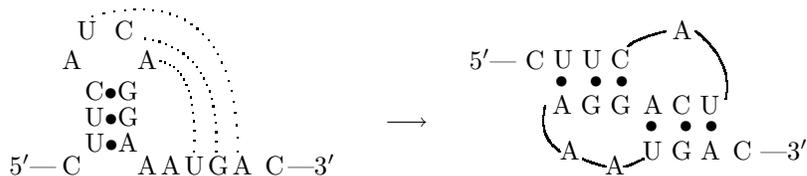

The thermodynamic model for non-pseudoknotted RNA secondary structure
includes some stereotypical interactions, such as stacked base-paired
stems, hairpins, bulges, internal loops, and multiloops.  Formally,
non pseudoknotted structures obey a ``nesting'' convention: that for
any two base pairs $i,j$ and $k,l$ (where $i<j$ and $k<l$), either $i
< k < l < j$ or $k < i < j < l$. It is precisely this ``nesting''
convention that the Zuker dynamic programming algorithm relies upon to
recursively calculate the minimal energy structure on progressively
longer subsequences. An RNA pseudoknot is defined as a structure
containing base pairs which violate the nesting convention. An example
of a simple pseudoknot is shown in Figure~\ref{fig:pk}.

RNA pseudoknots are functionally important in several known RNAs (ten
Dam {\it et al.}, 1992). For example, by comparative analysis, RNA
pseudoknots are conserved in ribosomal RNAs, the catalytic core of
group I introns, and RNase P RNAs. Plausible pseudoknotted structures
have been proposed (Florentz {\it et al.}, 1982), and recently
confirmed (Kolk {\it et al.}, 1998) for the 3' end of several plant
viral RNAs, where pseudoknots are apparently used to mimic tRNA
structure. In vitro RNA evolution (SELEX) experiments have yielded
families of RNA structures which appear to share a common
pseudoknotted structure, such as RNA ligands selected to bind HIV-1
reverse transcriptase (Tuerk {\it et al.}, 1992).

Most methods for RNA folding which are capable of folding pseudoknots
adopt heuristic search procedures and sacrifice optimality. Examples
of these approaches include quasi-Monte Carlo searches (Abrahams {\it
et al.}, 1990) and genetic algorithms (Gultyaev {\it et al.}, 1995;
van Batenburg et al., 1995). These approaches are inherently unable to
guarantee that they have found the ``best'' structure given the
thermodynamic model, and consequently unable to say how far a given
prediction is from optimality.

A different approach to pseudoknot prediction based on the maximum
weigh\-ted matching (MWM) algorithm (Edmonds, 1965; Gabow, 1976) was
introduced by Cary and Stormo (1995). Using the MWM algorithm, an
optimal structure is found, even in the presence of complicated
knotted interactions, in $O(N^3)$ time and $O(N^2)$ space. However,
MWM seems best suited to folding sequences for which a previous
alignment exists.  In the scoring system used by Cary and Stormo,
weights are assigned by comparative analysis.  It is not clear to us
that the MWM algorithm will be amenable to folding single sequences or
collections of sequences which present little variation respect to
each other.  However, we believe that this was the first work that
indicated that optimal RNA pseudoknot predictions can be made with
polynomial time algorithms. It had been widely believed, but never
proven, that pseudoknot prediction would be an NP problem (NP =
nondeterministic polynomial; e.g. only solvable by heuristic or brute
force approaches).

In this paper we describe a dynamic programming algorithm which finds
optimal pseudoknotted RNA structures. We describe the algorithm using
a diagrammatic representation borrowed from quantum field theory
(Feynman diagrams). We implement a version of the algorithm that finds
minimal energy RNA structures using the standard RNA secondary
structure thermodynamic model (Freier {\it et al.}, 1986, Serra {\it
et al.}, 1995), augmented by a few pseudoknot-specific parameters that
are not yet available in the standard folding parameters, and by
coaxial stacking energies (Walter {\it et al.}, 1994) for both
pseudoknotted and non-pseudoknotted structures.  We demonstrate the
properties of the algorithm by testing it on several small RNA
structures, including both structures thought to contain pseudoknots
and structures thought not to contain pseudoknots.

\section{ALGORITHM}
In this section we will introduce a diagrammatic way of representing
RNA folding algorithms. We will start by describing the Nussinov
algorithm (Nussinov {\it et al.}, 1978), and the Zuker-Sankoff
algorithm (Zuker \& Sankoff 1984; Sankoff 1985) in the context of this
representation.  Later on we will extend the diagrammatic
representation to include pseudoknots and coaxial stackings.  The
Nussinov and Zuker-Sankoff algorithms can be implemented without the
diagrammatic representation, but this representation is essential to
manage the complexity introduced by pseudoknots.

\subsection{Preliminaries}
From here on, unless otherwise stated, a flat solid line will
represent the backbone of a RNA sequence with its $5^\prime$ end
placed in the left hand side of the segment.  $N$ will represent the
length (in number of nucleotides) of the RNA.

\begin{figure}
\begin{center}
\begin{fmfgraph*}(150,60)
   \fmfleft{i1,d1}
   \fmfright{i46,d2}
   \fmfn{plain}{i}{46}
   \fmfvn{d.shape=circle, d.size=1thick}{i}{46}
   \fmffreeze
   \fmf{photon, left=0.7}{i3,i30}
   \fmf{photon, left=0.7}{i4,i29}
   \fmf{photon, left=0.7, label=multiloop, l.side=right}{i5,i27}

   \fmf{photon,left=0.6}{i6,i12}
   \fmf{photon,left=0.6}{i7,i11}

   \fmf{photon, left=0.65}{i14,i26}
   \fmf{photon, left=0.65, label=int loop, l.side=right}{i15,i25}
   \fmf{photon, left=0.6}{i17,i22}
   \fmf{photon, left=0.6}{i18,i21}

   \fmflabel{bulge}{i28}
   \fmflabel{ss}{i1}
   \fmflabel{stem}{i6}
   \fmflabel{hairpin}{i11}
   \fmflabel{hairpin}{i20}

   \fmf{photon,left=0.6}{i32,i41}
   \fmf{photon,left=0.6}{i33,i40}
   \fmf{photon,left=0.6}{i34,i39}

   \fmf{photon,right=0.6, label=pseudoknot, l.side=right}{i36,i46}
   \fmf{photon,right=0.6}{i37,i45}
   \fmf{photon,right=0.6}{i38,i44}

\end{fmfgraph*}
\bigskip\bigskip
\caption{\label{fig:nested-knot} Diagrammatic representation of RNA 
most relevant secondary structures, including a pseudoknot.}
\end{center}
\end{figure}
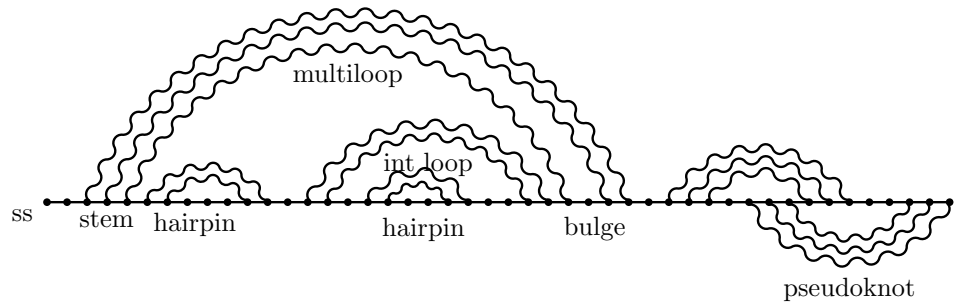

Secondary interactions will be represented by wavy lines connecting
the two interacting positions in the backbone chain, while the
backbone itself always remains flat.  No more than two bases are
allowed to interact at once.  This representation does not provide
insight about real (3D) spatial arrangements, but is very convenient
for algorithmic purposes.  When necessary for clarification single
stranded regions will be marked by dots, but when unambiguous, dots
will be omitted for simplicity. Using this representation
(figure~\ref{fig:nested-knot}) we can describe hairpins, bulges,
stems, internal loops and multiloops as simple nested structures; a
pseudoknot, on the other hand, corresponds to a non-nested structure.

\subsection{Diagrammatic representation of nested algorithms}
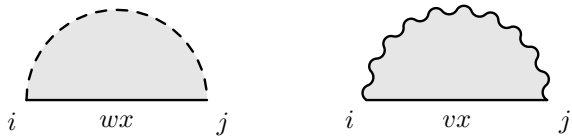
\begin{figure}[ht]
\begin{center}
\begin{fmfgraph*}(30,30)\fmfkeep{wx}
   \fmfleft{i,d1}
   \fmfright{j,d2}
   \fmffreeze
   \fmf{fillwx,left}{i,j}
   \fmf{plain,label=$wx$, l.side=right}{i,j}\fmflabel{$i$}{i}\fmflabel{$j$}{j}
   \fmf{dashes,left}{i,j}
\end{fmfgraph*}$\qquad\qquad$
\begin{fmfgraph*}(30,20)\fmfkeep{vx}
   \fmfleft{i,d1}
   \fmfright{j,d2}
   \fmf{fillvx,left}{i,j}
   \fmf{plain,label=$vx$, l.side=right}{i,j}\fmflabel{$i$}{i}\fmflabel{$j$}{j}
   \fmffreeze
   \fmf{photon,left}{i,j}
\end{fmfgraph*}
\bigskip
\caption{\label{fig:wx-vx} $Wx$ and $vx$ matrices.}
\end{center}
\end{figure}

In order to describe a nested algorithm we need to introduce two
triangular $N\times N$ matrices, to be called $vx$ and $wx$. These
matrices are defined in the following way: $vx(i,j)$ is the score of
the best folding between positions $i$ and $j$ provided that $i$ and
$j$ are paired to each other; whereas $wx(i,j)$ is the score of the
best folding between positions $i$ and $j$ regardless of whether $i$
and $j$ pair to each other or not.  These matrices are graphically
represented in the form indicated in fig.~\ref{fig:wx-vx}. The filled
inner space indicates that we do not know how many interactions (if
any) occur for the nucleotides inside, in contrast with a blank inner
space which indicates that the fragment inside is known to be single
stranded.  The wavy line in $vx$, as previously, indicates that $i$
and $j$ are definitely paired, and similarly the discontinuous line in
$wx$ indicates that the relation between $i$ and $j$ is unknown.  Also
part of our convention is that for a given fragment, nucleotide $i$ is
at the $5^\prime$-end, and nucleotide $j$ is at the $3^\prime$-end, so
that $i \leq j$.

The purpose of a dynamic programming algorithm is to fill the $vx$ and
$wx$ matrices with appropriate numerical weights by means of some sort
of recursive calculation.

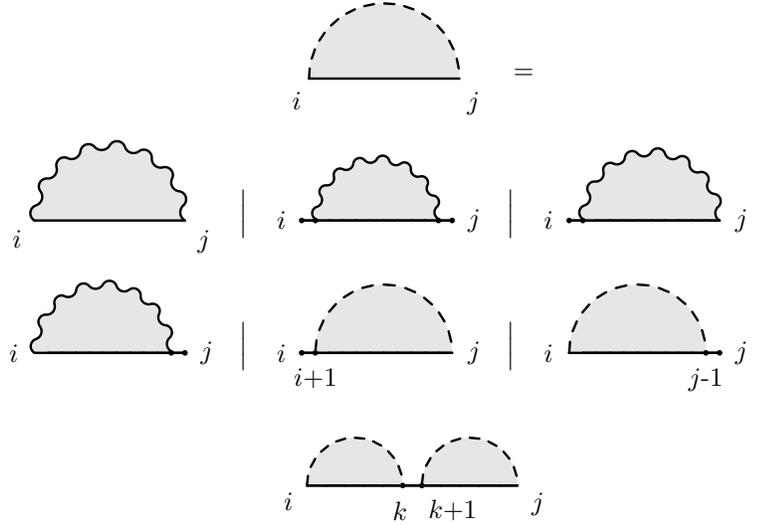
\begin{figure}[ht]
\begin{center}
\begin{fmfgraph*}(25,35)\fmfkeep{wx}
   \fmfleft{i,d1}
   \fmfright{j,d2}
   \fmffreeze
   \fmf{fillwx,left}{i,j}
   \fmf{plain}{i,j}\fmflabel{$i$}{i}\fmflabel{$j$}{j}
   \fmf{dashes,left}{i,j}
\end{fmfgraph*}
\quad=\quad 
\end{center}

\begin{center}
\begin{fmfgraph*}(25,15)\fmfkeep{vx}
   \fmfleft{i,d1}
   \fmfright{j,d2}
   \fmffreeze
   \fmf{fillvx,left}{i,j}
   \fmf{photon,left}{i,j}
   \fmf{plain}{i,j}\fmflabel{$i$}{i}\fmflabel{$j$}{j}
\end{fmfgraph*}
$\quad\Big |\quad$
\begin{fmfgraph*}(25,15)\fmfkeep{w2}
   \fmfleft{i,d1}\fmfv{l=$i$,l.a=180}{i}
   \fmfright{j,d2}
   \fmf{plain}{i,h1}
   \fmfn{plain}{h}{10}
   \fmf{plain}{h10,j}\fmfv{l=$j$,l.a=0}{j}
   \fmffreeze
   \fmf{fillvx,left}{h1,h10}
   \fmf{photon,left}{h1,h10}
   \fmfv{d.shape=circle, d.size=0.6thick}{i}
   \fmfv{d.shape=circle, d.size=0.6thick}{j}
   \fmfv{d.shape=circle, d.size=0.6thick}{h1}
   \fmfv{d.shape=circle, d.size=0.6thick}{h10}
   \fmf{plain}{i,h1}
   \fmfn{plain}{h}{10}
   \fmf{plain}{h10,j}
\end{fmfgraph*}
$\quad\Big |\quad$
\begin{fmfgraph*}(25,15)\fmfkeep{w3}
   \fmfleft{i,d1}\fmfv{l=$i$,l.a=180}{i}
   \fmfright{j,d2}
   \fmf{plain}{i,h1}
   \fmfn{plain}{h}{10}
   \fmf{plain}{h10,j}\fmfv{l=$j$,l.a=0}{j}
   \fmffreeze
   \fmf{fillvx,left}{h1,j}
   \fmf{photon,left}{h1,j}
   \fmfv{d.shape=circle, d.size=0.6thick}{i}
   \fmfv{d.shape=circle, d.size=0.6thick}{h1}
   \fmf{plain}{i,h1}
   \fmfn{plain}{h}{10}
   \fmf{plain}{h10,j}
\end{fmfgraph*}
$\quad\Big |\quad$

\begin{fmfgraph*}(25,15)\fmfkeep{w4}
   \fmfleft{i,d1}\fmfv{l=$i$,l.a=180}{i}
   \fmfright{j,d2}
   \fmf{plain}{i,h1}
   \fmfn{plain}{h}{10}
   \fmf{plain}{h10,j}\fmfv{l=$j$,l.a=0}{j}
   \fmffreeze
   \fmf{fillvx,left}{i,h10}
   \fmf{photon,left}{i,h10}
   \fmfv{d.shape=circle, d.size=0.6thick}{j}
   \fmfv{d.shape=circle, d.size=0.6thick}{h10}
   \fmf{plain}{i,h1}
   \fmfn{plain}{h}{10}
   \fmf{plain}{h10,j}
\end{fmfgraph*}
$\quad\Big |\quad$
\begin{fmfgraph*}(25,15)\fmfkeep{w5}
   \fmfleft{i,d1}\fmfv{l=$i$,l.a=180}{i}
   \fmfright{j,d2}
   \fmf{plain}{i,h1}
   \fmfn{plain}{h}{10}
   \fmf{plain}{h10,j}\fmfv{l=$j$,l.a=0}{j}
   \fmffreeze
   \fmf{fillwx,left}{h1,j}
   \fmf{dashes,left}{h1,j}
   \fmfv{d.shape=circle, d.size=0.6thick}{i}
   \fmfv{d.shape=circle, d.size=0.6thick,l=$i$+1,l.a=-90}{h1}
   \fmf{plain}{i,h1}
   \fmfn{plain}{h}{10}
   \fmf{plain}{h10,j}
\end{fmfgraph*}
$\quad\Big |\quad $
\begin{fmfgraph*}(25,15)\fmfkeep{w6}
   \fmfleft{i,d1}\fmfv{l=$i$,l.a=180}{i}
   \fmfright{j,d2}
   \fmf{plain}{i,h1}
   \fmfn{plain}{h}{10}
   \fmf{plain}{h10,j}\fmfv{l=$j$,l.a=0}{j}
   \fmffreeze
   \fmf{fillwx,left}{i,h10}
   \fmf{dashes,left}{i,h10}
   \fmfv{d.shape=circle, d.size=0.6thick}{j}
   \fmfv{d.shape=circle, d.size=0.6thick,l=$j$-1,l.a=-90}{h10}
   \fmf{plain}{i,h1}
   \fmfn{plain}{h}{10}
   \fmf{plain}{h10,j}
\end{fmfgraph*}
$\quad \Big |\quad $

\begin{fmfgraph*}(35,15)\fmfkeep{w7}
   \fmfleft{i,d1}
   \fmfright{j,d2}
   \fmf{plain}{i,k1}\fmflabel{$i$}{i}
   \fmfn{plain}{k}{10}
   \fmf{plain}{k10,j}\fmflabel{$j$}{j}
   \fmffreeze
   \fmf{fillwx,left}{i,k5}
   \fmf{fillwx,left}{k6,j}
   \fmf{dashes,left}{i,k5}
   \fmf{dashes,left}{k6,j}
   \fmf{plain}{i,k1}
   \fmfn{plain}{k}{10}
   \fmf{plain}{k10,j}
   \fmfv{d.shape=circle, d.size=0.6thick,l=$k$,l.a=-100}{k5}
   \fmfv{d.shape=circle, d.size=0.6thick,l=$k$+1,l.a=-65}{k6}
\end{fmfgraph*}
\end{center}
\caption{\label{fig:simplewxrec} Recursion for $wx$ in the nested algorithm}
\end{figure}

The recursion relations used to fill the $wx$ matrix include:
single-stranded nucleotides, external pairs, external dangling bases,
and bifurcations.  The actual recursion is easier to understand by
looking at the diagrams involved (given in fig.~\ref{fig:simplewxrec})
and the recursion can be expressed as,

\begin{equation}
wx(i,j) = optimal\left\{ 
\begin{array}{ll}

\vspace{2mm}

\left.
\begin{array}{l}
P + vx(i,j)
\end{array}
\hspace{4.7cm}\right]
& 
\mbox{paired}\\

\vspace{2mm}

\left.
\begin{array}{l}
L^i_{i+1,j-1} + R^j_{i+1,j-1} + P + vx(i+1,j-1)\\
L^i_{i+1,j} + P + vx(i+1,j)\\
R^j_{i,j-1} + P + vx(i,j-1)
\end{array}
\right]
&
\mbox{dangles}\\

\vspace{2mm}

\left.
\begin{array}{l}
Q + wx(i+1,j)\label{wxrecursion}\\ 
Q + wx(i,j-1)
\end{array}
\hspace{3.99cm}\right]
& 
\parbox[b][5mm][t]{80mm}{
single 
\vspace{-2mm}

stranded}\\

\vspace{2mm}

\left.
\begin{array}{l}
wx(i,k) + wx(k+1,j) \quad [\forall k,\quad i\leq k\leq j].
\end{array}
\hspace{0.06cm}\right]
& 
\mbox{bifurcations}

\end{array}
\right.
\end{equation}

\noindent
Each line gives the formal score of one of the diagrams in
fig.~\ref{fig:simplewxrec}. The diagram on the left is calculated as
the score of the best diagram on the right.  Here $P$ is some value
that represents the score for a base pair.  $Q$ represents the score
for a single stranded nucleotide, whereas $L^i_{i+1,j}$ and
$R^j_{i,j-1}$ stand for the score of a nucleotide dangling off a base
pair of nucleotides, at the $5^\prime$-end or the $3^\prime$-end
respectively.

The recursion for $vx$ includes hairpins, bulges, internal loops, and
multiloops. But what is special about hairpins, bulges, internal
loops, and multiloops in this diagrammatic representation?  To answer
this question we have to introduce two more definitions: {\it
Surfaces} (S) and {\it Irreducible Surfaces} (IS).  

Roughly speaking a Surface is any alternating sequence of solid and
wavy lines that closes on itself.  An Irreducible Surface is a Surface
such that if one of the H-bonds (or secondary interactions) is broken
there is no other surface contained inside, that is, an IS cannot be
``reduced'' to any other surface. (ISs are similar to the ``k-loops''
defined by Sankoff (1985).)  The {\it order}, ${\cal O}$, of an IS is
given by the number of wavy lines (secondary interactions), which is
equal to the number of solid-line intervals. It is easy to see that
hairpin loops constitute the ISs of ${\cal O}(1)$; stems, bulges and
internal loops are all the ISs of ${\cal O}(2)$, and what are referred
to in the literature as ``multiloops'' are the ISs of ${\cal O} > 2$.

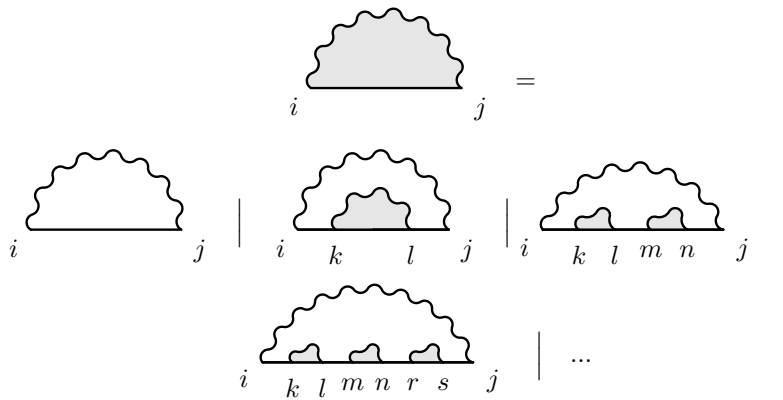
\begin{figure}[ht]
\begin{center}
\begin{fmfgraph*}(25,15)\fmfkeep{vx}
   \fmfleft{i,d1}
   \fmfright{j,d2}
   \fmffreeze
   \fmf{fillvx,left}{i,j}
   \fmf{photon,left}{i,j}
   \fmf{plain}{i,j}\fmflabel{$i$}{i}\fmflabel{$j$}{j}
\end{fmfgraph*}
\quad=\quad
\end{center}

\begin{center}
\begin{fmfgraph*}(25,15)\fmfkeep{vIS1}
   \fmfleft{i,d1}
   \fmfright{j,d2}
   \fmf{plain}{i,j}\fmflabel{$i$}{i}\fmflabel{$j$}{j}
   \fmffreeze
   \fmf{photon,left}{i,j}
\end{fmfgraph*}
$\quad\Big |\quad $
\begin{fmfgraph*}(25,15)\fmfkeep{vIS2}
   \fmfleft{i,d1}
   \fmfright{j,d2}
   \fmf{plain}{i,k,h,l,j}
   \fmflabel{$i$}{i}\fmflabel{$j$}{j}
   \fmfv{l=$k$,l.a=-90}{k}\fmfv{l=$l$,l.a=-90}{l}
   \fmffreeze
   \fmf{fillvx,left}{k,l}
   \fmf{photon,left}{i,j}
   \fmf{photon,left}{k,l}
   \fmf{plain}{i,k,h,l,j}
\end{fmfgraph*}
$\quad\Big |$
\begin{fmfgraph*}(30,15)\fmfkeep{vIS3}
   \fmfleft{i,d1}
   \fmfright{j,d2}
   \fmf{plain}{i,k,l,m,n,j}\fmflabel{$i$}{i}\fmflabel{$j$}{j}
   \fmfv{l=$k$,l.a=-90}{k}\fmfv{l=$l$,l.a=-90}{l}
   \fmfv{l=$m$,l.a=-90}{m}\fmfv{l=$n$,l.a=-90}{n}
   \fmffreeze
   \fmf{fillvx,left}{k,l}
   \fmf{fillvx,left}{m,n}
   \fmf{photon,left=0.7}{i,j}
   \fmf{photon,left}{k,l}
   \fmf{photon,left}{m,n}
   \fmf{plain}{i,k,l,m,n,j}
\end{fmfgraph*}
$\quad\Big |\quad $

\begin{fmfgraph*}(35,15)\fmfkeep{vIS4}
   \fmfleft{i,d1}
   \fmfright{j,d2}
   \fmf{plain}{i,k,l,m,n,r,s,j}\fmflabel{$i$}{i}\fmflabel{$j$}{j}
   \fmfv{l=$k$,l.a=-90}{k}\fmfv{l=$l$,l.a=-90}{l}
   \fmfv{l=$m$,l.a=-90}{m}\fmfv{l=$n$,l.a=-90}{n}
   \fmfv{l=$r$,l.a=-90}{r}\fmfv{l=$s$,l.a=-90}{s}
   \fmffreeze
   \fmf{fillvx,left}{k,l}
   \fmf{fillvx,left}{m,n}
   \fmf{fillvx,left}{r,s}
   \fmf{photon,left=0.7}{i,j}
   \fmf{photon,left}{k,l}
   \fmf{photon,left}{m,n}
   \fmf{photon,left}{r,s}
   \fmf{plain}{i,k,l,m,n,r,s,j}
\end{fmfgraph*}
$\quad\Big |\quad  ... $
\end{center}
\caption{\label{fig:simplevxrec} General recursion for $vx$ in the nested algorithm.}
\end{figure}
 
The actual recursion for $vx$ is given in fig.~\ref{fig:simplevxrec},
and can be expressed as,

\begin{equation}
vx(i,j) = optimal\left\{ 
\begin{array}{l}
IS^1(i,j)\\
IS^2(i,j:k,l) + vx(k,l)\nonumber\\
IS^3(i,j:k,l:m,n) + vx(k,l) + vx(m,n)\label{vxrecursion}\\
IS^4(i,j:k,l:m,n:r,s) + vx(k,l) \\
\qquad+ vx(m,n) + vx(r,s)\\
{\cal O} (IS^5)\\
\end{array}\right.
\end{equation}
\[
[\forall k,l,m,n,r,s,\quad i\leq k\leq l\leq r\leq s\leq m\leq n\leq j]
\]
This recursion is an expansion in ISs of successively higher order.
Here $IS^n(i_1,j_1:i_2,j_2:...:i_n,j_n)$ represent the score for an IS
of order n, in which $i_k$ is paired to $j_k$. This general algorithm
is quite impractical, because each IS of ${\cal O}(\gamma)$ adds a
complexity of ${\cal O}(N^{2\gamma})$ to the calculation. [An IS of
${\cal O}(N^{2\gamma})$ requires us to search through $2\gamma$
independent segments in the entire sequence of N nucleotides.]  To
make it useful we have to truncate the expansion in IS's at some order
in the recursion for $vx$ in fig.~\ref{fig:simplevxrec}.  The symbol
${\cal O} (IS^5)$ indicates the order of IS at which we truncate the
recursion. (Note that the recursion for $wx$ will remain always the
same.)

These recursions are equivalent to those proposed by Sankoff (1985) in
Theorem 2.  Notice also that in defining the recursive algorithm we
have not yet had to specify anything about the particular manner in
which the contribution from different IS's are calculated in order to
obtain the most optimal folding.

\begin{figure}[ht]
\begin{center}
\begin{fmfgraph*}(25,15)\fmfkeep{vx}
   \fmfleft{i,d1}
   \fmfright{j,d2}
   \fmf{plain}{i,j}\fmflabel{$i$}{i}\fmflabel{$j$}{j}
   \fmffreeze
   \fmf{fillvx,left}{i,j}
   \fmf{photon,left}{i,j}
   \fmf{plain}{i,j}
\end{fmfgraph*}
\quad =\quad
\begin{fmfgraph*}(25,15)\fmfkeep{vo(0)}
   \fmfleft{i,d1}
   \fmfright{j,d2}
   \fmf{plain}{i,h1}\fmfv{l=$i$,l.a=180}{i}
   \fmfn{plain}{h}{10}
   \fmf{plain}{h10,j}\fmfv{l=$j$,l.a=0}{j}
   \fmffreeze
   \fmf{fillwx,left}{h1,h10}
   \fmf{photon,left}{i,j}
   \fmf{dashes,left}{h1,h10}
   \fmfv{d.shape=circle, d.size=0.6thick}{i}
   \fmfv{d.shape=circle, d.size=0.6thick}{j}
   \fmfv{d.shape=circle, d.size=0.6thick,l=$i$+1,l.a=-70}{h1}
   \fmfv{d.shape=circle, d.size=0.6thick,l=$j$-1,l.a=-100}{h10}
   \fmf{plain}{i,h1}
   \fmfn{plain}{h}{10}
   \fmf{plain}{h10,j}
\end{fmfgraph*}
\end{center}
\caption{\label{fig:vo(0)} Recursion for $vx$ truncated at ${\cal O} (0)$}
\end{figure}
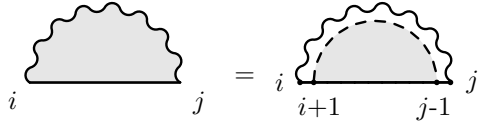
The simplest truncation is to stop at order zero. In
this approximation none of the ISs (hairpin, bulge, internal loop...)
are given any specialized scores. We only have to provide a specific
score for a base pair, B.  The recursion for $vx$ then simplifies to
fig.~\ref{fig:vo(0)}, and can be cast into the form,

\begin{equation}
vx(i,j) = B + wx(i+1,j-1).\label{vxnussinov}
\end{equation}

If we set $B=P=+1$, and $Q=0$ in equation (\ref{wxrecursion}) then we
have the Nussinov algorithm (Nussinov {\it et al.}, 1978).  This
simple algorithm calculates the folding with the maximum number of
base pairs.

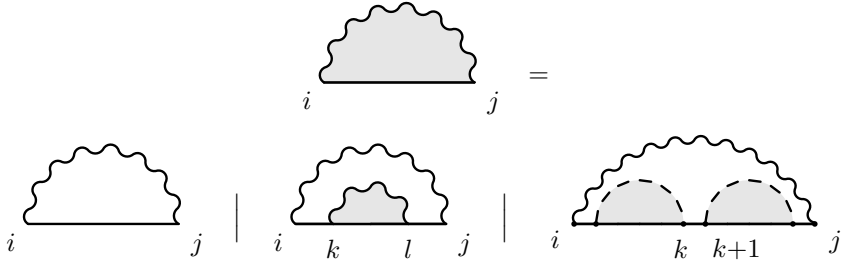
\begin{figure}[ht]
\begin{center}
\begin{fmfgraph*}(25,15)\fmfkeep{vx}
   \fmfleft{i,d1}
   \fmfright{j,d2}
   \fmf{plain}{i,j}
   \fmffreeze
   \fmf{fillvx,left}{i,j}
   \fmflabel{$i$}{i}\fmflabel{$j$}{j}
   \fmf{photon,left}{i,j}
   \fmf{plain}{i,j}
\end{fmfgraph*}
\quad=\quad
\end{center}

\begin{center}
\begin{fmfgraph*}(25,15)\fmfkeep{vis1}
   \fmfleft{i,d1}
   \fmfright{j,d2}
   \fmf{plain}{i,j}\fmflabel{$i$}{i}\fmflabel{$j$}{j}
   \fmffreeze
   \fmf{photon,left}{i,j}
\end{fmfgraph*}
$\quad\Big |\quad$
\begin{fmfgraph*}(25,15)\fmfkeep{vIS2}
   \fmfleft{i,d1}
   \fmfright{j,d2}
   \fmf{plain}{i,k,h,l,j}
   \fmflabel{$i$}{i}\fmflabel{$j$}{j}
   \fmfv{l=$k$,l.a=-90}{k}\fmfv{l=$l$,l.a=-90}{l}
   \fmffreeze
   \fmf{fillvx,left}{k,l}
   \fmf{photon,left}{i,j}
   \fmf{photon,left}{k,l}
   \fmf{plain}{i,k,h,l,j}
\end{fmfgraph*}
$\quad\Big |\quad$
\begin{fmfgraph*}(40,15)\fmfkeep{vrest}
   \fmfleft{i,d1}
   \fmfright{j,d2}
   \fmf{plain}{i,h1}\fmflabel{$i$}{i}
   \fmfn{plain}{h}{10}
   \fmf{plain}{h10,j}\fmflabel{$j$}{j}
   \fmffreeze
   \fmf{fillwx,left}{h1,h5}
   \fmf{fillwx,left}{h6,h10}
   \fmf{photon,left=0.7}{i,j}
   \fmf{dashes,left}{h1,h5}
   \fmf{dashes,left}{h6,h10}
   \fmfv{d.shape=circle, d.size=0.6thick}{i}
   \fmfv{d.shape=circle, d.size=0.6thick}{j}
   \fmfv{d.shape=circle, d.size=0.6thick}{h1}
   \fmfv{d.shape=circle, d.size=0.6thick}{h10}
   \fmfv{d.shape=circle, d.size=0.6thick,l=$k$,l.a=-100}{h5}
   \fmfv{d.shape=circle, d.size=0.6thick,l=$k$+1,l.a=-65}{h6}
   \fmf{plain}{i,h1}
   \fmfn{plain}{h}{10}
   \fmf{plain}{h10,j}
\end{fmfgraph*}
\end{center}
\caption{\label{fig:vo(2)} Recursion for $vx$ truncated at ${\cal O} (2)$.}
\end{figure}

The next order of complexity we explore corresponds to a truncation at
ISs of ${\cal O}(2)$. 
Hairpin loops, bulges, stems, and internal loops are treated with
precision by the scoring functions $IS^1$ and $IS^2$. The rest of ISs,
collected under the name of ``multiloops''---which are much less
frequent than the previous---are described in an approximate form.
The diagrams of this approximation are given in fig.~\ref{fig:vo(2)},
and correspond to,

\begin{equation}
vx(i,j) = optimal\left\{ 
\begin{array}{l}
IS^1(i,j) \\
IS^2(i,j:k,l) + vx(k,l)\label{vxIS2}\\
P_I + M + wx_I(i+1,k) + wx_I(k+1,j-1)\\
\end{array}\right.
\end{equation}
\[[\forall k,l\quad i\leq k\leq l\leq j]\]

\noindent
This is the algorithm described by Sankoff (1985) in theorem 3.  This
is the approximation that MFOLD (Zuker, 1981) and {\sl ViennaRNA}
(Schuster {\it et al.}, 1994) implement.  $P_I$ stands for the scoring
parameter for a pair in a multiloop, and parameter $M$ stands for the
score for a multiloop.  Note that the matrix $wx_I$ used to truncate
the recursion for $vx$ in (\ref{vxIS2}) does not have to be the same
as the one used in (\ref{wxrecursion}).  Matrix $wx_I$ is used
exclusively for diagrams which will be incorporated into multiloops.
Although both matrices $wx$ and $wx_I$ have similar recursions, the
parameters of these two recursions will have in general different
values (represent here by $P$, $Q$, $L$, $R$ and $P_I$, $Q_I$, $L_I$,
$R_I$ respectively). This feature is implemented both in MFOLD and in
our program.

Higher orders of specificity of the general algorithm are possible,
but are certainly more time consuming, and they have not been explored
so far. One reason for this relative lack of development is that there
is little information about the energetic properties of
multiloops. The generalized nested algorithm provides a way to unify
the currently available dynamic algorithms for RNA folding. At a given
order, the error of the approximation is given by the difference
between the assigned score to ``multiloops'' and the precise score
that one of those higher order ISs deserves.

\subsection{Description of the pseudoknot algorithm}

\begin{figure}
\begin{center}
\begin{picture}(50,60)

\put(0.0,26.0){$5^\prime$--}
\put(7.5,33){\line(0,-1){3}\makebox(0,0)[b]{29}}
\put(6.0,26.0){A}
\put(8.5,20.0){\shortstack{UUCCG\\
                           $\bullet\hspace{0.4mm}\bullet\hspace{0.4mm}
                            \bullet\hspace{0.4mm}\bullet\hspace{0.4mm}\bullet$\\
                            AGGGC}}
\put(20.5,29.0){\line(0,1){4}}
\put(20.5,33.0){\line(1,0){9}}
\put(32.5,33.0){\line(1,0){10}}
\put(42.5,21.0){\line(0,1){12}}
\put(42.5,21.0){\line(-1,0){2.5}}
\put(5.5,21.0){\line(1,0){3}}
\put(5.5,11.0){\line(0,1){10}}
\put(5.5,11.0){\line(1,0){6}}
\put(20.5,11.0){\line(1,0){5.5}}
\put(26.0,11.0){\line(0,1){2}}

\put(30.0,32){A}
\put(42.0,10){\line(0,1){3}\makebox(0,0)[s]{1}}
\put(22.0,13.5){\shortstack[l]{AACUCGA\\
                           $\hspace{3.0mm}\bullet\hspace{0.4mm}\bullet\hspace{0.4mm}
                            \bullet\hspace{0.4mm}\bullet\hspace{0.2mm}\bullet\hspace{0.4mm}\bullet$\\
                            $\hspace{3.0mm}$UGAGCUA -- $3^\prime$}}
\put(12.0,10.0){A}\put(15.0,10.0){A}\put(18.0,10.0){A}
\put(48,20){=}
\end{picture}
\end{center}

\begin{center}
\begin{picture}(45,35)
\put(0.0,26.0){$5^\prime$--}
\put(7.0,33){\line(0,-1){3}\makebox(0,0)[b]{29}}
\put(6.0,26.0){A}
\put(8.5,20.0){\shortstack{UUCCG\\
                           $\bullet\hspace{0.4mm}\bullet\hspace{0.4mm}
                            \bullet\hspace{0.4mm}\bullet\hspace{0.4mm}\bullet$\\
                            AGGGC}}
\put(20.5,29.0){\line(0,1){4}}
\put(20.5,33.0){\line(1,0){9}}
\put(32.5,33.0){\line(1,0){8}}
\put(40.5,21.0){\line(0,1){12}}
\put(5.5,21.0){\line(1,0){3}}

\put(30.0,32){A}
\put(31.0,28){\line(0,1){3}\makebox(0,0)[t]{23}}
\put(9.5,16){\line(0,1){3}\makebox(0,0)[t]{11}}
\put(20.5,16){\line(0,1){3}\makebox(0,0)[t]{15}}
\put(44,20){+}

\end{picture}
\begin{picture}(45,35)
\put(20.5,11.0){\line(1,0){5.5}}
\put(26.0,11.0){\line(0,1){2}}
\put(5.5,11.0){\line(0,1){10}}
\put(5.5,11.0){\line(1,0){6}}
\put(40.5,21.0){\line(1,0){3}}
\put(23.0,26){\line(0,-1){3}\makebox(0,0)[b]{16}}
\put(39.0,26){\line(0,-1){3}\makebox(0,0)[b]{22}}

\put(42.0,10){\line(0,1){3}\makebox(0,0)[s]{1}}
\put(22.0,13.5){\shortstack[l]{AACUCGA\\
                           $\hspace{3.0mm}\bullet\hspace{0.4mm}\bullet\hspace{0.4mm}
                            \bullet\hspace{0.4mm}\bullet\hspace{0.2mm}\bullet\hspace{0.4mm}\bullet$\\
                            $\hspace{3.0mm}$UGAGCUA -- $3^\prime$}}
\put(12.0,10.0){A}\put(15.0,10.0){A}\put(18.0,10.0){A}
\put(13.0,6.0){\line(0,1){3}\makebox(0,0)[t]{10}}
\put(50,20){=}

\end{picture}
\end{center}
\begin{center}

\begin{fmfgraph*}(45,35)
   \fmfleft{i1,d1}
   \fmfright{i19,d2}
   \fmfn{phantom}{i}{19}
   \fmffreeze
   \fmf{fillwx,left}{i1,i19}
   \fmf{unfillwx,left}{i7,i15}
   \fmf{plain}{i1,i7}\fmfv{l=$29$,l.a=-90}{i1}\fmfv{l=$23$,l.a=-70}{i6}
   \fmf{plain}{i15,i19}\fmfv{l=$15$,l.a=-90}{i15}\fmfv{l=$11$,l.a=-70}{i19}
   \fmf{dashes, left}{i1,i19}
   \fmf{dashes, left}{i7,i15}
\end{fmfgraph*}
$\quad +\quad$
\begin{fmfgraph*}(45,15)
   \fmfleft{i1,d1}
   \fmfright{i22,d2}
   \fmfn{phantom}{i}{22}
   \fmffreeze
   \fmf{fillwx,left}{i1,i22}
   \fmf{unfillwx,left}{i7,i13}
   \fmf{plain}{i1,i7}\fmfv{l=$22$,l.a=-90}{i1}\fmfv{l=$16$,l.a=-70}{i7}
   \fmf{plain}{i13,i22}\fmfv{l=$10$,l.a=-90}{i13}\fmfv{l=$1$,l.a=-70}{i22}
   \fmf{dashes, left}{i1,i22}
   \fmf{dashes, left}{i7,i13}
\end{fmfgraph*}
$\quad =$
\end{center}

\begin{center}
\begin{fmfgraph*}(45,25)
   \fmfleft{i1,d1}
   \fmfright{i29,d2}
   \fmfn{phantom}{i}{29}
   \fmffreeze
   \fmf{fillwx,left}{i1,i19}
   \fmf{unfillwx,left}{i7,i15}
   \fmf{fillwx,right}{i8,i29}
   \fmf{unfillwx,right}{i14,i20}
   \fmf{dashes, left}{i1,i19}
   \fmf{dashes, left}{i7,i15}
   \fmf{dashes, right}{i8,i29}
   \fmf{dashes, right}{i14,i20}
   \fmf{plain}{i1,i7}\fmfv{l=$29$,l.a=-90}{i1}\fmfv{l=$23$,l.a=-112}{i7}
   \fmf{plain}{i8,i14}
   \fmf{plain}{i15,i19}
   \fmf{plain}{i20,i29}\fmfv{l=$10$,l.a=60}{i20}\fmfv{l=$1$,l.a=60}{i29}
\end{fmfgraph*}
\bigskip\bigskip
\end{center}
\caption{\label{HIVRT_const}
Construction of a simple pseudoknot using two gap matrices.}
\end{figure}
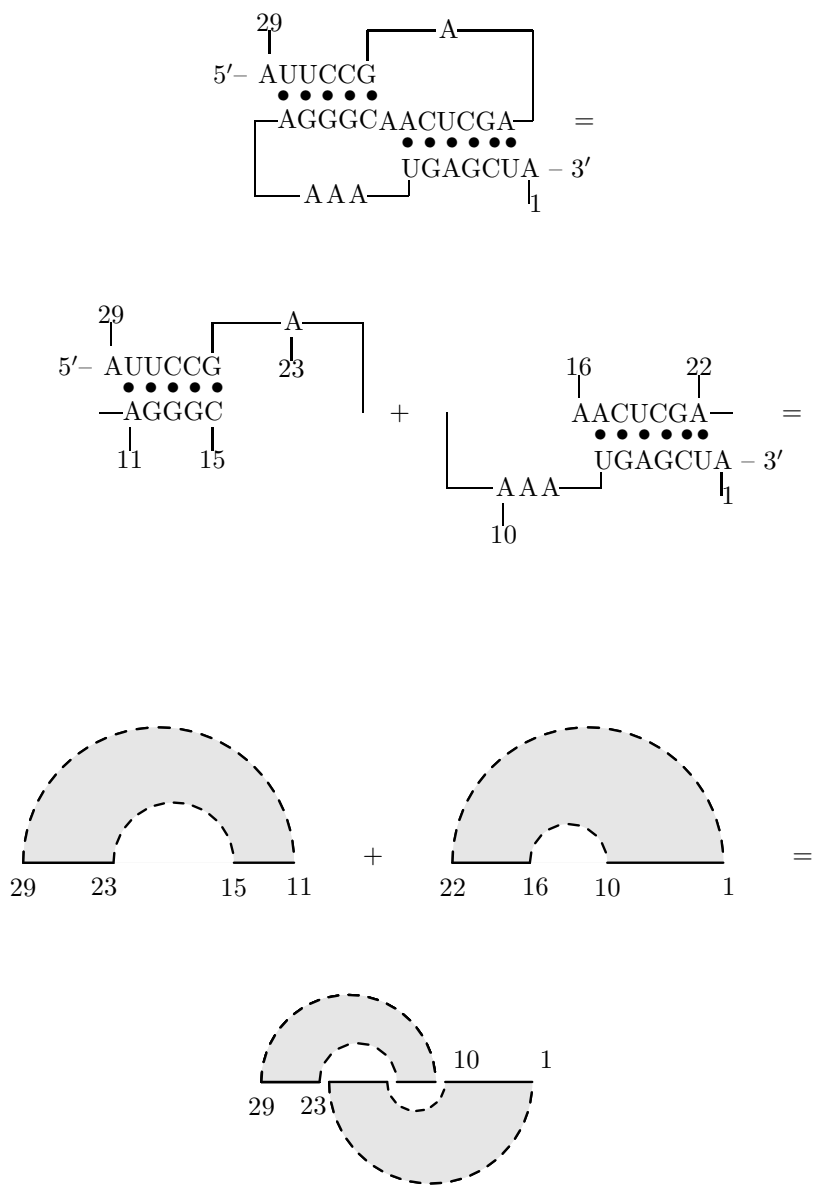

Pseudoknots are non-nested configurations and clearly cannot be
described with just the $wx$ and $vx$ matrices we introduced in the
previous section. The key point of the pseudoknot algorithm is the use
of {\it gap matrices} in addition to the $wx$ and $vx$ matrices.
Looking at the graphical representation of one of the simplest
pseudoknots, fig.~\ref{HIVRT_const}, we can see that we could describe
such a configuration by putting together two gap matrices with
complementary holes.

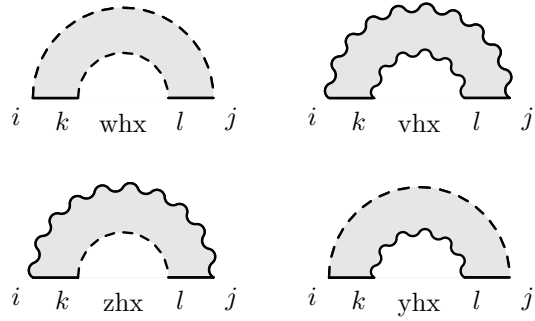
\begin{figure}
\begin{center}
\begin{fmfgraph*}(30,20)\fmfkeep{whx}
   \fmfleft{i,d1}
   \fmfright{j,d2}
   \fmf{plain}{i,k}\fmflabel{$i$}{i}\fmflabel{$k$}{k}
   \fmf{phantom}{k,m,l}
   \fmf{plain}{l,j}\fmflabel{$l$}{l}\fmflabel{$j$}{j}
   \fmffreeze
   \fmf{fillwx,left}{i,j}
   \fmf{unfillwx,left}{k,l}
   \fmf{dashes,left}{i,j}
   \fmf{dashes,left}{k,l}
   \fmfv{l=whx,l.a=-90}{m}
   \fmf{plain}{i,k}\fmf{plain}{l,j}
\end{fmfgraph*}
$\qquad$
\begin{fmfgraph*}(30,20)\fmfkeep{vhx}
   \fmfleft{i,d1}
   \fmfright{j,d2}
   \fmf{plain}{i,k}\fmflabel{$i$}{i}\fmflabel{$k$}{k}
   \fmf{phantom}{k,m,l}
   \fmf{plain}{l,j}\fmflabel{$l$}{l}\fmflabel{$j$}{j}
   \fmffreeze
   \fmf{fillvx,left}{i,j}
   \fmf{unfillvx,left}{k,l}
   \fmf{photon,left}{i,j}
   \fmf{photon,left}{k,l}
   \fmfv{l=vhx,l.a=-90}{m}
   \fmf{plain}{i,k}\fmf{plain}{l,j}
\end{fmfgraph*}
\end{center}

\begin{center}
\begin{fmfgraph*}(30,20)\fmfkeep{zhx}
   \fmfleft{i,d1}
   \fmfright{j,d2}
   \fmf{plain}{i,k}\fmflabel{$i$}{i}\fmflabel{$k$}{k}
   \fmf{phantom}{k,m,l}
   \fmf{plain}{l,j}\fmflabel{$l$}{l}\fmflabel{$j$}{j}
   \fmffreeze
   \fmf{fillvx,left}{i,j}
   \fmf{unfillwx,left}{k,l}
   \fmf{photon,left}{i,j}
   \fmf{dashes,left}{k,l}
   \fmfv{l=zhx,l.a=-90}{m}
   \fmf{plain}{i,k}\fmf{plain}{l,j}
\end{fmfgraph*}
$\qquad$
\begin{fmfgraph*}(30,20)\fmfkeep{yhx}
   \fmfleft{i,d1}
   \fmfright{j,d2}
   \fmf{plain}{i,k}\fmflabel{$i$}{i}\fmflabel{$k$}{k}
   \fmf{phantom}{k,m,l}
   \fmf{plain}{l,j}\fmflabel{$l$}{l}\fmflabel{$j$}{j}
   \fmffreeze
   \fmf{fillwx,left}{i,j}
   \fmf{unfillvx,left}{k,l}
   \fmf{dashes,left}{i,j}
   \fmf{photon,left}{k,l}
   \fmfv{l=yhx,l.a=-90}{m}
   \fmf{plain}{i,k}\fmf{plain}{l,j}
\end{fmfgraph*}
\end{center}
\caption{\label{fig:gapmatrices} Representation of the gap matrices used in the
algorithm for pseudoknots.}
\end{figure}

The pseudoknot dynamic programming algorithm uses one-hole or gap
matrices (fig.~\ref{fig:gapmatrices}) as a generalization of the $wx$
and $vx$ matrices.  Let us define $whx(i,j:k,l)$ as the graph that
describes the best folding that connects segments $[i,k]$ with
$[l,j]$, $i\leq k \leq l \leq j$ such that the relation between $i$
and $j$ and $k$ and $l$ is undetermined.  Similarly we define
$vhx(i,j:k,l)$ as the graph that describes the best folding that
connects segments $[i,k]$ with $[l,j]$, $i\leq k \leq l \leq j$ such
that $i$ and $j$ are base-paired and $k$ and $l$ are also base-paired.
For completeness we have to introduce also matrix $yhx(i,j:k,l)$ in
which $k$ and $l$ are paired, but the relation between $i$ and $j$ is
undetermined, and its counterpart $zhx(i,j:k,l)$ in which $i$ and $j$
are paired, but the relation between $k$ and $l$ is undetermined.

\begin{table}[ht]
\begin{center}
\begin{tabular}{|c|c|c|}\hline
matrix & relationship & relationship\\
($i\leq k\leq l\leq j$) &$i$, $j$ & $k$, $l$\\
\hline
$wx(i, j)$ & undetermined & ---\\
$vx(i, j)$ & paired & ---\\
\hline
$whx(i, j: k, l)$ & undetermined & undetermined\\
$vhx(i, j: k, l)$ & paired & paired\\
$zhx(i, j: k, l)$ & paired & undetermined\\
$yhx(i, j: k, l)$ & undetermined & paired\\
\hline
\end{tabular}
\medskip\end{center}
\caption{\label{matrices}
Specifications of the matrices used in the pseudoknot algorithm.}
\end{table}

The non-gap matrices $wx$, $vx$ are contained
as a particular case of the gap matrices.
When there is no hole, $k=l-1$, then by construction,
\begin{eqnarray}
whx(i,j:k,k+1) &=& wx(i,j)\label{gap-nongap}\\
zhx(i,j:k,k+1) &=& vx(i,j)\quad\forall k, \quad i\leq k\leq j.\nonumber
\end{eqnarray}

We have the gap matrices as the building blocks of the algorithm, but
how do we establish a consistent and complete recursion relation?
Here is where the analogy between the gap matrices and the Feynman
diagrams of quantum field theory was of great help (Bjorken \& Drell
1965).\footnote{
More precisely, the analogy is more cleanly expressed in terms of 
Schwinger-Dyson diagrams which in QFT are used to represent full
interacting vertices and propagators recursively in terms of 
elementary interactions.}

\begin{figure}
\begin{center}
\begin{fmfgraph*}(25,15)\fmfkeep{wx}
   \fmfleft{i,d1}
   \fmfright{j,d2}
   \fmffreeze
   \fmf{fillwx,left}{i,j}
   \fmf{plain}{i,j}\fmflabel{$i$}{i}\fmflabel{$j$}{j}
   \fmf{dashes,left}{i,j}
\end{fmfgraph*}
\quad=\quad 
\end{center}

\begin{center}
\begin{fmfgraph*}(25,15)\fmfkeep{vx}
   \fmfleft{i,d1}
   \fmfright{j,d2}
   \fmffreeze
   \fmf{fillvx,left}{i,j}
   \fmf{plain}{i,j}\fmflabel{$i$}{i}\fmflabel{$j$}{j}
   \fmf{photon,left}{i,j}
\end{fmfgraph*}
$\quad\Big |\quad$
\begin{fmfgraph*}(25,15)\fmfkeep{w2}
   \fmfleft{i,d1}\fmfv{l=$i$,l.a=180}{i}
   \fmfright{j,d2}
   \fmf{plain}{i,h1}
   \fmfn{plain}{h}{10}
   \fmf{plain}{h10,j}\fmfv{l=$j$,l.a=0}{j}
   \fmffreeze
   \fmf{fillvx,left}{h1,h10}
   \fmf{photon,left}{h1,h10}
   \fmfv{d.shape=circle, d.size=0.6thick}{i}
   \fmfv{d.shape=circle, d.size=0.6thick}{j}
   \fmfv{d.shape=circle, d.size=0.6thick,l=$i$+1,l.a=-70}{h1}
   \fmfv{d.shape=circle, d.size=0.6thick,l=$j$-1,l.a=-100}{h10}
\end{fmfgraph*}
$\quad\Big |\quad$
\begin{fmfgraph*}(25,15)\fmfkeep{w3}
   \fmfleft{i,d1}\fmfv{l=$i$,l.a=180}{i}
   \fmfright{j,d2}
   \fmf{plain}{i,h1}
   \fmfn{plain}{h}{10}
   \fmf{plain}{h10,j}\fmfv{l=$j$,l.a=0}{j}
   \fmffreeze
   \fmf{fillvx,left}{h1,j}
   \fmf{photon,left}{h1,j}
   \fmfv{d.shape=circle, d.size=0.6thick}{i}
   \fmfv{d.shape=circle, d.size=0.6thick,l=$i$+1,l.a=-70}{h1}
\end{fmfgraph*}
$\quad\Big |\quad$

\begin{fmfgraph*}(25,15)\fmfkeep{w4}
   \fmfleft{i,d1}\fmfv{l=$i$,l.a=180}{i}
   \fmfright{j,d2}
   \fmf{plain}{i,h1}
   \fmfn{plain}{h}{10}
   \fmf{plain}{h10,j}\fmfv{l=$j$,l.a=0}{j}
   \fmffreeze
   \fmf{fillvx,left}{i,h10}
   \fmf{photon,left}{i,h10}
   \fmf{plain}{i,h1}
   \fmfn{plain}{h}{10}
   \fmf{plain}{h10,j}
   \fmfv{d.shape=circle, d.size=0.6thick}{j}
   \fmfv{d.shape=circle, d.size=0.6thick,l=$j$-1,l.a=-100}{h10}
\end{fmfgraph*}
$\quad\Big |\quad$
\begin{fmfgraph*}(25,15)\fmfkeep{w5}
   \fmfleft{i,d1}\fmfv{l=$i$,l.a=180}{i}
   \fmfright{j,d2}
   \fmf{plain}{i,h1}
   \fmfn{plain}{h}{10}
   \fmf{plain}{h10,j}\fmfv{l=$j$,l.a=0}{j}
   \fmffreeze
   \fmf{fillwx,left}{h1,j}
   \fmf{dashes,left}{h1,j}
   \fmfv{d.shape=circle, d.size=0.6thick}{i}
   \fmfv{d.shape=circle, d.size=0.6thick,l=$i$+1,l.a=-70}{h1}
\end{fmfgraph*}
$\quad\Big |\quad $
\begin{fmfgraph*}(25,15)\fmfkeep{w6}
   \fmfleft{i,d1}\fmfv{l=$i$,l.a=180}{i}
   \fmfright{j,d2}
   \fmf{plain}{i,h1}
   \fmfn{plain}{h}{10}
   \fmf{plain}{h10,j}\fmfv{l=$j$,l.a=0}{j}
   \fmffreeze
   \fmf{fillwx,left}{i,h10}
   \fmf{dashes,left}{i,h10}
   \fmf{plain}{i,h1}
   \fmfn{plain}{h}{10}
   \fmf{plain}{h10,j}   
   \fmfv{d.shape=circle, d.size=0.6thick}{j}
   \fmfv{d.shape=circle, d.size=0.6thick,l=$j$-1,l.a=-70}{h10}
\end{fmfgraph*}
$\quad \Big |\quad $

\begin{fmfgraph*}(30,15)\fmfkeep{w7}
   \fmfleft{i,d1}
   \fmfright{j,d2}
   \fmf{plain}{i,k1}\fmflabel{$i$}{i}
   \fmfn{plain}{k}{10}
   \fmf{plain}{k10,j}\fmflabel{$j$}{j}
   \fmffreeze
   \fmf{fillwx,left}{i,k5}
   \fmf{fillwx,left}{k6,j}
   \fmf{dashes,left}{i,k5}
   \fmf{dashes,left}{k6,j}
   \fmf{plain}{i,k1}
   \fmfn{plain}{k}{10}
   \fmf{plain}{k10,j}
   \fmfv{d.shape=circle, d.size=0.6thick,l=$k$,l.a=-100}{k5}
   \fmfv{d.shape=circle, d.size=0.6thick,l=$k$+1,l.a=-65}{k6}
\end{fmfgraph*}
$\quad \Big |\quad $
\begin{fmfgraph*}(30,15)\fmfkeep{w9}
   \fmfleft{i,d1}
   \fmfright{j,d2}
   \fmf{plain}{i,k1}\fmflabel{$i$}{i}
   \fmfn{plain}{k}{20}
   \fmf{plain}{k20,j}\fmflabel{$j$}{j}
   \fmffreeze
   \fmf{fillwx,left=0.7}{i,k15}
   \fmf{unfillwx,left=0.7}{k5,k11}
   \fmf{fillwx,right=0.7}{k6,j}
   \fmf{unfillwx,right=0.7}{k10,k16}
   \fmf{dashes,left=0.7}{i,k15}
   \fmf{dashes,left=0.7}{k5,k11}
   \fmf{dashes,right=0.7}{k6,j}
   \fmf{dashes,right=0.7}{k10,k16}
   \fmf{plain}{i,k1}
   \fmfn{plain}{k}{20}
   \fmf{plain}{k20,j}
   \fmfv{d.shape=circle, d.size=0.6thick}{k5}
   \fmfv{d.shape=circle, d.size=0.6thick}{k6}
   \fmfv{d.shape=circle, d.size=0.6thick}{k10}
   \fmfv{d.shape=circle, d.size=0.6thick}{k11}
   \fmfv{d.shape=circle, d.size=0.6thick}{k15}
   \fmfv{d.shape=circle, d.size=0.6thick}{k16}
\end{fmfgraph*}
\end{center}
\caption{\label{fig:wxrec} Recursion for $wx$ in the pseudoknot algorithm
truncated at ${\cal O} (whx+whx+whx)$.}
\end{figure}
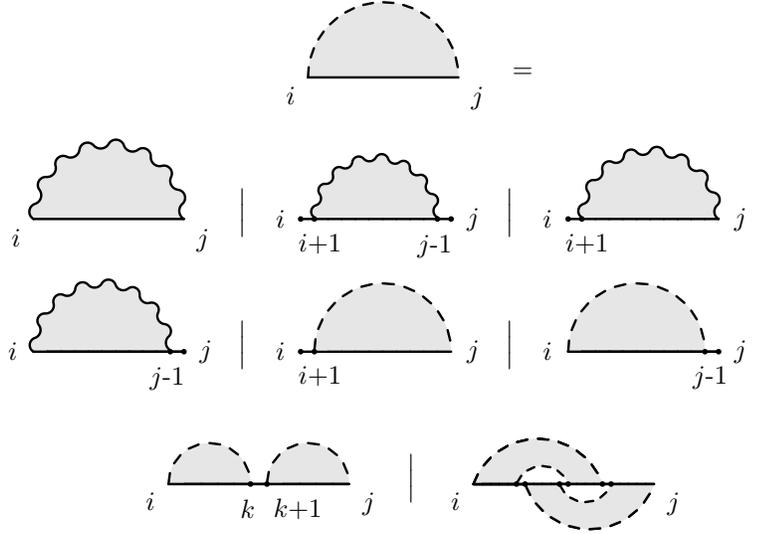

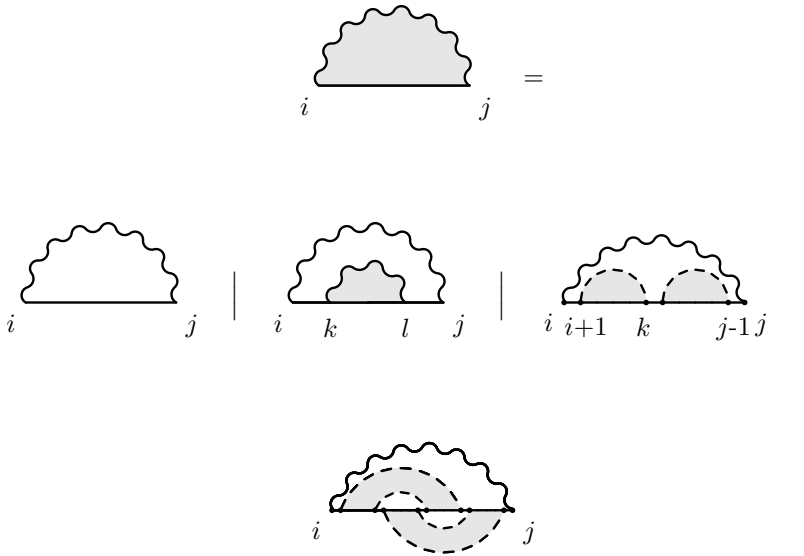
\begin{figure}
\begin{center}
\begin{fmfgraph*}(25,25)\fmfkeep{vx}
   \fmfleft{i,d1}
   \fmfright{j,d2}
   \fmf{plain}{i,j}\fmflabel{$i$}{i}\fmflabel{$j$}{j}
   \fmffreeze
   \fmf{fillvx,left}{i,j}
   \fmf{photon,left}{i,j}
   \fmf{plain}{i,j}
\end{fmfgraph*}
\quad=\quad
\end{center}

\begin{center}
\begin{fmfgraph*}(25,25)\fmfkeep{v1}
   \fmfleft{i,d1}
   \fmfright{j,d2}
   \fmf{plain}{i,j}\fmflabel{$i$}{i}\fmflabel{$j$}{j}
   \fmffreeze
   \fmf{photon,left}{i,j}
\end{fmfgraph*}
$\quad\Big |\quad$
\begin{fmfgraph*}(25,25)\fmfkeep{v2}
   \fmfleft{i,d1}
   \fmfright{j,d2}
   \fmf{plain}{i,k,h,l,j}
   \fmflabel{$i$}{i}\fmflabel{$j$}{j}
   \fmfv{l=$k$,l.a=-90}{k}\fmfv{l=$l$,l.a=-90}{l}
   \fmffreeze
   \fmf{fillvx,left}{k,l}
   \fmf{photon,left}{i,j}
   \fmf{photon,left}{k,l}
   \fmf{plain}{i,k,h,l,j}
\end{fmfgraph*}
$\quad\Big |\quad$
\begin{fmfgraph*}(30,25)\fmfkeep{v3}
   \fmfleft{i,d1}
   \fmfright{j,d2}
   \fmf{plain}{i,h1}\fmflabel{$i$}{i}
   \fmfn{plain}{h}{10}
   \fmf{plain}{h10,j}\fmflabel{$j$}{j}
   \fmffreeze
   \fmf{fillwx,left}{h1,h5}
   \fmf{fillwx,left}{h6,h10}
   \fmf{photon,left=0.7}{i,j}
   \fmf{dashes,left}{h1,h5}
   \fmf{dashes,left}{h6,h10}
   \fmfv{d.shape=circle, d.size=0.6thick}{i}
   \fmfv{d.shape=circle, d.size=0.6thick}{j}
   \fmfv{d.shape=circle, d.size=0.6thick,l=$i$+1,l.a=-70}{h1}
   \fmfv{d.shape=circle, d.size=0.6thick,l=$j$-1,l.a=-70}{h10}
   \fmfv{d.shape=circle, d.size=0.6thick,l=$k$,l.a=-100}{h5}
   \fmfv{d.shape=circle, d.size=0.6thick}{h6}
\end{fmfgraph*}
$\quad \Big |\quad $

\begin{fmfgraph*}(30,25)\fmfkeep{v7}
   \fmfleft{i,d1}
   \fmfright{j,d2}
   \fmf{plain}{i,k1}\fmflabel{$i$}{i}
   \fmfn{plain}{k}{20}
   \fmf{plain}{k20,j}\fmflabel{$j$}{j}
   \fmffreeze
   \fmf{fillwx,left=0.7}{k1,k15}
   \fmf{unfillwx,left=0.7}{k5,k11}
   \fmf{fillwx,right=0.7}{k6,k20}
   \fmf{unfillwx,right=0.7}{k10,k16}
   \fmf{photon,left=0.7}{i,j}
   \fmf{dashes,left=0.7}{k1,k15}
   \fmf{dashes,left=0.7}{k5,k11}
   \fmf{dashes,right=0.7}{k6,k20}
   \fmf{dashes,right=0.7}{k10,k16}
   \fmfv{d.shape=circle, d.size=0.6thick}{i}
   \fmfv{d.shape=circle, d.size=0.6thick}{j}
   \fmfv{d.shape=circle, d.size=0.6thick}{k1}
   \fmfv{d.shape=circle, d.size=0.6thick}{k5}
   \fmfv{d.shape=circle, d.size=0.6thick}{k6}
   \fmfv{d.shape=circle, d.size=0.6thick}{k10}
   \fmfv{d.shape=circle, d.size=0.6thick}{k11}
   \fmfv{d.shape=circle, d.size=0.6thick}{k15}
   \fmfv{d.shape=circle, d.size=0.6thick}{k16}
   \fmfv{d.shape=circle, d.size=0.6thick}{k20}
\end{fmfgraph*}
\end{center}
\bigskip
\caption{\label{fig:vxrec} Recursion for $vx$ in the pseudoknot algorithm
truncated at ${\cal O} (whx+whx+whx)$.}
\end{figure}

Let us start with the generalization of the recursions for $wx$ and
$vx$ in the presence of gap matrices. A non-gap matrix can be obtained
by combining two gap matrices together, therefore the recursions for
$wx$ and $vx$ add one more diagram with two gap matrices to recursions
(\ref{wxrecursion}) and (\ref{vxrecursion}).  Again the diagrammatic
representation (fig.~\ref{fig:wxrec},~\ref{fig:vxrec}) is more helpful
than words in explaining the recursion.  Note that the new term
introduced in both recursions involves two gap matrices. In fact the
recursion is an expansion in the number of gap matrices necessary at
each step of the recursion.

\begin{equation}
wx(i,j) = optimal\left\{ 
\begin{array}{l}
P + vx(i,j) \\
L^i_{i+1,j-1} + R^j_{i+1,j-1} + P + vx(i+1,j-1)\\
L^i_{i+1,j} + P + vx(i+1,j)\\
R^j_{i,j-1} + P + vx(i,j-1)\\
Q + wx(i+1,j) \\
Q + wx(i,j-i)\\
wx(i,k) + wx(k+1,j) \label{wxrecpskn}\\
{G}_w + whx(i,r:k,l) + whx(k+1,j:l-1,r+1) \\
{\cal O} (whx+whx+whx)
\end{array}\right.
\end{equation}
Where ${G}_w$ denotes the score for introducing a pseudoknot.

Similarly for $vx$,
\begin{equation}
vx(i,j) = optimal\left\{ 
\begin{array}{l}
IS^1(i,j) \\
IS^2(i,j:k,l) + vx(k,l)\label{vxrecpskn}\\
P_I + M + wx_I(i+1,k) + wx_I(k+1,j-1)\\
P_I + \widetilde {M} + G_{wI} + whx(i+1,r:k,l) \\
\hspace{23.5mm} + whx(k+1,j-1:l-1,r+1)\\
{\cal O} (whx+whx+whx)
\end{array}\right.
\end{equation}
\[[\forall i,k,l,r,j \quad i\leq k\leq l\leq r\leq j]\]
Here $\widetilde {M}$ stands for a generic score for generating a
non-nested multiloop, and $G_{wI}$ stands for the score for generating
an internal pseudoknot.

\begin{figure}
\begin{picture}(32,40)
  \put(5.0,0.0){\vector(0,1){20}}
  \put(20.0,0.0){\vector(0,1){20}}
  \put(5.0,3.0){\dashbox{0.4}(15,0)}
  \put(5.0,10.0){\dashbox{0.4}(15,0)}
  \put(5.0,17.0){\dashbox{0.4}(15,0)}
  \put(3.0,2){1}\put(3.0,9){5}\put(1.5,16){10}
  \put(20.0,2){15}\put(20.0,9){20}\put(20,16){25}
  \put(28.0,10.0){=}
\end{picture}
\begin{fmfgraph*}(40,21)
   \fmfleft{i1}
   \fmfright{i25}
   \fmfn{plain}{i}{25}
   \fmffreeze
   \fmf{photon, left=0.50, label=$I$, l.side=left}{i1,i15}
   \fmf{photon, right=0.50, label=$II/III$, l.side=right}{i5,i20}
   \fmf{photon, left=0.50, label=$IV$, l.side=left}{i10,i25}
\end{fmfgraph*}
\begin{picture}(4,30)
  \put(0.0,10.0){=}
\end{picture}
\begin{fmfgraph*}(40,21)
   \fmfleft{i1}
   \fmfright{i25}
   \fmfn{phantom}{i}{25}
   \fmffreeze
   \fmf{fillwx,left=0.60}{i1,i16}
   \fmf{unfillwx,left=0.60}{i3,i14}
   \fmf{fillwx,right=0.60}{i4,i25}
   \fmf{unfillwx,right=0.60}{i13,i17}
   \fmf{dashes, left=0.60}{i1,i16}
   \fmf{dashes, left=0.60}{i3,i14}
   \fmf{dashes, right=0.60}{i4,i25}
   \fmf{dashes, right=0.60}{i13,i17}
   \fmf{plain}{i1,i3}\fmf{plain}{i14,i16}
   \fmf{plain}{i4,i13}\fmf{plain}{i17,i25}
\end{fmfgraph*}

\begin{picture}(32,40)
  \put(28.0,10.0){=}
\end{picture}
\begin{fmfgraph*}(20,21)
   \fmfleft{i1}
   \fmfright{i16}
   \fmfn{phantom}{i}{16}
   \fmffreeze
   \fmf{fillwx,left=0.60}{i1,i16}
   \fmf{unfillwx,left=0.60}{i3,i14}
   \fmf{dashes, left=0.60}{i1,i16}
   \fmf{dashes, left=0.60}{i3,i14}
   \fmf{plain}{i1,i3}\fmf{plain}{i14,i16}
\end{fmfgraph*}
\begin{picture}(4,30)
  \put(0.0,10.0){+}
\end{picture}
\begin{fmfgraph*}(40,21)
   \fmfleft{i1}
   \fmfright{i21}
   \fmfn{phantom}{i}{21}
   \fmffreeze
   \fmf{fillwx,right=0.60}{i1,i17}
   \fmf{unfillwx,right=0.60}{i4,i13}
   \fmf{fillwx,left=0.60}{i5,i21}
   \fmf{unfillwx,left=0.60}{i9,i18}
   \fmf{dashes, right=0.60}{i1,i17}
   \fmf{dashes, right=0.60}{i4,i13}
   \fmf{dashes, left=0.60}{i5,i21}
   \fmf{dashes, left=0.60}{i9,i18}
   \fmf{plain}{i1,i4}\fmf{plain}{i13,i17}
   \fmf{plain}{i5,i9}\fmf{plain}{i18,i21}
\end{fmfgraph*}

\begin{picture}(32,40)
  \put(1.0,0.0){\vector(0,1){20}}
  \put(13.0,0.0){\vector(0,1){20}}
  \put(25.0,0.0){\vector(0,1){20}}
  \put(1.0,2.0){\dashbox{0.4}(12,0)}
  \put(1.0,10.0){\dashbox{0.4}(12,0)}
  \put(1.0,18.0){\dashbox{0.4}(12,0)}
  \put(13.0,6.0){\dashbox{0.4}(12,0)}
  \put(13.0,14.0){\dashbox{0.4}(12,0)}
  \put(-1.0,1){1}\put(-1.0,5){5}\put(-3.0,9){10}\put(-3.0,13){15}\put(-3.0,17){20}
  \put(13.0,1){25}\put(9.0,5){30}\put(13.0,9){35}\put(9.0,13){40}\put(13.0,17){45}
  \put(25.0,1){50}\put(25.0,5){55}\put(25.0,9){60}\put(25.0,13){65}\put(25.0,17){70}
  \put(29.0,10.0){=}
\end{picture}
\begin{fmfgraph*}(40,21)
   \fmfleft{i1}
   \fmfright{i65}
   \fmfn{plain}{i}{65}
   \fmffreeze
   \fmf{photon, left=0.60}{i1,i25}
   \fmf{photon, left=0.60}{i18,i47}
   \fmf{photon, left=0.60}{i40,i65}
   \fmf{photon, right=0.60}{i10,i35}
   \fmf{photon, right=0.60}{i28,i55}
\end{fmfgraph*}
\begin{picture}(4,30)
  \put(0.0,10.0){=}
\end{picture}
\begin{fmfgraph*}(40,21)
   \fmfleft{i1}
   \fmfright{i65}
   \fmfn{phantom}{i}{65}
   \fmffreeze
   \fmf{fillwx,left=0.60}{i1,i26}
   \fmf{unfillwx,left=0.60}{i6,i20}
   \fmf{fillwx,right=0.60}{i7,i65}
   \fmf{unfillwx,right=0.60}{i19,i27}
   \fmf{dashes, left=0.60}{i1,i26}
   \fmf{dashes, left=0.60}{i6,i20}
   \fmf{dashes, right=0.60}{i7,i65}
   \fmf{dashes, right=0.60}{i19,i27}
   \fmf{plain}{i1,i6}\fmf{plain}{i20,i26}
   \fmf{plain}{i7,i19}\fmf{plain}{i27,i65}
\end{fmfgraph*}

\begin{picture}(32,40)
  \put(28.0,10.0){=}
\end{picture}
\begin{fmfgraph*}(20,21)
   \fmfleft{i1}
   \fmfright{i12}
   \fmfn{phantom}{i}{12}
   \fmffreeze
   \fmf{fillwx,left=0.60}{i1,i12}
   \fmf{unfillwx,left=0.60}{i3,i10}
   \fmf{dashes, left=0.60}{i1,i12}
   \fmf{dashes, left=0.60}{i3,i10}
   \fmf{plain}{i1,i3}\fmf{plain}{i10,i12}
\end{fmfgraph*}
\begin{picture}(4,30)
  \put(0.0,10.0){+}
\end{picture}
\begin{fmfgraph*}(40,21)
   \fmfleft{i1}
   \fmfright{i57}
   \fmfn{phantom}{i}{57}
   \fmffreeze
   \fmf{fillwx,right=0.60}{i1,i31}
   \fmf{unfillwx,right=0.60}{i5,i27}
   \fmf{fillwx,left=0.60}{i7,i44}
   \fmf{unfillwx,left=0.60}{i12,i40}
   \fmf{fillwx,right=0.60}{i20,i50}
   \fmf{unfillwx,right=0.60}{i24,i46}
   \fmf{fillwx,left=0.60}{i34,i57}
   \fmf{unfillwx,left=0.60}{i38,i53}
   \fmf{plain}{i1,i5}\fmf{plain}{i27,i31}
   \fmf{plain}{i7,i12}\fmf{plain}{i40,i44}
   \fmf{plain}{i20,i24}\fmf{plain}{i46,i50}
   \fmf{plain}{i34,i38}\fmf{plain}{i53,i57}
   \fmf{dashes, right=0.60}{i1,i31}
   \fmf{dashes, right=0.60}{i5,i27}
   \fmf{dashes, left=0.60}{i7,i44}
   \fmf{dashes, left=0.60}{i12,i40}
   \fmf{dashes, right=0.60}{i20,i50}
   \fmf{dashes, right=0.60}{i24,i46}
   \fmf{dashes, left=0.60}{i34,i57}
   \fmf{dashes, left=0.60}{i38,i53}
\end{fmfgraph*}

\caption{\label{fig:trunc} Top: The non-planar
pseudoknot presented in $\alpha$ mRNA and how to build it with gap
matrices.  The Roman numbers correspond to the numbering of stems
introduced by Gluick {\it et al.} (1994).  Bottom: An example of a
pseudoknot that the algorithm cannot handle: interlaced interactions
as seen in proteins in parallel $\beta$-sheet.  The assembly of this
interaction using gap matrices would require us to use four gap
matrices at once which is not allowed by the approximation at hand.}
\end{figure}
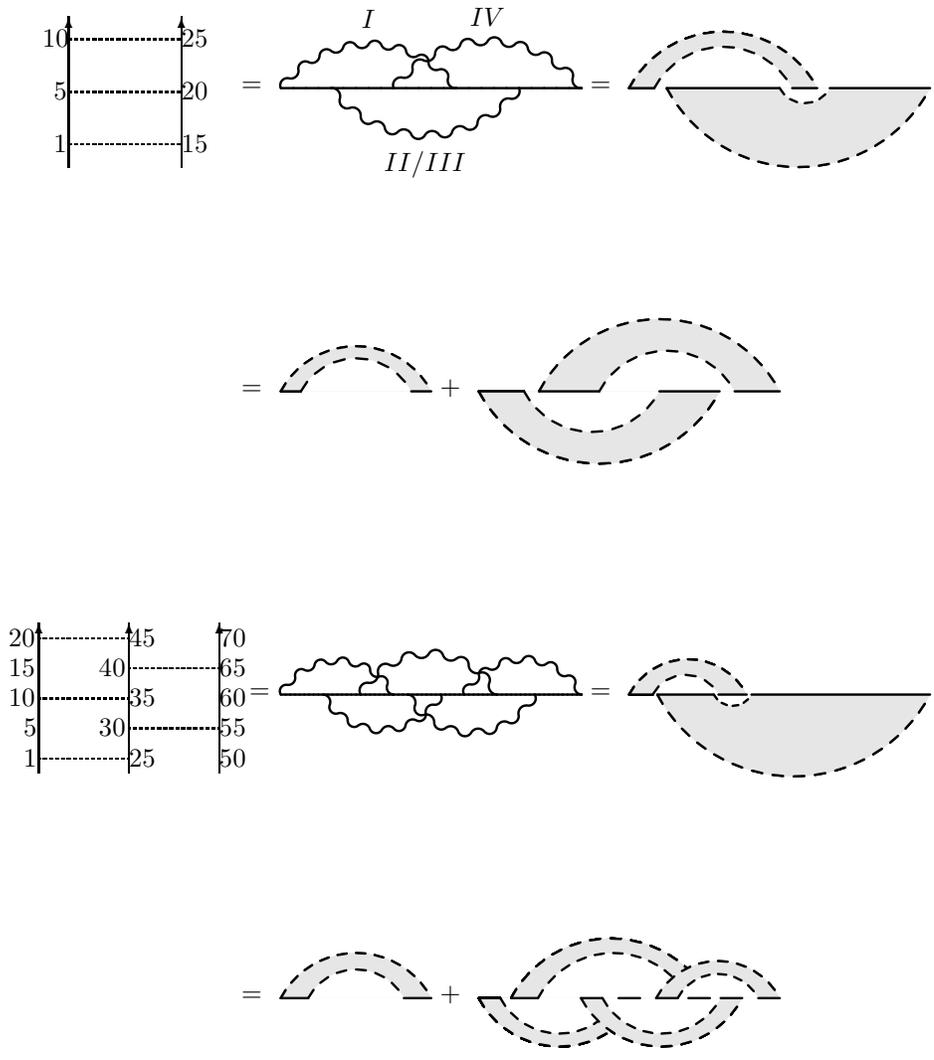

Practical considerations make us truncate the expansion at this point,
so we will not include diagrams that require three or more gap
matrices. This statement should not mislead one into thinking that we
cannot deal with complicated pseudoknots.  The recursive nature of the
approximation allows us to describe overlapping pseudoknots (defined
as those pseudoknots for which a planar representation does not
require crossing lines) as well as non-planar pseudoknots (for which a
planar representation requires crossing lines).  The {\it Escherichia
coli} $\alpha$ mRNA presented by Gluick {\it et al.} (1994) is an
example of a non-planar RNA pseudoknot that can be parsed using the
pseudoknot algorithm.  However the algorithm is not able to find all
possible knotted configurations (fig.~\ref{fig:trunc}). Nevertheless,
the approximation seems to be adequate for the currently known
pseudoknots in RNA folding.  

Note that two approximations are involved in the algorithm.  Apart
from that just mentioned (how to truncate the infinite expansion in
gap matrices to make the algorithm polynomial), we also use the
approximation previously introduced for the nested algorithm (that
IS's of ${\cal O} > 2$ or multiloops are described in some
approximated form).

The algorithm is not complete until we provide the full recursive
expressions to calculate the gap matrices.  For a given gap matrix, we
have to consider in how many different ways that diagram can be
assembled using one or two matrices at a time. The full description of
those diagrams is given in Subsection \ref{sec:methods.appenb} . (Again
Feynman diagrams are of great use here.)


\subsection{Coaxial stacking}

It is quite frequent in RNA folding to create a more stable
configuration when two independent configurations stack coaxially.
That occurs for instance, when two hairpin loops with their respective
stems are contiguous, so that one of them can fall on top of the other
creating a more stable configuration than when the two hairpins just
coexist without interaction of any kind.

The algorithm implements coaxial energies for both nested and
non-nested structures.  We adopt the coaxial energies provided by
Walter {\it et al.} (1994) for coaxial stacking of nested
structures. For coaxial stacking of non-nested structures we multiply
these previous energies by an estimated (ad hoc) weighting parameter
$g<1$.

Using our diagrammatic representation it is possible to be systematic
in describing the possible coaxial stacking that can occur.  In the
general recursion one has to look for contiguous nucleotides and allow
them to be explicitly paired---but not to each other.  This is best
understood with an example.  Consider the recursion for $wx$ in
fig.~\ref{fig:wxrec}, in particular the bifurcation diagram
\[
wx(i,j)\quad\longrightarrow\quad wx(i,k) + wx(k+1,j),
\quad\quad \forall k, i\le k\le j.
\]

\noindent
In order to allow for the possibility of coaxial stacking such diagram
has to be complemented with another one in which the nucleotides of
the bifurcation are base-paired   
\[
wx(i,j)\quad\longrightarrow\quad vx(i,k) + vx(k+1,j) + C(k,i:k+1,j),
\quad\quad \forall k, i\le k\le j.
\]

\noindent
This new diagram indicates that if nucleotides $k$ and $k+1$ are
paired to nucleotides $i$ and $j$ respectively, that configuration is
specially favored by an amount $C(k,i:k+1,j)$ (presumably negative in
energy units) because both sub-structures, $vx(i,k)$ and $vx(k+1,j)$,
will stack onto each other.

Notice also that the new diagram really corresponds to four new
diagrams because once we allow pairing, dangling bases have also to be
considered, so the full nearest-neighbour interaction is taken into
account.  For purposes of clarity we will not explicitly specify any
of the extra diagrams with dangling involved.  The rest of the
additional diagrams to be included in the recursions to take care of
coaxial stackings are also given in Subsection
\ref{sec:methods.appenb} along with the full set of diagrams. Coaxial
diagrams can be recognized by the empty dots representing the
contiguous coaxially-stacking nucleotides.

\subsection{Minimum-energy implementation. Thermodynamic parameters}

We have implemented the pseudoknot algorithm using thermodynamic
parameters in order to fill the scoring matrices, both gapped and
ungapped.  For the relevant nested structures: hairpin loops, bulges,
stems, internal loops and multiloops we have used the same set of
energies as used in MFOLD.\footnote{Since the implementation of
the pseudoknot algorithm the Turner group has produced a new complete
and more accurate list of parameters (Matthews {\it et al.}, 1998)
which we have not yet implemented.} Free energies for coaxial stacking
$C$ where obtained from Walter {\it et al.},
(1994). Table~\ref{nestedparam} provides a list of the parameters used
for nested conformations.

\begin{table}[ht]
\begin{center}
\begin{tabular}{|c|l|c|}\hline
Symbol & Scoring parameter for & Value(Kcal/mol)\\
\hline
$IS^1$ & hairpin loops & varies\\
\hline
$IS^2$ & bulges, stems and int loops & varies\\
\hline
$C$ & coaxial stacking & varies\\
\hline
$P$ & external pair & 0\\
\hline
$Q$ & single stranded base & $0$\\
\hline
$R,L$ & base dangling off an external pair & $dangle+Q$\\
\hline
$P_{I}$ & pair in a multiloop & $0.1$\\
\hline
$Q_{I}$ & not paired base inside multiloop & $0.4$\\
\hline
$R_{I},L_{I}$ & base dangling off a multiloop pair & $dangle+Q_I$\\
\hline
$M$ & nested multiloop & $4.6$\\
\hline
\end{tabular}
\end{center}
\caption{\label{nestedparam}
This table includes all the parameters for which there is
thermodynamic information provided by the Turner group.  This
parameters are identical to those used in MFOLD
(http://www.ibc.wustl.edu/\~{}zuker/rna).}
\end{table}

For the non-nested configurations, there is not much thermodynamic
information available (Wyatt {\it et al.}, 1990; Gluick {\it et al.},
1994).  This is not an untypical situation; there is already very
little thermodynamic information available for regular multiloops, let
alone for pseudoknots. We had to tune by hand the parameters related
to pseudoknots.  For some non-nested structures we multiplied the
nested parameters by an estimated weighting parameter $g<1$. It would
be very useful, in order to improve the accuracy of this thermodynamic
implementation of the pseudoknot algorithm, to have more accurate
experimentally-based determinations of these parameters.
Table~\ref{pkparam} provides a list of the parameters we used for
pseudoknot-related conformations.

\begin{table}[ht]
\begin{center}
\begin{tabular}{|c|l|c|c|}\hline
Symbol & Scoring parameter for & Value(Kcal/mol)\\
\hline
$\widetilde {IS}^2$ & $IS^2$ in a gap matrix & $IS^2*g(0.83)$ \\
\hline
$\widetilde C$ & coaxial stacking in pseudoknots & $C*g$ \\
\hline
$\widetilde P$ & pair in a pseudoknot & $0.1$\\
\hline
$\widetilde Q$ & not paired base in pseudoknot & $0.2$\\
\hline
$\widetilde R, \widetilde L$ & base dangling off a pseudoknot pair 
                             & $dangle*g+\widetilde Q$\\
\hline
$\widetilde {M}$ & non-nested multiloop & $8.43$\\
\hline
${G}_{w}$ & generating a new pseudoknot & $7.0$\\
\hline
${G}_{wI}$ & generating a pseudoknot in a multiloop & $13.0$\\
\hline
${G}_{wh}$ & overlapping pseudoknots & $6.0$\\
\hline
\end{tabular}
\end{center}
\caption{\label{pkparam}
In this table we introduce the new thermodynamic parameters specific for
pseudoknot configurations what we had to estimate.}
\end{table}

\section{RESULTS}

The main purpose of this paper is to present an algorithm that solves
optimal pseudoknotted RNA structures by dynamic programming.  RNA
structure prediction of single sequences with nested algorithms
already involves some approximation and inaccuracy (Zuker, 1995;
Huynen {\it et al}, 1997). We expect this inaccuracy to increase in
our case since the algorithm now allows a much larger configuration
space.  Therefore our limited objective here is to show that on a few
small RNAs that are thought to conserve pseudoknots, our program (a
minimal-energy implementation of the pseudoknot algorithm using a
thermodynamic model) will actually find the pseudoknots; and for a few
small RNAs that do not conserve pseudoknots, our program finds results
similar to MFOLD, and does not introduce spurious pseudoknots.

\subsection{tRNAs}
Almost all transfer RNAs (tRNA) share a common cloverleaf structure.
We have tested the algorithm on a group of $25$ tRNAs selected at
random from the Sprinzl tRNA database (Steinberg {\it et al.}, 1993).
The program finds no spurious pseudoknot for any of the tested
sequences.  All tRNAs fold into a cloverleaf configuration but one
(DT5090). Of the $24$ cloverleaf foldings, $15$ are completely
consistent with their proposed structures (that is, each helical
region has at least 3 base pairs in common with its proposed
folding). The remaining $9$ cloverleaf foldings misplace one ($6$
sequences) or two ($3$ sequences) of the helical regions.  On the
other hand, MFOLD's lowest--energy prediction for the same set of tRNA
sequences includes only $19$ cloverleaf foldings of which $14$ are
completely consistent with their proposed structures. Performance for
our program is therefore at least comparable to MFOLD; the
inaccuracies found are the result of the approximations in the
thermodynamic model, not a problem with the pseudoknot algorithm per
se.  The relevant result in relation to the pseudoknot algorithm is
that its implementation predicts no spurious pseudoknots for tRNAs.

One should not think of this result as a trivial one, because when
knots are allowed, the configuration space available becomes much
larger than the observed class of conformations. This problem is
particularly relevant for ``maxi\-mum-pairing-like'' algorithms, such
as the MWM algorithm presented by Cary \& Stormo (1995) or a Nussinov
implementation of our pseudoknot algorithm (fig~\ref{fig:vo(0)}). In
both cases, the result is almost universal pairing because there is
enough freedom to be able to coordinate any position with another one
in the sequence.

Another important aspect of tRNA folding is coaxial energies.  Most
tRNAs gain stability by stacking coaxially two of the hairpin loops,
and the third one with the acceptor stem. This aspect of tRNA folding
is very important and in some cases crucial to determine the right
structure.  There are situations like tRNA DA0260 in which MFOLD does
not assign the lowest energy to the correct structure (the MFOLD $3.0$
prediction for DA0260 misses the acceptor stem, and has a free energy
of $-22.0$ Kcal/mol). Our algorithm, on the other hand, implements
coaxial energies; as a result the cloverleaf configuration becomes the
most stable folding for tRNA DA0260 ($\Delta G = -24.3$ Kcal/mol).
The implementation of coaxial energies explains why we found more
cloverleaf structures for tRNAs than MFOLD does.

\subsection{HIV-1-RT-ligand RNA pseudoknots}
High-affinity ligands of the reverse transcriptase of human
immunodeficiency virus type 1 (HIV-1) isolated by a SELEX procedure by
Tuerk {\it et al.} (1992) seem to have a pseudoknot consensus
secondary structure.  These oligonucleotides have between $34$ and
$47$ bases and fold into a simple pseudoknot. Of a total of $63$
SELEX-selected pseudoknotted sequences available from Tuerk {\it et
al.} (1992), we found $54$ foldings that agreed exactly with the
structures derived by comparative analysis.  ($\Delta G = -10.9$
Kcal/mol for sequence pattern I (3-2)). As expected, MFOLD predicts
only one of the two stems ($\Delta G = -7.5$ Kcal/mol for the same
sequence).

\subsection{Viral RNAs}
Some virus RNA genomes [such as turnip yellow mosaic virus (TYMV)
(Guilley {\it et al}, 1979)] present a tRNA-like structure at their
$3^\prime$ end that includes a pseudoknot in the aminoacyl acceptor
arm very close to the $3^\prime$ end (Kolk {\it et al}, 1998; Florentz
{\it et al.}, 1982; Dumas {\it et al}, 1987).  Our program predicts
correctly the TYMV tRNA-like structure with its pseudoknot for the
last $86$ bases at the $3^\prime$ end with $\Delta G = -30.4$ Kcal/mol
(the MFOLD $3.0$ prediction for TYMV has a free energy of $\Delta G =
-28.9$ Kcal/mol).  The tRNA-like $3^\prime$ terminal structure is
conserved among tymoviruses and also for the tobacco mosaic virus
cowpea strain (CcTMV) another valine acceptor. Of the seven
valine-acceptor tRNA-like structures proposed to date (Van Belkum {\it
et al}, 1987) we reproduce six of them, except for kennedya yellow
mosaic virus (KYMV).

Finally we have considered the last $189$ bases of the $3^\prime$
terminal of the Tobacco mosaic virus (TMV) (Van Belkum {\it et al},
1985).  TMV also has a tRNA-like structure at the end, but it may have
additional upstream pseudoknots, up to a total of five, forming a long
quasi-continuous helix. We folded the upstream and downstream regions
separately in a piece of 84 nucleotides (the folding requires 52
minutes and 9.8 Mb) and 105 nucleotides (the folding requires 246
minutes and 22.5 Mb) respectively. Our program predicts the 105
nucleotides downstream region exactly with $\Delta G = -32.5$
Kcal/mol.  For the 84 nucleotides upstream region we find four of the
five helical regions with $\Delta G = -19.0$ Kcal/mol.

\section{DISCUSSION}

In this paper we present an algorithm able to predict pseudoknots by
dynamic programming.  This algorithm demonstrates that using certain
approximations consistent with the accepted Turner thermodynamic
model, the prediction of pseudoknotted structures is a problem of
polynomial complexity (although admittedly high). Having an optimal
dynamic programming algorithm will enable extending other dynamic
programming based methods that rigorously explore the conformational
space for RNA folding (McCaskill, 1990; Bonhoeffer {\it et al.}, 1993)
to pseudoknotted structures.

Apart from the usefulness of the algorithm in predicting pseudoknots,
we also include coaxial energies (when two stems stack coaxially), a
very common feature of RNA folding. We expect MFOLD will also
include coaxial energies in the near future (Matthews {\it et al.},
1998).

Our algorithm is presented in the context of a general framework in
which a generic folding is expressed in terms of its elementary
secondary interactions (which we have identified as the irreducible
surfaces). This is a further generalization of Sankoff's result
(1985). The calculation of an optimal folding becomes an expansion in
ISs of increasingly higher order. Our formalization incorporates all
current dynamic programming RNA folding algorithms in addition to our
pseudoknot algorithm. It also establishes the limitations of each
approximation by determining at which order the expansion is
truncated.

As for the thermodynamic implementation presented in this paper, one
of our major problems is the almost complete lack of thermodynamic
information about pseudoknot configurations.  The thermodynamic
algorithm is also sensitive to the accuracy of the existing
thermodynamic parameters.  We expect to improve this aspect by
implementing the more complete set of parameters provided by the
Turner group (Matthews {\it et al.}, 1998).

The principal drawback is the time and memory constraints imposed by
the computational complexity of the algorithm. At this early stage, we
cannot analyze sequences much larger than 130-140 bases. For now the
program is adequate for folding small RNAs. A 100 nucleotide RNA takes
about 4 hours and 22.5 MB to fold on an SGI R10K Origin200.

Due to practical limitations, at a given point in the recursion we
only allow the incorporation of two gap matrices. However, since each
of those gap matrices can in turn be assembled by other two of those
matrices, it implies that the algorithm includes in its configuration
space a large variety of knotted motifs.  The limitations of this
truncation appeared when we considered applying this approach to
describe pairwise residue interactions in protein folding.  A parallel
$\beta$-sheet configuration provides an example of a complicated
knotted folding that cannot be handled by the pseudoknot algorithm
presented in this paper.  However, all known RNA pseudoknots can be
handled by the algorithm, which makes the approximation useful enough
for RNA secondary structure.

Although we implemented the algorithm for energy minimization,
extending MFOLD to pseudoknotted structures, the algorithm is not
limited to energy minimization. Our algorithm can be converted into
a probabilistic model for pseudoknot-containing RNA folding.
Probabilistic models of RNA secondary structure based on ``stochastic
context free grammar'' (SCFG) formalisms (Eddy {\it et al.}, 1994;
Sakakibara {\it et al.}, 1994; Lefebvre, 1996) have been introduced
both for RNA single-sequence folding and for RNA structural alignment
and structural similarity searches. The Inside and CYK dynamic
programming algorithms used for SCFG-based structural alignment are
fundamentally similar to the Zuker algorithm (Durbin {\it et al.},
1998), and have consequently also been unable to deal with
pseudoknots. Heuristic approaches to applying SCFG-like structural
alignment models to pseudoknots have been introduced (Brown, 1996;
Notredame {\it et al.}, 1997). An SCFG-like probabilistic version of
our pseudoknot algorithm could be designed to obtain optimal
structural alignment of pseudoknot-containing RNAs.

\section{METHODS}

\subsection{Implementation}
The algorithm was implemented on a Silicon Graphics Origin200.  The
algorithm has a theoretical worst-case complexity of ${\cal O}
(N^{6})$ in time and ${\cal O} (N^{4})$ in storage.  At its present
stage, the program is empirically observed to run ${\cal O} (N^{6.8})$
in time and ${\cal O} (N^{3.8})$ in memory.  For instance, a tRNA of
$75$ nucleotides takes 24 minutes and uses $6.6$ Mb of memory. The
$3^\prime$ end of tobacco mosaic virus has $105$ nucleotides and takes
$246$ minutes and uses $22.5$ Mb.  The program empirically scales
above the theoretical complexity in time of the algorithm. This effect
seems to have to do with the way the machine allocates memory for
larger RNAs.  The software and parameter sets are available by request
from E. Rivas (elena@genetics.wustl.edu).

\subsection{Complete set of diagrams for the pseudoknot algorithm}
\label{sec:methods.appenb} 

In this section we provide the complete recursion relations for all the 
matrices used in the pseudoknot algorithm. 

The recursion for the non-gap matrix $wx$ is given by
(cf. fig.~\ref{fig:wxrecur}):	

\begin{equation}
wx(i,j) = optimal\left\{ 
\begin{array}{ll}

\vspace{2mm}

\left.
\begin{array}{l}
P + vx(i,j) 
\end{array}
\hspace{57.2mm}\right]
&
\mbox{paired}\\

\vspace{2mm}

\left.
\begin{array}{l}
L^i_{i+1,j-1} + R^j_{i+1,j-1} + P + vx(i+1,j-1)\\
L^i_{i+1,j} + P + vx(i+1,j)\\
R^j_{i,j-1} + P + vx(i,j-1)
\end{array}
\hspace{10.2mm}\right]
&
\mbox{dangles}\\

\vspace{2mm}

\left.
\begin{array}{l}
Q + wx(i+1,j) \\
Q + wx(i,j-i)
\end{array}
\hspace{50mm}\right]
&
\parbox[b][5mm][t]{80mm}{
single 
\vspace{-2mm}

stranded}\\

\vspace{2mm}

\left.
\begin{array}{l}
wx(i,k) + wx(k+1,j) \label{wxrecur}\\
C(k,i:k+1,j) + vx(i,k) + vx(k+1,j) 
\end{array}
\hspace{13mm}\right]
&
\parbox[b][5mm][t]{80mm}{
nested 
\vspace{-2mm}

bifurcations}\\

\vspace{2mm}

\left.
\begin{array}{l}
{G}_w + whx(i,r:k,l) + whx(k+1,j:l-1,r+1) \\

2*P + {G}_w + C(l-1,r+1:l,k) \\
\hspace{6.0mm}+ yhx(i,r:k,l) + yhx(k+1,j:l-1,r+1)
\end{array}
\right]
&
\parbox[b][5mm][t]{80mm}{
non-nested 
\vspace{-2mm}

bifurcations}\\

\end{array}
\right.
\end{equation}
Here ${G}_w$ denotes a score for introducing a
pseudoknot, and $C(l,r:l+1,k)$ is a special score for a coaxial
stacking of pairs $(l,r)$ and $(l+1,k)$.

We should also remember that the algorithm uses two different $wx$
matrices depending on whether the subset $i...j$ is free-standing
($wx$) or appears inside a multiloop (in which case we use
$wx_I$). Both recursions are identical apart from having different
coefficients as described in table~\ref{nestedparam}.

\begin{figure}
\begin{center}
\begin{fmfgraph*}(30,30)\fmfkeep{wx}
   \fmfleft{i,d1}
   \fmfright{j,d2}
   \fmffreeze
   \fmf{fillwx,left}{i,j}
   \fmf{plain}{i,j}\fmflabel{$i$}{i}\fmflabel{$j$}{j}
   \fmf{dashes,left}{i,j}
\end{fmfgraph*}
\quad=
\end{center}

\begin{center}
\begin{fmfgraph*}(30,25)\fmfkeep{w1}
   \fmfleft{i,d1}
   \fmfright{j,d2}
   \fmf{plain}{i,j}\fmflabel{$i$}{i}\fmflabel{$j$}{j}
   \fmffreeze
   \fmf{fillvx,left}{i,j}
   \fmf{photon,left}{i,j}
   \fmf{plain}{i,j}
\end{fmfgraph*}
$\quad\Big |\quad$
\begin{fmfgraph*}(30,25)\fmfkeep{w2}
   \fmfleft{i,d1}\fmfv{l=$i$,l.a=180}{i}
   \fmfright{j,d2}
   \fmf{plain}{i,h1}
   \fmfn{plain}{h}{10}
   \fmf{plain}{h10,j}\fmfv{l=$j$,l.a=0}{j}
   \fmffreeze
   \fmf{fillvx,left}{h1,h10}
   \fmf{photon,left}{h1,h10}
   \fmfv{d.shape=circle, d.size=0.6thick}{i}
   \fmfv{d.shape=circle, d.size=0.6thick}{j}
   \fmfv{d.shape=circle, d.size=0.6thick,l=$i$+1,l.a=-70}{h1}
   \fmfv{d.shape=circle, d.size=0.6thick,l=$j$-1,l.a=-100}{h10}
\end{fmfgraph*}
$\quad\Big |\quad$
\begin{fmfgraph*}(30,25)\fmfkeep{w3}
   \fmfleft{i,d1}\fmfv{l=$i$,l.a=180}{i}
   \fmfright{j,d2}
   \fmf{plain}{i,h1}
   \fmfn{plain}{h}{10}
   \fmf{plain}{h10,j}\fmfv{l=$j$,l.a=0}{j}
   \fmffreeze
   \fmf{fillvx,left}{h1,j}
   \fmf{photon,left}{h1,j}
   \fmfv{d.shape=circle, d.size=0.6thick}{i}
   \fmfv{d.shape=circle, d.size=0.6thick,l=$i$+1,l.a=-70}{h1}
\end{fmfgraph*}
$\quad\Big |\quad$
\end{center}

\begin{center}
\begin{fmfgraph*}(30,25)\fmfkeep{w4}
   \fmfleft{i,d1}\fmfv{l=$i$,l.a=180}{i}
   \fmfright{j,d2}
   \fmf{plain}{i,h1}
   \fmfn{plain}{h}{10}
   \fmf{plain}{h10,j}\fmfv{l=$j$,l.a=0}{j}
   \fmffreeze
   \fmf{fillvx,left}{i,h10}
   \fmf{photon,left}{i,h10}
   \fmf{plain}{i,h1}
   \fmfn{plain}{h}{10}
   \fmf{plain}{h10,j}
   \fmfv{d.shape=circle, d.size=0.6thick}{j}
   \fmfv{d.shape=circle, d.size=0.6thick,l=$j$-1,l.a=-100}{h10}
\end{fmfgraph*}
$\quad\Big |\quad$
\begin{fmfgraph*}(30,25)\fmfkeep{w5}
   \fmfleft{i,d1}\fmfv{l=$i$,l.a=180}{i}
   \fmfright{j,d2}
   \fmf{plain}{i,h1}
   \fmfn{plain}{h}{10}
   \fmf{plain}{h10,j}\fmfv{l=$j$,l.a=0}{j}
   \fmffreeze
   \fmf{fillwx,left}{h1,j}
   \fmf{dashes,left}{h1,j}
   \fmfv{d.shape=circle, d.size=0.6thick}{i}
   \fmfv{d.shape=circle, d.size=0.6thick,l=$i$+1,l.a=-70}{h1}
\end{fmfgraph*}
$\quad\Big |\quad $
\begin{fmfgraph*}(30,25)\fmfkeep{w6}
   \fmfleft{i,d1}\fmfv{l=$i$,l.a=180}{i}
   \fmfright{j,d2}
   \fmf{plain}{i,h1}
   \fmfn{plain}{h}{10}
   \fmf{plain}{h10,j}\fmfv{l=$j$,l.a=0}{j}
   \fmffreeze
   \fmf{fillwx,left}{i,h10}
   \fmf{dashes,left}{i,h10}
   \fmf{plain}{i,h1}
   \fmfn{plain}{h}{10}
   \fmf{plain}{h10,j}   
   \fmfv{d.shape=circle, d.size=0.6thick}{j}
   \fmfv{d.shape=circle, d.size=0.6thick,l=$j$-1,l.a=-70}{h10}
\end{fmfgraph*}
$\quad \Big |\quad $
\end{center}

\begin{center}
\begin{fmfgraph*}(35,25)\fmfkeep{w7}
   \fmfleft{i,d1}
   \fmfright{j,d2}
   \fmf{plain}{i,k1}\fmflabel{$i$}{i}
   \fmfn{plain}{k}{10}
   \fmf{plain}{k10,j}\fmflabel{$j$}{j}
   \fmffreeze
   \fmf{fillwx,left}{i,k5}
   \fmf{fillwx,left}{k6,j}
   \fmf{dashes,left}{i,k5}
   \fmf{dashes,left}{k6,j}
   \fmf{plain}{i,k1}
   \fmfn{plain}{k}{10}
   \fmf{plain}{k10,j}
   \fmfv{d.shape=circle, d.size=0.6thick,l=$k$,l.a=-100}{k5}
   \fmfv{d.shape=circle, d.size=0.6thick,l=$k$+1,l.a=-65}{k6}
\end{fmfgraph*}
$\quad \Big |\quad $
\begin{fmfgraph*}(35,25)\fmfkeep{w8}
   \fmfleft{i,d1}
   \fmfright{j,d2}
   \fmf{plain}{i,k1}\fmflabel{$i$}{i}
   \fmfn{plain}{k}{10}
   \fmf{plain}{k10,j}\fmflabel{$j$}{j}
   \fmffreeze
   \fmf{fillvx,left}{i,k5}
   \fmf{fillvx,left}{k6,j}
   \fmf{photon,left}{i,k5}
   \fmf{photon,left}{k6,j}
   \fmf{plain}{i,k1}
   \fmfn{plain}{k}{10}
   \fmf{plain}{k10,j}
   \fmfv{d.shape=circle, d.filled=empty, d.size=3pt,l=$k$,l.a=-100}{k5}
   \fmfv{d.shape=circle, d.filled=empty, d.size=3pt,l=$k$+1,l.a=-65}{k6}
\end{fmfgraph*}
$\quad \Big |\quad $
\end{center}

\begin{center}
\begin{fmfgraph*}(35,25)\fmfkeep{w9}
   \fmfleft{i,d1}
   \fmfright{j,d2}
   \fmf{plain}{i,k1}\fmflabel{$i$}{i}
   \fmfn{plain}{k}{20}
   \fmf{plain}{k20,j}\fmflabel{$j$}{j}
   \fmffreeze
   \fmf{fillwx,left=0.7}{i,k15}
   \fmf{unfillwx,left=0.7}{k5,k11}
   \fmf{fillwx,right=0.7}{k6,j}
   \fmf{unfillwx,right=0.7}{k10,k16}
   \fmf{dashes,left=0.7}{i,k15}
   \fmf{dashes,left=0.7}{k5,k11}
   \fmf{dashes,right=0.7}{k6,j}
   \fmf{dashes,right=0.7}{k10,k16}
   \fmf{plain}{i,k1}
   \fmfn{plain}{k}{20}
   \fmf{plain}{k20,j}
   \fmfv{d.shape=circle, d.size=0.6thick}{k5}
   \fmfv{d.shape=circle, d.size=0.6thick}{k6}
   \fmfv{d.shape=circle, d.size=0.6thick}{k10}
   \fmfv{d.shape=circle, d.size=0.6thick}{k11}
   \fmfv{d.shape=circle, d.size=0.6thick}{k15}
   \fmfv{d.shape=circle, d.size=0.6thick}{k16}
\end{fmfgraph*}
$\quad \Big |\quad $
\begin{fmfgraph*}(35,25)\fmfkeep{w12}
   \fmfleft{i,d1}
   \fmfright{j,d2}
   \fmf{plain}{i,k1}\fmflabel{$i$}{i}
   \fmfn{plain}{k}{20}
   \fmf{plain}{k20,j}\fmflabel{$j$}{j}
   \fmffreeze
   \fmf{fillwx,left=0.75}{i,k15}
   \fmf{unfillvx,left=0.75}{k5,k11}
   \fmf{fillwx,right=0.75}{k6,j}
   \fmf{unfillvx,right=0.75}{k10,k16}
   \fmf{dashes,left=0.75}{i,k15}
   \fmf{photon,left=0.75}{k5,k11}
   \fmf{dashes,right=0.75}{k6,j}
   \fmf{photon,right=0.75}{k10,k16}
   \fmf{plain}{i,k1}
   \fmfn{plain}{k}{20}
   \fmf{plain}{k20,j}
   \fmfv{d.shape=circle, d.size=0.6thick}{k5}
   \fmfv{d.shape=circle, d.size=0.6thick}{k6}
   \fmfv{d.shape=circle, d.filled=empty, d.size=3pt}{k10}
   \fmfv{d.shape=circle, d.filled=empty, d.size=3pt}{k11}
   \fmfv{d.shape=circle, d.size=0.6thick}{k15}
   \fmfv{d.shape=circle, d.size=0.6thick}{k16}
\end{fmfgraph*}
\end{center}\bigskip
\caption{\label{fig:wxrecur}
Recursion for the $wx$ matrix.}
\end{figure}
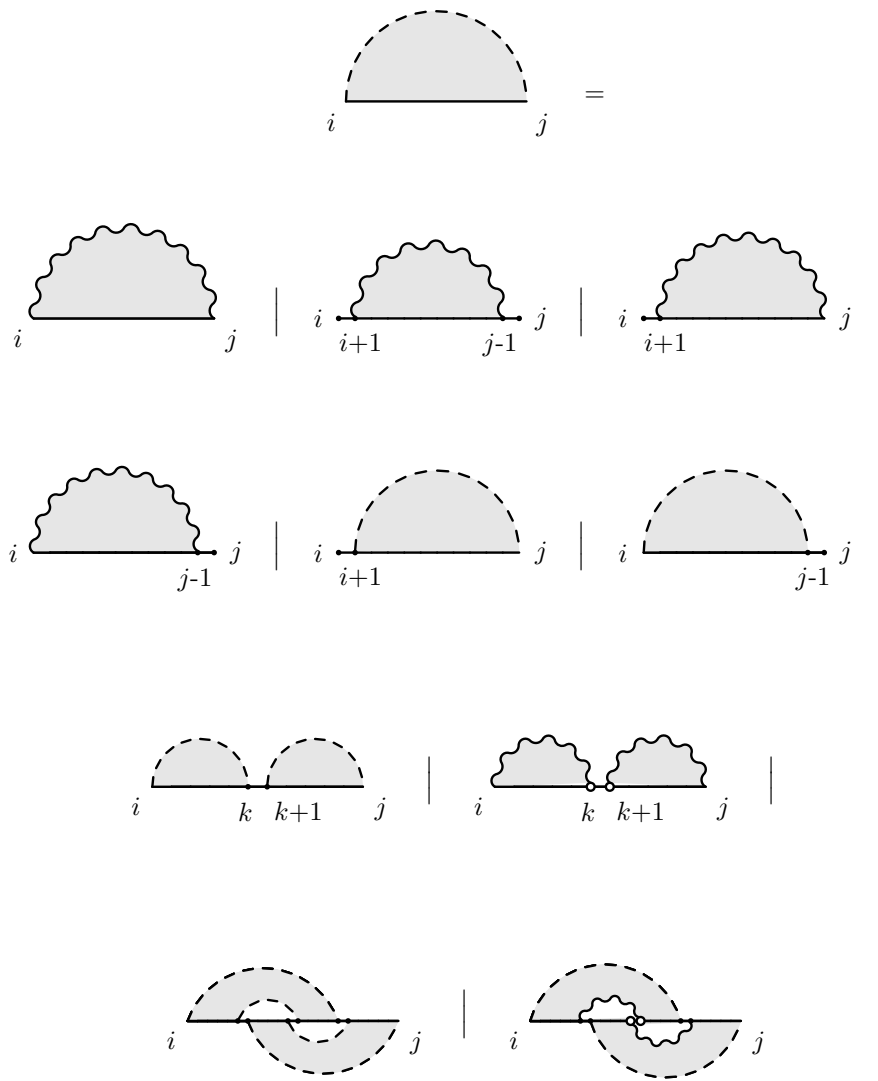

\begin{figure}
\begin{center}
\begin{fmfgraph*}(30,30)\fmfkeep{vx}
   \fmfleft{i,d1}
   \fmfright{j,d2}
   \fmf{plain}{i,j}\fmflabel{$i$}{i}\fmflabel{$j$}{j}
   \fmffreeze
   \fmf{fillvx,left}{i,j}
   \fmf{photon,left}{i,j}
   \fmf{plain}{i,j}
\end{fmfgraph*}
\quad=
\end{center}

\begin{center}
\begin{fmfgraph*}(30,25)\fmfkeep{v1}
   \fmfleft{i,d1}
   \fmfright{j,d2}
   \fmf{plain}{i,j}\fmflabel{$i$}{i}\fmflabel{$j$}{j}
   \fmffreeze
   \fmf{photon,left}{i,j}
\end{fmfgraph*}
$\quad\Big |\quad$
\begin{fmfgraph*}(30,25)\fmfkeep{v2}
   \fmfleft{i,d1}
   \fmfright{j,d2}
   \fmf{plain}{i,k,h,l,j}
   \fmflabel{$i$}{i}\fmflabel{$j$}{j}
   \fmfv{l=$k$,l.a=-90}{k}\fmfv{l=$l$,l.a=-90}{l}
   \fmffreeze
   \fmf{fillvx,left}{k,l}
   \fmf{photon,left}{i,j}
   \fmf{photon,left}{k,l}
   \fmf{plain}{i,k,h,l,j}
\end{fmfgraph*}
$\quad\Big |\quad$
\begin{fmfgraph*}(30,25)\fmfkeep{v3}
   \fmfleft{i,d1}
   \fmfright{j,d2}
   \fmf{plain}{i,h1}\fmfv{l=$i$,l.a=-180}{i}
   \fmfn{plain}{h}{10}
   \fmf{plain}{h10,j}\fmfv{l=$j$,l.a=0}{j}
   \fmffreeze
   \fmf{fillwx,left}{h1,h5}
   \fmf{fillwx,left}{h6,h10}
   \fmf{photon,left=0.7}{i,j}
   \fmf{dashes,left}{h1,h5}
   \fmf{dashes,left}{h6,h10}
   \fmfv{d.shape=circle, d.size=0.6thick}{i}
   \fmfv{d.shape=circle, d.size=0.6thick}{j}
   \fmfv{d.shape=circle, d.size=0.6thick, l=$i$+1,l.a=-70}{h1}
   \fmfv{d.shape=circle, d.size=0.6thick, l=$j$-1,l.a=-70}{h10}
   \fmfv{d.shape=circle, d.size=0.6thick,l=$k$,l.a=-100}{h5}
   \fmfv{d.shape=circle, d.size=0.6thick}{h6}
\end{fmfgraph*}
$\quad \Big |\quad $
\end{center}

\begin{center}
\begin{fmfgraph*}(30,25)\fmfkeep{v4}
   \fmfleft{i,d1}
   \fmfright{j,d2}
   \fmf{plain}{i,h1}\fmfv{l=$i$, l.a=-180}{i}
   \fmfn{plain}{h}{10}
   \fmf{plain}{h10,j}\fmfv{l=$j$, l.a=0}{j}
   \fmffreeze
   \fmf{fillvx,left}{h1,h5}
   \fmf{fillwx,left}{h6,h10}
   \fmf{photon,left=0.75}{i,j}
   \fmf{photon,left}{h1,h5}
   \fmf{dashes,left}{h6,h10}
   \fmfv{d.shape=circle, d.filled=empty, d.size=3pt}{i}
   \fmfv{d.shape=circle, d.size=0.6thick}{j}
   \fmfv{d.shape=circle, d.filled=empty, d.size=3pt, l=$i$+1, l.a=-70}{h1}
   \fmfv{d.shape=circle, d.size=0.6thick, l=$j$-1, l.a=-70}{h10}
   \fmfv{d.shape=circle, d.size=0.6thick, l=$k$, l.a=-100}{h5}
   \fmfv{d.shape=circle, d.size=0.6thick}{h6}
\end{fmfgraph*}
$\quad \Big |\quad $
\begin{fmfgraph*}(30,25)\fmfkeep{v5}
   \fmfleft{i,d1}
   \fmfright{j,d2}
   \fmf{plain}{i,h1}\fmfv{l=$i$, l.a=-180}{i}
   \fmfn{plain}{h}{10}
   \fmf{plain}{h10,j}\fmfv{l=$j$, l.a=0}{j}
   \fmffreeze
   \fmf{fillwx,left}{h1,h5}
   \fmf{fillvx,left}{h6,h10}
   \fmf{photon,left=0.75}{i,j}
   \fmf{dashes,left}{h1,h5}
   \fmf{photon,left}{h6,h10}
   \fmfv{d.shape=circle, d.size=0.6thick}{i}
   \fmfv{d.shape=circle, d.filled=empty, d.size=3pt}{j}
   \fmfv{d.shape=circle, d.size=0.6thick, l=$i$+1, l.a=-70}{h1}
   \fmfv{d.shape=circle, d.filled=empty, d.size=3pt, l=$j$-1,l.a=-70}{h10}
   \fmfv{d.shape=circle, d.size=0.6thick, l=$k$, l.a=-100}{h5}
   \fmfv{d.shape=circle, d.size=0.6thick}{h6}
\end{fmfgraph*}
$\quad \Big |\quad $
\begin{fmfgraph*}(30,25)\fmfkeep{v6}
   \fmfleft{i,d1}
   \fmfright{j,d2}
   \fmf{plain}{i,h1}\fmfv{l=$i$, l.a=-180}{i}
   \fmfn{plain}{h}{10}
   \fmf{plain}{h10,j}\fmfv{l=$j$, l.a=0}{j}
   \fmffreeze
   \fmf{fillvx,left}{h2,h5}
   \fmf{fillvx,left}{h6,h9}
   \fmf{photon,left=0.75}{i,j}
   \fmf{photon,left}{h2,h5}
   \fmf{photon,left}{h6,h9}
   \fmf{plain}{i,h1}
   \fmfn{plain}{h}{10}
   \fmf{plain}{h10,j}
   \fmfv{l=$i^\prime$,l.a=-70}{h2}
   \fmfv{d.shape=circle, d.filled=empty, d.size=3pt,l=$k$,l.a=-90}{h5}
   \fmfv{d.shape=circle, d.filled=empty, d.size=3pt}{h6}
   \fmfv{l=$j^\prime$,l.a=-70}{h9}
\end{fmfgraph*}
$\quad \Big |\quad $
\end{center}

\begin{center}
\begin{fmfgraph*}(35,25)\fmfkeep{v7}
   \fmfleft{i,d1}
   \fmfright{j,d2}
   \fmf{plain}{i,k1}\fmfv{l=$i$,l.a=-180}{i}
   \fmfn{plain}{k}{20}
   \fmf{plain}{k20,j}\fmflabel{$j$}{j}
   \fmffreeze
   \fmf{fillwx,left=0.7}{k1,k15}
   \fmf{unfillwx,left=0.7}{k5,k11}
   \fmf{fillwx,right=0.7}{k6,k20}
   \fmf{unfillwx,right=0.7}{k10,k16}
   \fmf{photon,left=0.7}{i,j}
   \fmf{dashes,left=0.7}{k1,k15}
   \fmf{dashes,left=0.7}{k5,k11}
   \fmf{dashes,right=0.7}{k6,k20}
   \fmf{dashes,right=0.7}{k10,k16}
   \fmfv{d.shape=circle, d.size=0.6thick}{i}
   \fmfv{d.shape=circle, d.size=0.6thick}{j}
   \fmfv{d.shape=circle, d.size=0.6thick}{k1}
   \fmfv{d.shape=circle, d.size=0.6thick}{k5}
   \fmfv{d.shape=circle, d.size=0.6thick}{k6}
   \fmfv{d.shape=circle, d.size=0.6thick}{k10}
   \fmfv{d.shape=circle, d.size=0.6thick}{k11}
   \fmfv{d.shape=circle, d.size=0.6thick}{k15}
   \fmfv{d.shape=circle, d.size=0.6thick}{k16}
   \fmfv{d.shape=circle, d.size=0.6thick}{k20}
\end{fmfgraph*}
$\quad \Big |\quad $
\begin{fmfgraph*}(35,25)\fmfkeep{v8}
   \fmfleft{i,d1}
   \fmfright{j,d2}
   \fmf{plain}{i,k1}\fmflabel{$i$}{i}
   \fmfn{plain}{k}{20}
   \fmf{plain}{k20,j}\fmflabel{$j$}{j}
   \fmffreeze
   \fmf{fillvx,left=0.75}{k1,k15}
   \fmf{unfillwx,left=0.75}{k5,k11}
   \fmf{fillwx,right=0.75}{k6,k20}
   \fmf{unfillwx,right=0.75}{k10,k16}
   \fmf{photon,left=0.75}{i,j}
   \fmf{photon,left=0.75}{k1,k15}
   \fmf{dashes,left=0.75}{k5,k11}
   \fmf{dashes,right=0.75}{k6,k20}
   \fmf{dashes,right=0.75}{k10,k16}
   \fmfv{d.shape=circle, d.filled=empty, d.size=3pt}{i}
   \fmfv{d.shape=circle, d.size=0.6thick}{j}
   \fmfv{d.shape=circle, d.filled=empty, d.size=3pt}{k1}
   \fmfv{d.shape=circle, d.size=0.6thick}{k5}
   \fmfv{d.shape=circle, d.size=0.6thick}{k6}
   \fmfv{d.shape=circle, d.size=0.6thick}{k10}
   \fmfv{d.shape=circle, d.size=0.6thick}{k11}
   \fmfv{d.shape=circle, d.size=0.6thick}{k15}
   \fmfv{d.shape=circle, d.size=0.6thick}{k16}
   \fmfv{d.shape=circle, d.size=0.6thick}{k20}
\end{fmfgraph*}
$\quad \Big |\quad $
\end{center}

\begin{center}
\begin{fmfgraph*}(35,25)\fmfkeep{v9}
   \fmfleft{i,d1}
   \fmfright{j,d2}
   \fmf{plain}{i,k1}\fmflabel{$i$}{i}
   \fmfn{plain}{k}{20}
   \fmf{plain}{k20,j}\fmflabel{$j$}{j}
   \fmffreeze
   \fmf{fillwx,left=0.75}{k1,k15}
   \fmf{unfillwx,left=0.75}{k5,k11}
   \fmf{fillvx,right=0.75}{k6,k20}
   \fmf{unfillwx,right=0.75}{k10,k16}
   \fmf{photon,left=0.75}{i,j}
   \fmf{dashes,left=0.75}{k1,k15}
   \fmf{dashes,left=0.75}{k5,k11}
   \fmf{photon,right=0.75}{k6,k20}
   \fmf{dashes,right=0.75}{k10,k16}
   \fmfv{d.shape=circle, d.size=0.6thick}{i}
   \fmfv{d.shape=circle, d.filled=empty, d.size=3pt}{j}
   \fmfv{d.shape=circle, d.size=0.6thick}{k1}
   \fmfv{d.shape=circle, d.size=0.6thick}{k5}
   \fmfv{d.shape=circle, d.size=0.6thick}{k6}
   \fmfv{d.shape=circle, d.size=0.6thick}{k10}
   \fmfv{d.shape=circle, d.size=0.6thick}{k11}
   \fmfv{d.shape=circle, d.size=0.6thick}{k15}
   \fmfv{d.shape=circle, d.size=0.6thick}{k16}
   \fmfv{d.shape=circle, d.filled=empty, d.size=3pt}{k20}
\end{fmfgraph*}
$\quad \Big |\quad $
\begin{fmfgraph*}(35,25)\fmfkeep{v10}
   \fmfleft{i,d1}
   \fmfright{j,d2}
   \fmf{plain}{i,k1}\fmflabel{$i$}{i}
   \fmfn{plain}{k}{20}
   \fmf{plain}{k20,j}\fmflabel{$j$}{j}
   \fmffreeze
   \fmf{fillwx,left=0.75}{k1,k15}
   \fmf{unfillvx,left=0.75}{k5,k11}
   \fmf{fillwx,right=0.75}{k6,k20}
   \fmf{unfillvx,right=0.75}{k10,k16}
   \fmf{photon,left=0.75}{i,j}
   \fmf{dashes,left=0.75}{k1,k15}
   \fmf{photon,left=0.75}{k5,k11}
   \fmf{dashes,right=0.75}{k6,k20}
   \fmf{photon,right=0.75}{k10,k16}
   \fmfv{d.shape=circle, d.size=0.6thick}{i}
   \fmfv{d.shape=circle, d.size=0.6thick}{j}
   \fmfv{d.shape=circle, d.size=0.6thick}{k1}
   \fmfv{d.shape=circle, d.size=0.6thick}{k5}
   \fmfv{d.shape=circle, d.size=0.6thick}{k6}
   \fmfv{d.shape=circle, d.filled=empty, d.size=3pt}{k10}
   \fmfv{d.shape=circle, d.filled=empty, d.size=3pt}{k11}
   \fmfv{d.shape=circle, d.size=0.6thick}{k15}
   \fmfv{d.shape=circle, d.size=0.6thick}{k16}
   \fmfv{d.shape=circle, d.size=0.6thick}{k20}
\end{fmfgraph*}
\end{center}\bigskip
\caption{\label{fig:vxrecur}
Recursion for the $vx$ matrix.}
\end{figure}
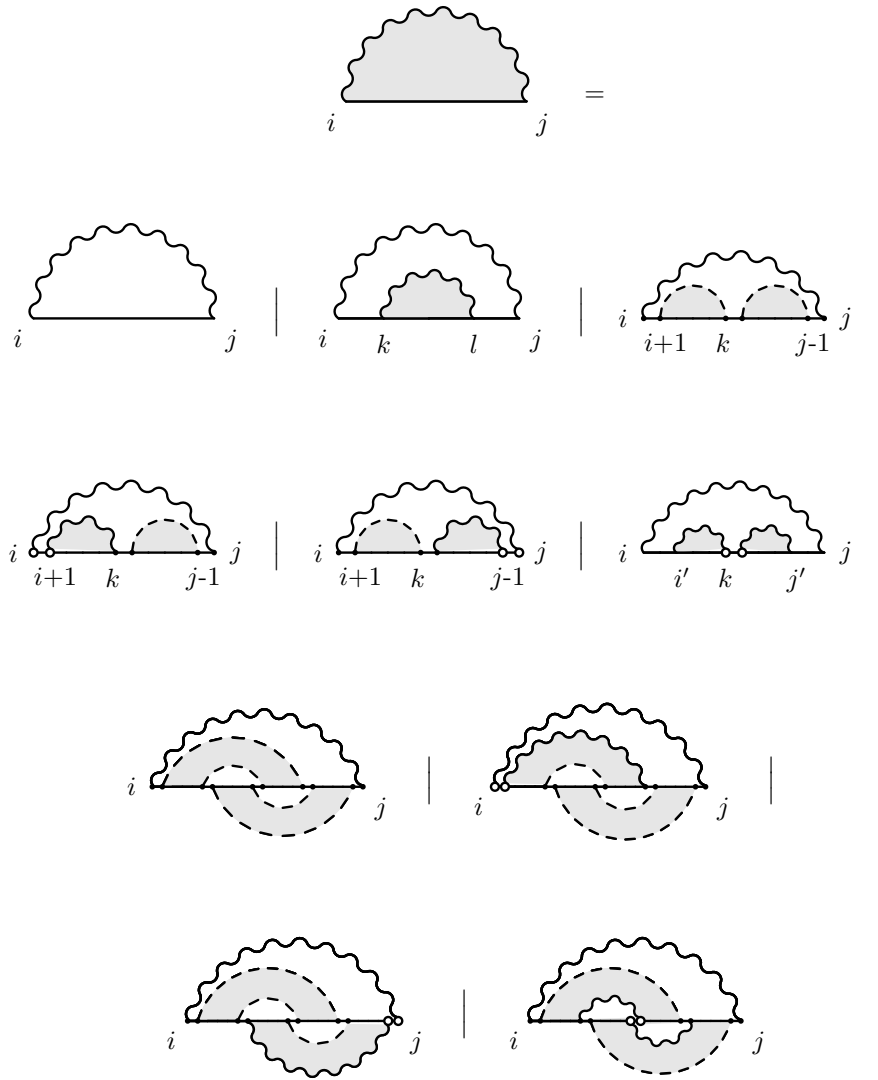

The recursion for the non-gap matrix $vx$ is given by
(cf. fig.~\ref{fig:vxrecur}):

\begin{equation}
vx(i,j) = optimal\left\{ 
\begin{array}{ll}

\vspace{2mm}

\left.
\begin{array}{l}
IS^1(i,j) 
\end{array}
\hspace{90.8mm}\right]
&
\mbox{IS(1)}\\

\vspace{2mm}

\left.
\begin{array}{l}
IS^2(i,j:k,l) + vx(k,l)\label{vxrecur}
\end{array}
\hspace{67.8mm}\right]
&
\mbox{IS(2)}\\

\vspace{2mm}

\left.
\begin{array}{l}
P_I + M + wx_I(i+1,k) + wx_I(k+1,j-1)\\
2*P_I + C(i,j:i+1,k) + M + vx(i+1,k) + wx_I(k+1,j-1)\\
2*P_I + C(j-1,k+1:j,i) + M + wx_I(i+1,k) + vx(k+1,j-1)\\
3*P_I + C(k,i^\prime:k+1,j^\prime) + M + vx(i^\prime,k) + vx(k+1,j^\prime)\\
\end{array}
\right]
&
\parbox[b][5mm][t]{80mm}{
nested 
\vspace{-2mm}

multiloops}\\

\vspace{2mm}

\left.
\begin{array}{l}
P_I + \widetilde{M} + G_{wI} + whx(i+1,r:k,l)\\ 
\hspace{6.0mm} + whx(k+1,j-1:l-1,r+1)\\

2*P_I + \widetilde{M} + G_{wI} + C(i,j,:i+1,r)\\ 
\hspace{6.0mm} + zhx(i+1,r:k,l) + whx(k+1,j-1:l-1,r+1)\\

2*P_I + \widetilde{M} + G_{wI} + C(j-1,k+1:j,i)\\ 
\hspace{6.0mm} + whx(i+1,r:k,l) + zhx(k+1,j-1:l-1,r+1)\\

3*P_I + \widetilde{M} + G_{wI} + C(l-1,r+1:l,k)\\ 
\hspace{6.0mm} + yhx(i+1,r:k,l) + yhx(k+1,j-1:l-1,r+1)
\end{array}
\hspace{17.7mm}\right]
&
\parbox[b][5mm][t]{80mm}{
non-nested 
\vspace{-2mm}

multiloops}\\

\end{array}
\right.
\end{equation}

\[[\forall i,i^\prime,k,l,r,j^\prime,j \quad 
i\leq i^\prime\leq k\leq l\leq r \leq j^\prime\leq j]\]

\noindent 
Parameters $P_I$, $\widetilde{M}$, and $G_{wI}$ are defined
in table~\ref{nestedparam}.

The initialization conditions are
\begin{eqnarray}
wx(i,i) &=& 0,\label{nestedinit}\\
vx(i,i) &=& + \infty.\nonumber
\end{eqnarray}
\[[\forall i\quad 1\leq i\leq N]\]

The recursion for the vhx matrix in the pseudoknot algorithm 
is given by (cf. fig.~\ref{fig:vhxrecur}):

\begin{equation}
vhx(i,j:k,l) = optimal\left\{ 
\begin{array}{l}
\widetilde {IS}^2(i,j:k,l) \\
\widetilde {IS}^2(i,j:r,s) + vhx(r,s:k,l)\label{vhxrecur}\\
\widetilde {IS}^2(r,s:k,l) + vhx(i,j:r,s)\\
2*\widetilde {P} + \widetilde {M} + whx(i+1,j-1:k-1,l+1)\\
\end{array}\right.
\end{equation}
\[[\forall i,r,k,l,s,j\quad i\leq r\leq k\leq l\leq s\leq j]\]

\noindent
Here $\widetilde {M}$ correspond to the score given to a multiloop in
a $vhx$ gap matrix.  It could be equal to $M$, the score defined for
multiloops in the $vx$ matrix, but it does not have to be.  Similarly,
the score for an irreducible surface of ${\cal O} (2)$, $\widetilde
{IS}^2$, could be the same as the one given for nested structures,
${IS}^2$, but again, it does not have to be. We found the best fits by
giving them values different to the ones used for nested foldings
(cf. table~\ref{nestedparam} and table~\ref{pkparam}).

The recursions for the gap matrices $zhx$ and $yhx$ in the pseudoknot
algorithm are complementary and given by 
(cf. fig.~\ref{fig:zhxrecur} and fig.~\ref{fig:yhxrecur}):

\begin{equation}
zhx(i,j:k,l) = optimal\left\{ 
\begin{array}{ll}

\vspace{2mm}

\left.
\begin{array}{l}
\widetilde P + vhx(i,j:k,l) 
\end{array}
\hspace{55.4mm}\right]
&
\mbox{paired}\\

\vspace{2mm}

\left.
\begin{array}{l}
\widetilde L + \widetilde R + \widetilde P 
+ vhx(i,j:k-1,l+1) \label{zhxrecur}\\
\widetilde R + \widetilde P + vhx(i,j:k-1,l)\\
\widetilde L + \widetilde P + vhx(i,j:k,l+1)
\end{array}
\hspace{29.3mm}\right]
&
\mbox{dangles}\\

\vspace{2mm}

\left.
\begin{array}{l}
\widetilde Q + zhx(i,j:k-1,l)\\
\widetilde Q + zhx(i,j:k,l+1)
\end{array}
\hspace{49mm}\right]
&
\parbox[b][5mm][t]{80mm}{
single 
\vspace{-2mm}

stranded}\\

\vspace{2mm}

\left.
\begin{array}{l}
zhx(i,j:r,l) + wx_I(r+1,k)\\
2*\widetilde P + C(r,l:r+1,k) + vhx(i,j:r,l) + vx(r+1,k)\\
zhx(i,j:k,s) + wx_I(l,s-1)\\
2*\widetilde P + C(s-1,l:s,k) + vhx(i,j:k,s) + vx(l,s-1)\\
\widetilde {IS}^2(i,j:r,s) + zhx(r,s:k,l)\\
\widetilde P + \widetilde M + whx(i+1,j-1:k,l)
\end{array}
\right]
&
\parbox[b][5mm][t]{80mm}{
nested 
\vspace{-2mm}

bifurcations}\\

\end{array}
\right.
\end{equation}

\begin{equation}
yhx(i,j:k,l) = optimal\left\{ 
\begin{array}{ll}

\vspace{2mm}

\left.
\begin{array}{l}
\widetilde P + vhx(i,j:k,l) 
\end{array}
\hspace{56.6mm}\right]
&
\mbox{paired}\\

\vspace{2mm}

\left.
\begin{array}{l}
\widetilde L + \widetilde R + \widetilde P + vhx(i+1,j-1:k,l)\label{yhxrecur}\\
\widetilde L + \widetilde P + vhx(i+1,j:k,l)\\
\widetilde R + \widetilde P + vhx(i,j-1:k,l)
\end{array}
\hspace{30.5mm}\right]
&
\mbox{dangles}\\

\vspace{2mm}

\left.
\begin{array}{l}
\widetilde Q + yhx(i+1,j:k,l)\\
\widetilde Q + yhx(i,j-1:k,l)
\end{array}
\hspace{50.2mm}\right]
&
\parbox[b][5mm][t]{80mm}{
single 
\vspace{-2mm}

stranded}\\

\left.
\begin{array}{l}
wx_I(i,r) + yhx(r+1,j:k,l)\\
2*\widetilde P + C(r,i:r+1,j) + vx(i,r) + vhx(r+1,j:k,l)\\
yhx(i,s:k,l) + wx_I(s+1,j)\\
2*\widetilde P + C(s,i:s+1,j)) + vhx(i,s:k,l) + vx(s+1,j)\\
yhx(i,j:r,s) + \widetilde{IS}^2(r,s:k,l)\\
\widetilde P + \widetilde M + whx(i,j:k-1,l+1)
\end{array}
\right]
&
\parbox[b][5mm][t]{80mm}{
nested 
\vspace{-2mm}

bifurcations}\\

\vspace{2mm}

\end{array}
\right.
\end{equation}
\[[\forall i,r,k,l,s,j\quad i\leq r\leq k\leq l\leq s\leq j]\]

Finally, the recursion for the gap matrix $whx$ appears in
fig.~\ref{fig:whxrecur}, and is given by:

\begin{equation}
whx(i,j:k,l) = optimal\left\{ 
\begin{array}{ll}

\vspace{2mm}

\left.
\begin{array}{l}
2*\widetilde P + vhx(i,j:k,l) \\
\widetilde P + zhx(i,j:k,l) \\
\widetilde P + yhx(i,j:k,l) 
\end{array}
\hspace{50.8mm}\right]
&
\mbox{paired}\\

\vspace{2mm}

\left.
\begin{array}{l}
\widetilde L + \widetilde R + 2*\widetilde P + vhx(i+1,j:k-1,l)\\
\widetilde L + \widetilde R + 2*\widetilde P + vhx(i,j-1:k,l+1)\\
2*\widetilde L + 2*\widetilde P + vhx(i+1,j:k,l+1)\\
2*\widetilde R + 2*\widetilde P + vhx(i,j-1:k-1,l)\\

\widetilde L + 2*\widetilde R + 2*\widetilde P + vhx(i+1,j-1:k-1,l)\\
2*\widetilde L + \widetilde R + 2*\widetilde P + vhx(i+1,j:k-1,l+1)\\
2*\widetilde L + \widetilde R + 2*\widetilde P + vhx(i+1,j-1:k,l+1)\\
\widetilde L + 2*\widetilde R + 2*\widetilde P + vhx(i,j-1:k-1,l+1)\\

2*\widetilde L +2*\widetilde R + 2*\widetilde P + + vhx(i+1,j-1:k-1,l+1)\\

\widetilde L + \widetilde R + \widetilde P + zhx(i+1,j-1:k,l)\\
\widetilde L + \widetilde P + zhx(i+1,j:k,l)\\
\widetilde R + \widetilde P + zhx(i,j-1:k,l)\\

\widetilde L + \widetilde R + \widetilde P + yhx(i,j:k-1,l+1)\\
\widetilde R + \widetilde P + yhx(i,j:k-1,l)\\
\widetilde L + \widetilde P + yhx(i,j:k,l+1)
\end{array}
\right]
&
\mbox{dangles}\\

\vspace{2mm}

\left.
\begin{array}{l}
\widetilde Q + whx(i+1,j:k,l)\\
\widetilde Q + whx(i,j-1:k,l)\\
\widetilde Q + whx(i,j:k-1,l)\\
\widetilde Q + whx(i,j:k,l+1)
\end{array}
\hspace{49.0mm}\right]
&
\parbox[b][5mm][t]{80mm}{
single 
\vspace{-2mm}

stranded}\\

\end{array}\right.
\end{equation}

\rightline{\it (cont.)}
\newpage

\begin{equation}
whx(i,j:k,l) = optimal\left\{ 
\begin{array}{ll}

\vspace{2mm}

\left.
\begin{array}{l}
wx_I(i,k) + wx_I(l,j)\label{whxrec}\\

wx_I(i,r) + whx(r+1,j:k,l)\\
2*\widetilde P + C(r,i:r+1,j) + vx(i,r) + zhx(r+1,j:k,l)\\

whx(i,j:r,l) + wx_I(r+1,k)\\
2*\widetilde P + C(r,l:r+1,k) + yhx(i,j:r,l) + vx(r+1,k)\\

whx(i,s:k,l) + wx_I(s+1,j)\\
2*\widetilde P + C(s,i:s+1,j) + zhx(i,s:k,l) + vx(s+1,j)\\

whx(i,j:k,s+1) + wx_I(l,s)\\
2*\widetilde P + C(s,l:s+1,k) + yhx(i,j:k,s+1) + vx(l,s)\\

yhx(i,j:r,s) + zhx(r,s:k,l)\\

\widetilde M + whx(i,j:r,s) + whx(r+1,s-1:k,l)
\end{array}
\right]
&
\parbox[b][5mm][t]{80mm}{
nested 
\vspace{-2mm}

bifurcations}\\

\left.
\begin{array}{l}
{G}_{wh} + whx(i,s:r,l) + whx(r+1,j:k,s+1)\\

{G}_{wh} + whx(i,s^\prime:k,s) + whx(l,j:s-1,s^\prime+1)\\

2*\widetilde P + {G}_{wh} + C(s^\prime,i:s^\prime+1,s-1) \\
         \hspace{7.5mm} + zhx(i,s^\prime:k,s) + yhx(l,j:s-1,s^\prime+1)\\

2*\widetilde P + {G}_{wh} + C(s-1,s^\prime+1:s,k) \\ 
         \hspace{7.5mm} + yhx(i,s^\prime:k,s) + yhx(l,j:s-1,s^\prime+1)\\

{G}_{wh} + whx(r,j:r^\prime,l) + whx(i,k:r-1,r^\prime+1)\\

2*\widetilde P + {G}_{wh} + C(r-1,r^\prime+1:r,j) \\ 
         \hspace{7.5mm} + zhx(r,j:r^\prime,l) + yhx(i,k:r-1,r^\prime+1)\\

2*\widetilde P + {G}_{wh} + C(r^\prime,l:r^\prime+1,r-1) \\ 
         \hspace{7.5mm} + yhx(r,j:r^\prime,l) + yhx(i,k:r-1,r^\prime+1)

\end{array}
\hspace{10mm}\right]
&
\parbox[b][5mm][t]{80mm}{
non-nested 
\vspace{-2mm}

bifurcations}\\

\end{array}\right.
\end{equation}
\[[\forall i,r,r^\prime,k,l,s,s^\prime,j\quad 
i\leq r\leq r^\prime\leq k\leq l\leq s\leq s^\prime\leq j]\]

Here ${G}_{wh}$ stands for the score given for finding overlapping
pseudoknots, that is pseudoknots that appear within already existing
pseudoknots. 

The initialization conditions are
\begin{eqnarray}
whx(i,j:i,j) &=& + \infty,\nonumber\\
vhx(i,j:k,k) &=& + \infty,\label{gapinit}\\
yhx(i,j:k,k) &=& + \infty,\nonumber\\
whx(i,j:k,k) &=& whx(i,j:k,k+1) = wx(i,j),\nonumber\\
zhx(i,j:k,k) &=& zhx(i,j:k,k+1) = vx(i,j).\nonumber
\end{eqnarray}
\[[\forall i,k,j \quad 1\leq i\leq k\leq j\leq N]\]


\bigskip
\Acknowledgements 

This work was supported by NIH grant HG01363 and by a gift from Eli
Lilly. The idea for the algorithm came from a discussion with Gary
Stormo at a meeting at the Aspen Center for Physics. Tim Hubbard
suggested parallel $\beta$-strands in proteins as an example of a set
of pairwise interactions that the algorithm cannot handle.


\begin{figure}[h]
\begin{center}
\begin{fmfgraph*}(30,30)\fmfkeep{vhx}
   \fmfleft{i,d1}
   \fmfright{j,d2}
   \fmf{plain}{i,k}
   \fmf{phantom}{k,m,l}
   \fmf{plain}{l,j}
   \fmffreeze
   \fmf{fillvx,left}{i,j}
   \fmf{unfillvx,left}{k,l}
   \fmfv{l=$k$,l.a=-90}{k}\fmflabel{$i$}{i} 
   \fmfv{l=$l$,l.a=-90}{l}\fmflabel{$j$}{j}
   \fmf{photon,left}{i,j}
   \fmf{photon,left}{k,l}
   \fmf{plain}{i,k}
\end{fmfgraph*}
\quad=\quad
\end{center}
\bigskip

\begin{fmfgraph*}(30,25)\fmfkeep{vh1}
   \fmfleft{i,d1}
   \fmfright{j,d2}
   \fmf{phantom}{i,k,h,l,j}
   \fmfv{l=$k$,l.a=-90}{k}\fmflabel{$i$}{i} 
   \fmfv{l=$l$,l.a=-90}{l}\fmflabel{$j$}{j}
   \fmffreeze 
   \fmf{plain}{i,k} \fmf{plain}{l,j}
   \fmf{photon,left}{i,j}
   \fmf{photon,left}{k,l}
\end{fmfgraph*}
$\quad\Big |\quad$
\begin{fmfgraph*}(30,25)\fmfkeep{vh2}
   \fmfleft{i,d1}
   \fmfright{j,d2}
   \fmf{phantom}{i,m,k,h,l,n,j}
   \fmfv{l=$k$,l.a=-65}{k}\fmflabel{$i$}{i} 
   \fmfv{l=$l$,l.a=-110}{l}\fmflabel{$j$}{j}
   \fmfv{l=$r$,l.a=-90}{m}\fmfv{l=$s$,l.a=-90}{n}
   \fmffreeze 
   \fmf{fillvx,left}{m,n}
   \fmf{unfillvx,left}{k,l}
   \fmf{plain}{i,m,k} \fmf{plain}{l,n,j}
   \fmf{photon,left}{i,j}
   \fmf{photon,left}{k,l}
   \fmf{photon,left}{m,n}
\end{fmfgraph*}
$\quad\Big |\quad$
\begin{fmfgraph*}(30,25)\fmfkeep{vh3}
   \fmfleft{i,d1}
   \fmfright{j,d2}
   \fmf{phantom}{i,m,k,h,l,n,j}
   \fmfv{l=$k$,l.a=-65}{k}\fmflabel{$i$}{i} 
   \fmfv{l=$l$,l.a=-110}{l}\fmflabel{$j$}{j}
   \fmfv{l=$r$,l.a=-90}{m}\fmfv{l=$s$,l.a=-90}{n}
   \fmffreeze 
   \fmf{fillvx,left}{i,j}
   \fmf{unfillvx,left}{m,n}
   \fmf{plain}{i,m,k} \fmf{plain}{l,n,j}
   \fmf{photon,left}{i,j}
   \fmf{photon,left}{k,l}
   \fmf{photon,left}{m,n}
\end{fmfgraph*}
$\quad\Big |\quad$

\begin{center}
\begin{fmfgraph*}(30,25)\fmfkeep{vh4}
   \fmfleft{i,d1}
   \fmfright{j,d2}
   \fmf{phantom}{i,h1}
   \fmfn{phantom}{h}{16}
   \fmf{phantom}{h16,j}
   \fmfv{l=$k$,l.a=-90}{h5}\fmflabel{$i$}{i} 
   \fmfv{l=$l$,l.a=-90}{h12}\fmflabel{$j$}{j}
   \fmffreeze 
   \fmf{fillwx,left}{h1,h16}
   \fmf{unfillwx,left}{h4,h13}
   \fmf{dashes,left}{h1,h16}
   \fmf{dashes,left}{h4,h13}
   \fmf{plain}{i,h5} \fmf{plain}{h12,j}
   \fmf{photon,left}{i,j}
   \fmf{photon,left}{h5,h12}
   \fmfv{d.shape=circle, d.size=0.6thick}{i}
   \fmfv{d.shape=circle, d.size=0.6thick}{j}
   \fmfv{d.shape=circle, d.size=0.6thick}{h1}
   \fmfv{d.shape=circle, d.size=0.6thick}{h4}
   \fmfv{d.shape=circle, d.size=0.6thick}{h5}
   \fmfv{d.shape=circle, d.size=0.6thick}{h12}
   \fmfv{d.shape=circle, d.size=0.6thick}{h13}
   \fmfv{d.shape=circle, d.size=0.6thick}{h16}
\end{fmfgraph*}
\end{center}\bigskip
\caption{\label{fig:vhxrecur}
Recursion for the $vhx$ matrix.}
\end{figure}
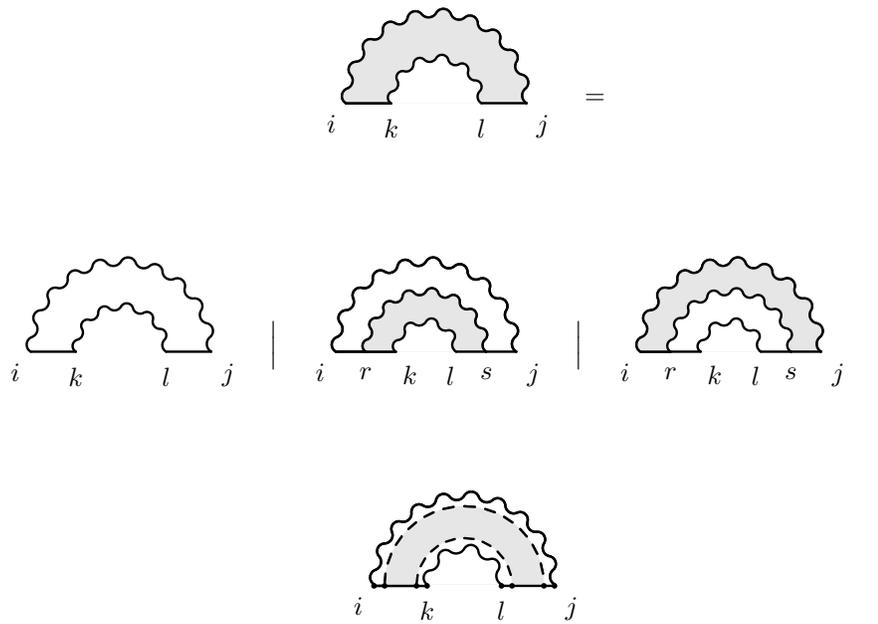

\begin{figure}[h]
\begin{center}
\begin{fmfgraph*}(30,30)\fmfkeep{zhx}
   \fmfleft{i,d1}
   \fmfright{j,d2}
   \fmf{plain}{i,k}
   \fmf{phantom}{k,m,l}
   \fmf{plain}{l,j}
   \fmfv{l=$k$,l.a=-90}{k}\fmflabel{$i$}{i} 
   \fmfv{l=$l$,l.a=-90}{l}\fmflabel{$j$}{j}
   \fmffreeze
   \fmf{fillvx,left}{i,j}
   \fmf{unfillwx,left}{k,l}
   \fmf{photon,left}{i,j}
   \fmf{dashes,left}{k,l}
   \fmf{plain}{i,k}\fmf{plain}{l,j}
\end{fmfgraph*}
\quad=\quad
\end{center}

\begin{fmfgraph*}(30,25)\fmfkeep{zh1}
   \fmfleft{i,d1}
   \fmfright{j,d2}
   \fmf{plain}{i,k}
   \fmf{phantom}{k,m,l}
   \fmf{plain}{l,j}
   \fmfv{l=$k$,l.a=-90}{k}\fmflabel{$i$}{i} 
   \fmfv{l=$l$,l.a=-90}{l}\fmflabel{$j$}{j}
   \fmffreeze
   \fmf{fillvx,left}{i,j}
   \fmf{unfillvx,left}{k,l}
   \fmf{photon,left}{i,j}
   \fmf{photon,left}{k,l}
   \fmf{plain}{i,k}\fmf{plain}{l,j}
\end{fmfgraph*}
$\quad\Big |\quad$
\begin{fmfgraph*}(30,25)\fmfkeep{zh2}
   \fmfleft{i,d1}
   \fmfright{j,d2}
   \fmf{phantom}{i,h1}
   \fmfn{phantom}{h}{16}
   \fmf{phantom}{h16,j}
   \fmffreeze 
   \fmf{fillvx,left}{h4,h13}
   \fmf{fillvx,left}{i,j}
   \fmf{unfillvx,left}{h4,h13}
   \fmf{photon,left}{i,j}
   \fmf{photon,left}{h4,h13}
   \fmf{plain}{i,h5} \fmf{plain}{h12,j}
   \fmfv{l=$k$,l.a=-90}{h5}\fmfv{d.shape=,d.size=0.2thick,l=$i$}{i} 
   \fmfv{l=$l$,l.a=-90}{h12}\fmfv{d.shape=,d.size=0.2thick,l=$j$}{j}
   \fmfv{d.shape=circle, d.size=0.6thick}{h5}
   \fmfv{d.shape=circle, d.size=0.6thick}{h4}
   \fmfv{d.shape=circle, d.size=0.6thick}{h12}
   \fmfv{d.shape=circle, d.size=0.6thick}{h13}
\end{fmfgraph*}
$\quad\Big |\quad$
\begin{fmfgraph*}(30,25)\fmfkeep{zh3}
   \fmfleft{i,d1}
   \fmfright{j,d2}
   \fmf{phantom}{i,h1}
   \fmfn{phantom}{h}{16}
   \fmf{phantom}{h16,j}
   \fmfv{l=$k$,l.a=-90}{h5}\fmfv{d.shape=,d.size=0.2thick,l=$i$}{i} 
   \fmfv{l=$l$,l.a=-90}{h12}\fmfv{d.shape=,d.size=0.2thick,l=$j$}{j}
   \fmffreeze 
   \fmf{fillvx,left}{i,j}
   \fmf{unfillvx,left}{h4,h12}
   \fmf{plain}{i,h5} \fmf{plain}{h12,j}
   \fmf{photon,left}{i,j}
   \fmf{photon,left}{h4,h12}
   \fmfv{d.shape=circle, d.size=0.6thick}{h5}
   \fmfv{d.shape=circle, d.size=0.6thick}{h4}
\end{fmfgraph*}
$\quad\Big |\quad$

\begin{fmfgraph*}(30,25)\fmfkeep{zh4}
   \fmfleft{i,d1}
   \fmfright{j,d2}
   \fmf{phantom}{i,h1}
   \fmfn{phantom}{h}{16}
   \fmf{phantom}{h16,j}
   \fmfv{l=$k$,l.a=-90}{h5}\fmfv{d.shape=,d.size=0.2thick,l=$i$}{i}
   \fmfv{l=$l$,l.a=-90}{h12}\fmfv{d.shape=,d.size=0.2thick,l=$j$}{j}
   \fmffreeze 
   \fmf{fillvx,left}{i,j}
   \fmf{unfillvx,left}{h5,h13}
   \fmf{plain}{i,h5} \fmf{plain}{h12,j}
   \fmf{photon,left}{i,j}
   \fmf{photon,left}{h5,h13}
   \fmfv{d.shape=circle, d.size=0.6thick}{h13}
   \fmfv{d.shape=circle, d.size=0.6thick}{h12}
\end{fmfgraph*}
$\quad\Big |\quad$
\begin{fmfgraph*}(30,25)\fmfkeep{zh5}
   \fmfleft{i,d1}
   \fmfright{j,d2}
   \fmf{phantom}{i,h1}
   \fmfn{phantom}{h}{16}
   \fmf{phantom}{h16,j}
   \fmfv{l=$k$,l.a=-90}{h5}\fmfv{d.shape=,d.size=0.2thick,l=$i$}{i}
   \fmfv{l=$l$,l.a=-90}{h12}\fmfv{d.shape=,d.size=0.2thick,l=$j$}{j}
   \fmffreeze 
   \fmf{fillvx,left}{i,j}
   \fmf{unfillwx,left}{h4,h12}
   \fmf{plain}{i,h5} \fmf{plain}{h12,j}
   \fmf{photon,left}{i,j}
   \fmf{dashes,left}{h4,h12}
   \fmfv{d.shape=circle, d.size=0.6thick}{h5}
   \fmfv{d.shape=circle, d.size=0.6thick}{h4}
\end{fmfgraph*}
$\quad\Big |\quad$
\begin{fmfgraph*}(30,25)\fmfkeep{zh6}
   \fmfleft{i,d1}
   \fmfright{j,d2}
   \fmf{phantom}{i,h1}
   \fmfn{phantom}{h}{16}
   \fmf{phantom}{h16,j}
   \fmfv{l=$k$,l.a=-90}{h5}\fmfv{d.shape=,d.size=0.2thick,l=$i$}{i}
   \fmfv{l=$l$,l.a=-90}{h12}\fmfv{d.shape=,d.size=0.2thick,l=$j$}{j}
   \fmffreeze 
   \fmf{fillvx,left}{i,j}
   \fmf{unfillwx,left}{h5,h13}
   \fmf{plain}{i,h5} \fmf{plain}{h12,j}
   \fmf{photon,left}{i,j}
   \fmf{dashes,left}{h5,h13}
   \fmfv{d.shape=circle, d.size=0.6thick}{h13}
   \fmfv{d.shape=circle, d.size=0.6thick}{h12}
\end{fmfgraph*}
$\quad\Big |$

\begin{fmfgraph*}(30,25)\fmfkeep{zh7}
   \fmfleft{i,d1}
   \fmfright{j,d2}
   \fmf{phantom}{i,h1}
   \fmfn{phantom}{h}{16}
   \fmf{phantom}{h16,j}
   \fmfv{l=$k$,l.a=-90}{h7}\fmfv{d.shape=,d.size=0.2thick,l=$i$}{i}
   \fmfv{l=$l$,l.a=-90}{h13}\fmfv{d.shape=,d.size=0.2thick,l=$j$}{j}
   \fmfv{l=$r$,l.a=-90}{h3}
   \fmffreeze 
   \fmf{fillvx,left}{i,j}
   \fmf{unfillwx,left}{h3,h13}
   \fmf{fillwx,left}{h4,h7}
   \fmf{photon,left}{i,j}
   \fmf{dashes,left}{h3,h13}
   \fmf{dashes,left}{h4,h7}
   \fmfv{d.shape=circle, d.size=0.6thick}{h3}
   \fmfv{d.shape=circle, d.size=0.6thick}{h4}
   \fmf{plain}{i,h1,h2,h3,h4,h5,h6,h7} \fmf{plain}{h13,j}
\end{fmfgraph*}
$\quad\Big |\quad$
\begin{fmfgraph*}(30,25)\fmfkeep{zh8}
   \fmfleft{i,d1}
   \fmfright{j,d2}
   \fmf{phantom}{i,h1}
   \fmfn{phantom}{h}{16}
   \fmf{phantom}{h16,j}
   \fmfv{l=$k$,l.a=-90}{h7}\fmfv{d.shape=,d.size=0.2thick,l=$i$}{i}
   \fmfv{l=$l$,l.a=-90}{h13}\fmfv{d.shape=,d.size=0.2thick,l=$j$}{j}
   \fmfv{l=$r$,l.a=-90}{h3}
   \fmffreeze 
   \fmf{fillvx,left}{i,j}
   \fmf{unfillvx,left}{h3,h13}
   \fmf{fillvx,left}{h4,h7}
   \fmf{plain}{i,h3}\fmf{plain}{h3,h4,h5,h6,h7} \fmf{plain}{h13,j}
   \fmf{photon,left}{i,j}
   \fmf{photon,left}{h3,h13}
   \fmf{photon,left}{h4,h7}
   \fmfv{d.shape=circle, d.filled=empty, d.size=3pt}{h3}
   \fmfv{d.shape=circle, d.filled=empty, d.size=3pt}{h4}
\end{fmfgraph*}
$\quad\Big |\quad$
\begin{fmfgraph*}(30,25)\fmfkeep{zh9}
   \fmfleft{i,d1}
   \fmfright{j,d2}
   \fmf{phantom}{i,h1}
   \fmfn{phantom}{h}{16}
   \fmf{phantom}{h16,j}
   \fmfv{l=$k$,l.a=-90}{h5}\fmfv{d.shape=,d.size=0.2thick,l=$i$}{i}
   \fmfv{l=$l$,l.a=-90}{h10}\fmfv{d.shape=,d.size=0.2thick,l=$j$}{j}
   \fmfv{l=$s$,l.a=-90}{h14}
   \fmffreeze 
   \fmf{fillvx,left}{i,j}
   \fmf{unfillwx,left}{h4,h14}
   \fmf{fillwx,left}{h10,h13}
   \fmf{plain}{i,h4} \fmf{plain}{h10,j}
   \fmf{photon,left}{i,j}
   \fmf{dashes,left}{h4,h14}
   \fmf{dashes,left}{h10,h13}
   \fmfv{d.shape=circle, d.size=0.6thick}{h13}
   \fmfv{d.shape=circle, d.size=0.6thick}{h14}
\end{fmfgraph*}
$\quad\Big |\quad$

\begin{fmfgraph*}(30,25)\fmfkeep{zh10}
   \fmfleft{i,d1}
   \fmfright{j,d2}
   \fmf{phantom}{i,h1}
   \fmfn{phantom}{h}{16}
   \fmf{phantom}{h16,j}
   \fmfv{l=$k$,l.a=-90}{h4}\fmfv{d.shape=,d.size=0.2thick,l=$i$}{i}
   \fmfv{l=$l$,l.a=-90}{h10}\fmfv{d.shape=,d.size=0.2thick,l=$j$}{j}
   \fmfv{l=$s$,l.a=-90}{h14}
   \fmffreeze 
   \fmf{fillvx,left}{i,j}
   \fmf{unfillvx,left}{h4,h14}
   \fmf{fillvx,left}{h10,h13}
   \fmf{plain}{i,h4} \fmf{plain}{h10,j}
   \fmf{photon,left}{i,j}
   \fmf{photon,left}{h4,h14}
   \fmf{photon,left}{h10,h13}
   \fmfv{d.shape=circle, d.filled=empty, d.size=3pt}{h13}
   \fmfv{d.shape=circle, d.filled=empty, d.size=3pt}{h14}
\end{fmfgraph*}
$\quad\Big |\quad$
\begin{fmfgraph*}(30,25)\fmfkeep{zh11}
   \fmfleft{i,d1}
   \fmfright{j,d2}
   \fmf{phantom}{i,r,k,h,l,s,j}
   \fmfv{l=$k$,l.a=-90}{k}\fmflabel{$i$}{i} 
   \fmfv{l=$l$,l.a=-90}{l}\fmflabel{$j$}{j}
   \fmfv{l=$r$,l.a=-90}{r}\fmfv{l=$s$,l.a=-90}{s}
   \fmffreeze 
   \fmf{fillvx,left}{r,s}
   \fmf{unfillwx,left}{k,l}
   \fmf{photon,left}{i,j}
   \fmf{photon,left}{r,s}
   \fmf{dashes,left}{k,l}
   \fmf{plain}{i,r,k}\fmf{plain}{l,s,j}
\end{fmfgraph*}
$\quad\Big |\quad$
\begin{fmfgraph*}(30,25)\fmfkeep{zh12}
   \fmfleft{i,d1}
   \fmfright{j,d2}
   \fmf{phantom}{i,h1}
   \fmfn{phantom}{h}{16}
   \fmf{phantom}{h16,j}
   \fmfv{l=$k$,l.a=-90}{h5}\fmfv{d.shape=,d.size=0.2thick,l=$i$}{i}
   \fmfv{l=$l$,l.a=-90}{h12}\fmfv{d.shape=,d.size=0.2thick,l=$j$}{j}
   \fmffreeze 
   \fmf{fillwx,left}{h1,h16}
   \fmf{unfillwx,left}{h5,h12}
   \fmf{plain}{i,h5} \fmf{plain}{h12,j}
   \fmf{photon,left}{i,j}
   \fmf{dashes,left}{h1,h16}
   \fmf{dashes,left}{h5,h12}
\end{fmfgraph*}\bigskip
\caption{\label{fig:zhxrecur}
Recursion for the $zhx$ matrix.}
\end{figure}
\clearpage

\begin{figure}[h]
\begin{center}
\begin{fmfgraph*}(30,30)\fmfkeep{yhx}
   \fmfleft{i,d1}
   \fmfright{j,d2}
   \fmf{plain}{i,k}
   \fmf{phantom}{k,m,l}
   \fmf{plain}{l,j}
   \fmfv{l=$k$,l.a=-90}{k}\fmflabel{$i$}{i} 
   \fmfv{l=$l$,l.a=-90}{l}\fmflabel{$j$}{j}
   \fmffreeze
   \fmf{fillwx,left}{i,j}
   \fmf{unfillvx,left}{k,l}
   \fmf{dashes,left}{i,j}
   \fmf{photon,left}{k,l}
   \fmf{plain}{i,k}\fmf{plain}{l,j}
\end{fmfgraph*}
\quad=\quad
\end{center}

\begin{fmfgraph*}(30,25)\fmfkeep{yh1}
   \fmfleft{i,d1}
   \fmfright{j,d2}
   \fmf{plain}{i,k}
   \fmf{phantom}{k,m,l}
   \fmf{plain}{l,j}
   \fmfv{l=$k$,l.a=-90}{k}\fmflabel{$i$}{i} 
   \fmfv{l=$l$,l.a=-90}{l}\fmflabel{$j$}{j}
   \fmffreeze
   \fmf{fillvx,left}{i,j}
   \fmf{unfillvx,left}{k,l}
   \fmf{photon,left}{i,j}
   \fmf{photon,left}{k,l}
   \fmf{plain}{i,k}\fmf{plain}{l,j}
\end{fmfgraph*}
$\quad\Big |\quad$
\begin{fmfgraph*}(30,25)\fmfkeep{yh2}
   \fmfleft{i,d1}
   \fmfright{j,d2}
   \fmf{phantom}{i,h1}
   \fmfn{phantom}{h}{16}
   \fmf{phantom}{h16,j}
   \fmfv{l=$k$,l.a=-90}{h5}\fmflabel{$i$}{i} 
   \fmfv{l=$l$,l.a=-90}{h12}\fmflabel{$j$}{j}
   \fmffreeze 
   \fmf{fillvx,left}{h1,h16}
   \fmf{unfillvx,left}{h5,h12}
   \fmf{plain}{i,h5} \fmf{plain}{h12,j}
   \fmf{photon,left}{h1,h16}
   \fmf{photon,left}{h5,h12}
   \fmfv{d.shape=circle, d.size=0.6thick}{i}
   \fmfv{d.shape=circle, d.size=0.6thick}{h1}
   \fmfv{d.shape=circle, d.size=0.6thick}{j}
   \fmfv{d.shape=circle, d.size=0.6thick}{h16}
\end{fmfgraph*}
$\quad\Big |\quad$
\begin{fmfgraph*}(30,25)\fmfkeep{yh3}
   \fmfleft{i,d1}
   \fmfright{j,d2}
   \fmf{phantom}{i,h1}
   \fmfn{phantom}{h}{16}
   \fmf{phantom}{h16,j}
   \fmfv{l=$k$,l.a=-90}{h5}\fmflabel{$i$}{i} 
   \fmfv{l=$l$,l.a=-90}{h12}\fmflabel{$j$}{j}
   \fmffreeze 
   \fmf{fillvx,left}{h1,j}
   \fmf{unfillvx,left}{h5,h12}
   \fmf{plain}{i,h5} \fmf{plain}{h12,j}
   \fmf{photon,left}{h1,j}
   \fmf{photon,left}{h5,h12}
   \fmfv{d.shape=circle, d.size=0.6thick}{i}
   \fmfv{d.shape=circle, d.size=0.6thick}{h1}
\end{fmfgraph*}
$\quad\Big |\quad$

\begin{fmfgraph*}(30,25)\fmfkeep{yh4}
   \fmfleft{i,d1}
   \fmfright{j,d2}
   \fmf{phantom}{i,h1}
   \fmfn{phantom}{h}{16}
   \fmf{phantom}{h16,j}
   \fmfv{l=$k$,l.a=-90}{h5}\fmflabel{$i$}{i} 
   \fmfv{l=$l$,l.a=-90}{h12}\fmflabel{$j$}{j}
   \fmffreeze 
   \fmf{fillvx,left}{i,h16}
   \fmf{unfillvx,left}{h5,h12}
   \fmf{plain}{i,h5} \fmf{plain}{h12,j}
   \fmf{photon,left}{h5,h12}
   \fmf{photon,left}{i,h16}
   \fmfv{d.shape=circle, d.size=0.6thick}{j}
   \fmfv{d.shape=circle, d.size=0.6thick}{h16}
\end{fmfgraph*}
$\quad\Big |\quad$
\begin{fmfgraph*}(30,25)\fmfkeep{yh5}
   \fmfleft{i,d1}
   \fmfright{j,d2}
   \fmf{phantom}{i,h1}
   \fmfn{phantom}{h}{16}
   \fmf{phantom}{h16,j}
   \fmfv{l=$k$,l.a=-90}{h5}\fmflabel{$i$}{i} 
   \fmfv{l=$l$,l.a=-90}{h12}\fmflabel{$j$}{j}
   \fmffreeze 
   \fmf{fillwx,left}{h1,j}
   \fmf{unfillvx,left}{h5,h12}
   \fmf{plain}{i,h5} \fmf{plain}{h12,j}
   \fmf{photon,left}{h5,h12}
   \fmf{dashes,left}{h1,j}
   \fmfv{d.shape=circle, d.size=0.6thick}{i}
   \fmfv{d.shape=circle, d.size=0.6thick}{h1}
\end{fmfgraph*}
$\quad\Big |\quad$
\begin{fmfgraph*}(30,25)\fmfkeep{yh6}
   \fmfleft{i,d1}
   \fmfright{j,d2}
   \fmf{phantom}{i,h1}
   \fmfn{phantom}{h}{16}
   \fmf{phantom}{h16,j}
   \fmfv{l=$k$,l.a=-90}{h5}\fmflabel{$i$}{i} 
   \fmfv{l=$l$,l.a=-90}{h12}\fmflabel{$j$}{j}
   \fmffreeze 
   \fmf{fillwx,left}{i,h16}
   \fmf{unfillvx,left}{h5,h12}
   \fmf{plain}{i,h5} \fmf{plain}{h12,j}
   \fmf{photon,left}{h5,h12}
   \fmf{dashes,left}{i,h16}
   \fmfv{d.shape=circle, d.size=0.6thick}{j}
   \fmfv{d.shape=circle, d.size=0.6thick}{h16}
\end{fmfgraph*}
$\quad\Big |$

\begin{fmfgraph*}(30,25)\fmfkeep{yh7}
   \fmfleft{i,d1}
   \fmfright{j,d2}
   \fmf{phantom}{i,h1}
   \fmfn{phantom}{h}{16}
   \fmf{phantom}{h16,j}
   \fmfv{l=$k$,l.a=-90}{h7}\fmflabel{$i$}{i} 
   \fmfv{l=$l$,l.a=-90}{h13}\fmflabel{$j$}{j}
   \fmfv{l=$r$,l.a=-90}{h3}
   \fmffreeze 
   \fmf{fillwx,left}{i,h3}
   \fmf{fillwx,left}{h4,j}
   \fmf{unfillvx,left}{h7,h13}
   \fmf{plain}{i,h7} \fmf{plain}{h13,j}
   \fmf{photon,left}{h7,h13}
   \fmf{dashes,left}{h4,j}
   \fmf{dashes,left}{i,h3}
   \fmfv{d.shape=circle, d.size=0.6thick}{h3}
   \fmfv{d.shape=circle, d.size=0.6thick}{h4}
\end{fmfgraph*}
$\quad\Big |\quad$
\begin{fmfgraph*}(30,25)\fmfkeep{yh8}
   \fmfleft{i,d1}
   \fmfright{j,d2}
   \fmf{phantom}{i,h1}
   \fmfn{phantom}{h}{16}
   \fmf{phantom}{h16,j}
   \fmfv{l=$k$,l.a=-90}{h7}\fmflabel{$i$}{i} 
   \fmfv{l=$l$,l.a=-90}{h13}\fmflabel{$j$}{j}
   \fmfv{l=$r$,l.a=-90}{h3}
   \fmffreeze 
   \fmf{fillvx,left}{i,h3}
   \fmf{fillvx,left}{h4,j}
   \fmf{unfillvx,left}{h7,h13}
   \fmf{plain}{i,h7} \fmf{plain}{h13,j}
   \fmf{photon,left}{h7,h13}
   \fmf{photon,left}{h4,j}
   \fmf{photon,left}{i,h3}
   \fmfv{d.shape=circle, d.filled=empty, d.size=3pt}{h3}
   \fmfv{d.shape=circle, d.filled=empty, d.size=3pt}{h4}
\end{fmfgraph*}
$\quad\Big |\quad$
\begin{fmfgraph*}(30,25)\fmfkeep{yh9}
   \fmfleft{i,d1}
   \fmfright{j,d2}
   \fmf{phantom}{i,h1}
   \fmfn{phantom}{h}{16}
   \fmf{phantom}{h16,j}
   \fmfv{l=$k$,l.a=-90}{h4}\fmflabel{$i$}{i} 
   \fmfv{l=$l$,l.a=-90}{h10}\fmflabel{$j$}{j}
   \fmfv{l=$s$,l.a=-90}{h13}
   \fmffreeze 
   \fmf{fillwx,left}{h14,j}
   \fmf{fillwx,left}{i,h13}
   \fmf{unfillvx,left}{h4,h10}
   \fmf{plain}{i,h4} \fmf{plain}{h10,h11,h12,h13,h14,h16,j}
   \fmf{photon,left}{h4,h10}
   \fmf{dashes,left}{i,h13}
   \fmf{dashes,left}{h14,j}
   \fmfv{d.shape=circle, d.size=0.6thick}{h13}
   \fmfv{d.shape=circle, d.size=0.6thick}{h14}
\end{fmfgraph*}
$\quad\Big |\quad$

\begin{fmfgraph*}(30,25)\fmfkeep{yh10}
   \fmfleft{i,d1}
   \fmfright{j,d2}
   \fmf{phantom}{i,h1}
   \fmfn{phantom}{h}{16}
   \fmf{phantom}{h16,j}
   \fmfv{l=$k$,l.a=-90}{h4}\fmflabel{$i$}{i} 
   \fmfv{l=$l$,l.a=-90}{h10}\fmflabel{$j$}{j}
   \fmfv{l=$s$,l.a=-90}{h13}
   \fmffreeze 
   \fmf{fillvx,left}{h14,j}
   \fmf{fillvx,left}{i,h13}
   \fmf{unfillvx,left}{h4,h10}
   \fmf{plain}{i,h4} \fmf{plain}{h10,j}
   \fmf{photon,left}{h4,h10}
   \fmf{photon,left}{i,h13}
   \fmf{photon,left}{h14,j}
   \fmfv{d.shape=circle, d.filled=empty, d.size=3pt}{h13}
   \fmfv{d.shape=circle, d.filled=empty, d.size=3pt}{h14}
\end{fmfgraph*}
$\quad\Big |\quad$
\begin{fmfgraph*}(30,25)\fmfkeep{yh11}
   \fmfleft{i,d1}
   \fmfright{j,d2}
   \fmf{phantom}{i,r,k,h,l,s,j}
   \fmfv{l=$k$,l.a=-90}{k}\fmflabel{$i$}{i} 
   \fmfv{l=$l$,l.a=-90}{l}\fmflabel{$j$}{j}
   \fmfv{l=$r$,l.a=-90}{r}\fmfv{l=$s$,l.a=-90}{s}
   \fmffreeze 
   \fmf{fillwx,left}{i,j}
   \fmf{unfillvx,left}{r,s}
   \fmf{plain}{i,r,k}\fmf{plain}{l,s,j}
   \fmf{photon,left}{k,l}
   \fmf{photon,left}{r,s}
   \fmf{dashes,left}{i,j}
\end{fmfgraph*}
$\quad\Big |\quad$
\begin{fmfgraph*}(30,25)\fmfkeep{yh12}
   \fmfleft{i,d1}
   \fmfright{j,d2}
   \fmf{phantom}{i,h1}
   \fmfn{phantom}{h}{16}
   \fmf{phantom}{h16,j}
   \fmfv{l=$k$,l.a=-90}{h5}\fmfv{d.shape=,d.size=0.2thick,l=$i$}{i}
   \fmfv{l=$l$,l.a=-90}{h12}\fmfv{d.shape=,d.size=0.2thick,l=$j$}{j}
   \fmffreeze 
   \fmf{fillwx,left}{i,j}
   \fmf{unfillwx,left}{h4,h13}
   \fmf{plain}{i,h5} \fmf{plain}{h12,j}
   \fmf{photon,left}{h5,h12}
   \fmf{dashes,left}{i,j}
   \fmf{dashes,left}{h4,h13}
\end{fmfgraph*}\bigskip
\caption{\label{fig:yhxrecur}
Recursion for the $yhx$ matrix.}
\end{figure}

\begin{figure}[h]
\begin{center}
\begin{fmfgraph*}(30,30)\fmfkeep{whx}
   \fmfleft{i,d1}
   \fmfright{j,d2}
   \fmf{plain}{i,k}
   \fmf{phantom}{k,m,l}
   \fmf{plain}{l,j}
   \fmfv{l=$k$,l.a=-90}{k}\fmflabel{$i$}{i} 
   \fmfv{l=$l$,l.a=-90}{l}\fmflabel{$j$}{j}
   \fmffreeze
   \fmf{fillwx,left}{i,j}
   \fmf{unfillwx,left}{k,l}
   \fmf{dashes,left}{i,j}
   \fmf{dashes,left}{k,l}
   \fmf{plain}{i,k} \fmf{plain}{l,j}
\end{fmfgraph*}
\quad=
\end{center}

\begin{fmfgraph*}(30,25)\fmfkeep{wh1(1)}
   \fmfleft{i,d1}
   \fmfright{j,d2}
   \fmf{plain}{i,k}
   \fmf{phantom}{k,m,l}
   \fmf{plain}{l,j}
   \fmfv{l=$k$,l.a=-90}{k}\fmflabel{$i$}{i} 
   \fmfv{l=$l$,l.a=-90}{l}\fmflabel{$j$}{j}
   \fmffreeze
   \fmf{fillvx,left}{i,j}
   \fmf{unfillvx,left}{k,l}
   \fmf{photon,left}{i,j}
   \fmf{photon,left}{k,l}
   \fmf{plain}{i,k} \fmf{plain}{l,j}
\end{fmfgraph*}
$\quad\Big |\quad$
\begin{fmfgraph*}(30,25)\fmfkeep{wh2(1)}
   \fmfleft{i,d1}
   \fmfright{j,d2}
   \fmf{plain}{i,k}
   \fmf{phantom}{k,m,l}
   \fmf{plain}{l,j}
   \fmfv{l=$k$,l.a=-90}{k}\fmflabel{$i$}{i} 
   \fmfv{l=$l$,l.a=-90}{l}\fmflabel{$j$}{j}
   \fmffreeze
   \fmf{fillvx,left}{i,j}
   \fmf{unfillwx,left}{k,l}
   \fmf{photon,left}{i,j}
   \fmf{dashes,left}{k,l}
   \fmf{plain}{i,k}\fmf{plain}{l,j}
\end{fmfgraph*}
$\quad\Big |\quad$
\begin{fmfgraph*}(30,25)\fmfkeep{wh3(1)}
   \fmfleft{i,d1}
   \fmfright{j,d2}
   \fmf{plain}{i,k}
   \fmf{phantom}{k,m,l}
   \fmf{plain}{l,j}
   \fmfv{l=$k$,l.a=-90}{k}\fmflabel{$i$}{i} 
   \fmfv{l=$l$,l.a=-90}{l}\fmflabel{$j$}{j}
   \fmffreeze
   \fmf{fillwx,left}{i,j}
   \fmf{unfillvx,left}{k,l}
   \fmf{dashes,left}{i,j}
   \fmf{photon,left}{k,l}
   \fmf{plain}{i,k}\fmf{plain}{l,j}
\end{fmfgraph*}
$\quad\Big |\quad$

\begin{fmfgraph*}(30,25)\fmfkeep{wh1(2)}
   \fmfleft{i,d1}
   \fmfright{j,d2}
   \fmf{phantom}{i,h1}
   \fmfn{phantom}{h}{16}
   \fmf{phantom}{h16,j}
   \fmfv{l=$k$,l.a=-90}{h5}\fmflabel{$i$}{i} 
   \fmfv{l=$l$,l.a=-90}{h12}\fmflabel{$j$}{j}
   \fmffreeze 
   \fmf{fillvx,left}{h1,j}
   \fmf{unfillvx,left}{h4,h12}
   \fmf{plain}{i,h5} \fmf{plain}{h12,j}
   \fmf{photon,left}{h1,j}
   \fmf{photon,left}{h4,h12}
   \fmfv{d.shape=circle, d.size=0.6thick}{i}
   \fmfv{d.shape=circle, d.size=0.6thick}{h1}
   \fmfv{d.shape=circle, d.size=0.6thick}{h5}
   \fmfv{d.shape=circle, d.size=0.6thick}{h4}
\end{fmfgraph*}
$\quad\Big |\quad$
\begin{fmfgraph*}(30,25)\fmfkeep{wh1(3)}
   \fmfleft{i,d1}
   \fmfright{j,d2}
   \fmf{phantom}{i,h1}
   \fmfn{phantom}{h}{16}
   \fmf{phantom}{h16,j}
   \fmfv{l=$k$,l.a=-90}{h5}\fmflabel{$i$}{i} 
   \fmfv{l=$l$,l.a=-90}{h12}\fmflabel{$j$}{j}
   \fmffreeze 
   \fmf{fillvx,left}{i,h16}
   \fmf{unfillvx,left}{h5,h13}
   \fmf{plain}{i,h5} \fmf{plain}{h12,j}
   \fmf{photon,left}{i,h16}
   \fmf{photon,left}{h5,h13}
   \fmfv{d.shape=circle, d.size=0.6thick}{j}
   \fmfv{d.shape=circle, d.size=0.6thick}{h16}
   \fmfv{d.shape=circle, d.size=0.6thick}{h12}
   \fmfv{d.shape=circle, d.size=0.6thick}{h13}
\end{fmfgraph*}
$\quad\Big |\quad$
\begin{fmfgraph*}(30,25)\fmfkeep{wh1(4)}
   \fmfleft{i,d1}
   \fmfright{j,d2}
   \fmf{phantom}{i,h1}
   \fmfn{phantom}{h}{16}
   \fmf{phantom}{h16,j}
   \fmfv{l=$k$,l.a=-90}{h5}\fmflabel{$i$}{i} 
   \fmfv{l=$l$,l.a=-90}{h12}\fmflabel{$j$}{j}
   \fmffreeze 
   \fmf{fillvx,left}{h1,j}
   \fmf{unfillvx,left}{h5,h13}
   \fmf{plain}{i,h5} \fmf{plain}{h12,j}
   \fmf{photon,left}{h1,j}
   \fmf{photon,left}{h5,h13}
   \fmfv{d.shape=circle, d.size=0.6thick}{i}
   \fmfv{d.shape=circle, d.size=0.6thick}{h1}
   \fmfv{d.shape=circle, d.size=0.6thick}{h12}
   \fmfv{d.shape=circle, d.size=0.6thick}{h13}
\end{fmfgraph*}
$\quad\Big |\quad$

\begin{fmfgraph*}(30,25)\fmfkeep{wh1(5)}
   \fmfleft{i,d1}
   \fmfright{j,d2}
   \fmf{phantom}{i,h1}
   \fmfn{phantom}{h}{16}
   \fmf{phantom}{h16,j}
   \fmfv{l=$k$,l.a=-90}{h5}\fmflabel{$i$}{i} 
   \fmfv{l=$l$,l.a=-90}{h12}\fmflabel{$j$}{j}
   \fmffreeze 
   \fmf{fillvx,left}{i,h16}
   \fmf{unfillvx,left}{h4,h12}
   \fmf{plain}{i,h5} \fmf{plain}{h12,j}
   \fmf{photon,left}{i,h16}
   \fmf{photon,left}{h4,h12}
   \fmfv{d.shape=circle, d.size=0.6thick}{j}
   \fmfv{d.shape=circle, d.size=0.6thick}{h16}
   \fmfv{d.shape=circle, d.size=0.6thick}{h4}
   \fmfv{d.shape=circle, d.size=0.6thick}{h5}
\end{fmfgraph*}
$\quad\Big |\quad$
\begin{fmfgraph*}(30,25)\fmfkeep{wh1(6)}
   \fmfleft{i,d1}
   \fmfright{j,d2}
   \fmf{phantom}{i,h1}
   \fmfn{phantom}{h}{16}
   \fmf{phantom}{h16,j}
   \fmfv{l=$k$,l.a=-90}{h5}\fmflabel{$i$}{i} 
   \fmfv{l=$l$,l.a=-90}{h12}\fmflabel{$j$}{j}
   \fmffreeze 
   \fmf{fillvx,left}{h1,h16}
   \fmf{unfillvx,left}{h4,h12}
   \fmf{plain}{i,h5} \fmf{plain}{h12,j}
   \fmf{photon,left}{h1,h16}
   \fmf{photon,left}{h4,h12}
   \fmfv{d.shape=circle, d.size=0.6thick}{j}
   \fmfv{d.shape=circle, d.size=0.6thick}{h16}
   \fmfv{d.shape=circle, d.size=0.6thick}{i}
   \fmfv{d.shape=circle, d.size=0.6thick}{h1}
   \fmfv{d.shape=circle, d.size=0.6thick}{h4}
   \fmfv{d.shape=circle, d.size=0.6thick}{h5}
\end{fmfgraph*}
$\quad\Big |\quad$
\begin{fmfgraph*}(30,25)\fmfkeep{wh1(7)}
   \fmfleft{i,d1}
   \fmfright{j,d2}
   \fmf{phantom}{i,h1}
   \fmfn{phantom}{h}{16}
   \fmf{phantom}{h16,j}
   \fmfv{l=$k$,l.a=-90}{h5}\fmflabel{$i$}{i} 
   \fmfv{l=$l$,l.a=-90}{h12}\fmflabel{$j$}{j}
   \fmffreeze 
   \fmf{fillvx,left}{h1,j}
   \fmf{unfillvx,left}{h4,h13}
   \fmf{plain}{i,h5} \fmf{plain}{h12,j}
   \fmf{photon,left}{h1,j}
   \fmf{photon,left}{h4,h13}
   \fmfv{d.shape=circle, d.size=0.6thick}{h12}
   \fmfv{d.shape=circle, d.size=0.6thick}{h13}
   \fmfv{d.shape=circle, d.size=0.6thick}{i}
   \fmfv{d.shape=circle, d.size=0.6thick}{h1}
   \fmfv{d.shape=circle, d.size=0.6thick}{h4}
   \fmfv{d.shape=circle, d.size=0.6thick}{h5}
\end{fmfgraph*}
$\quad\Big |\quad$

\begin{fmfgraph*}(30,25)\fmfkeep{wh1(8)}
   \fmfleft{i,d1}
   \fmfright{j,d2}
   \fmf{phantom}{i,h1}
   \fmfn{phantom}{h}{16}
   \fmf{phantom}{h16,j}
   \fmfv{l=$k$,l.a=-90}{h5}\fmflabel{$i$}{i} 
   \fmfv{l=$l$,l.a=-90}{h12}\fmflabel{$j$}{j}
   \fmffreeze 
   \fmf{fillvx,left}{h1,h16}
   \fmf{unfillvx,left}{h5,h13}
   \fmf{plain}{i,h5} \fmf{plain}{h12,j}
   \fmf{photon,left}{h1,h16}
   \fmf{photon,left}{h5,h13}
   \fmfv{d.shape=circle, d.size=0.6thick}{i}
   \fmfv{d.shape=circle, d.size=0.6thick}{h1}
   \fmfv{d.shape=circle, d.size=0.6thick}{j}
   \fmfv{d.shape=circle, d.size=0.6thick}{h16}
   \fmfv{d.shape=circle, d.size=0.6thick}{h12}
   \fmfv{d.shape=circle, d.size=0.6thick}{h13}
\end{fmfgraph*}
$\quad\Big |\quad$
\begin{fmfgraph*}(30,25)\fmfkeep{wh1(9)}
   \fmfleft{i,d1}
   \fmfright{j,d2}
   \fmf{phantom}{i,h1}
   \fmfn{phantom}{h}{16}
   \fmf{phantom}{h16,j}
   \fmfv{l=$k$,l.a=-90}{h5}\fmflabel{$i$}{i} 
   \fmfv{l=$l$,l.a=-90}{h12}\fmflabel{$j$}{j}
   \fmffreeze 
   \fmf{fillvx,left}{i,h16}
   \fmf{unfillvx,left}{h4,h13}
   \fmf{plain}{i,h5} \fmf{plain}{h12,j}
   \fmf{photon,left}{i,h16}
   \fmf{photon,left}{h4,h13}
   \fmfv{d.shape=circle, d.size=0.6thick}{j}
   \fmfv{d.shape=circle, d.size=0.6thick}{h16}
   \fmfv{d.shape=circle, d.size=0.6thick}{h12}
   \fmfv{d.shape=circle, d.size=0.6thick}{h13}
   \fmfv{d.shape=circle, d.size=0.6thick}{h4}
   \fmfv{d.shape=circle, d.size=0.6thick}{h5}
\end{fmfgraph*}
$\quad\Big |\quad$
\begin{fmfgraph*}(30,25)\fmfkeep{wh1(10)}
   \fmfleft{i,d1}
   \fmfright{j,d2}
   \fmf{phantom}{i,h1}
   \fmfn{phantom}{h}{16}
   \fmf{phantom}{h16,j}
   \fmfv{l=$k$,l.a=-90}{h5}\fmflabel{$i$}{i} 
   \fmfv{l=$l$,l.a=-90}{h12}\fmflabel{$j$}{j}
   \fmffreeze 
   \fmf{plain}{i,h5} \fmf{plain}{h12,j}
   \fmf{fillvx,left}{h1,h16}
   \fmf{unfillvx,left}{h4,h13}
   \fmf{photon,left}{h1,h16}
   \fmf{photon,left}{h4,h13}
   \fmfv{d.shape=circle, d.size=0.6thick}{h12}
   \fmfv{d.shape=circle, d.size=0.6thick}{h13}
   \fmfv{d.shape=circle, d.size=0.6thick}{i}
   \fmfv{d.shape=circle, d.size=0.6thick}{h1}
   \fmfv{d.shape=circle, d.size=0.6thick}{j}
   \fmfv{d.shape=circle, d.size=0.6thick}{h16}
   \fmfv{d.shape=circle, d.size=0.6thick}{h4}
   \fmfv{d.shape=circle, d.size=0.6thick}{h5}
\end{fmfgraph*}
$\quad\Big |\quad$

\begin{fmfgraph*}(30,25)\fmfkeep{wh2(2)}
   \fmfleft{i,d1}
   \fmfright{j,d2}
   \fmf{phantom}{i,h1}
   \fmfn{phantom}{h}{16}
   \fmf{phantom}{h16,j}
   \fmfv{l=$k$,l.a=-90}{h5}\fmflabel{$i$}{i} 
   \fmfv{l=$l$,l.a=-90}{h12}\fmflabel{$j$}{j}
   \fmffreeze 
   \fmf{fillvx,left}{h1,h16}
   \fmf{unfillwx,left}{h5,h12}
   \fmf{plain}{i,h5} \fmf{plain}{h12,j}
   \fmf{photon,left}{h1,h16}
   \fmf{dashes,left}{h5,h12}
   \fmfv{d.shape=circle, d.size=0.6thick}{i}
   \fmfv{d.shape=circle, d.size=0.6thick}{h1}
   \fmfv{d.shape=circle, d.size=0.6thick}{j}
   \fmfv{d.shape=circle, d.size=0.6thick}{h16}
\end{fmfgraph*}
$\quad\Big |\quad$
\begin{fmfgraph*}(30,25)\fmfkeep{wh2(3)}
   \fmfleft{i,d1}
   \fmfright{j,d2}
   \fmf{phantom}{i,h1}
   \fmfn{phantom}{h}{16}
   \fmf{phantom}{h16,j}
   \fmfv{l=$k$,l.a=-90}{h5}\fmflabel{$i$}{i} 
   \fmfv{l=$l$,l.a=-90}{h12}\fmflabel{$j$}{j}
   \fmffreeze 
   \fmf{fillvx,left}{h1,j}
   \fmf{unfillwx,left}{h5,h12}
   \fmf{plain}{i,h5} \fmf{plain}{h12,j}
   \fmf{photon,left}{h1,j}
   \fmf{dashes,left}{h5,h12}
   \fmfv{d.shape=circle, d.size=0.6thick}{i}
   \fmfv{d.shape=circle, d.size=0.6thick}{h1}
\end{fmfgraph*}
$\quad\Big |\quad$
\begin{fmfgraph*}(30,25)\fmfkeep{wh2(4)}
   \fmfleft{i,d1}
   \fmfright{j,d2}
   \fmf{phantom}{i,h1}
   \fmfn{phantom}{h}{16}
   \fmf{phantom}{h16,j}
   \fmfv{l=$k$,l.a=-90}{h5}\fmflabel{$i$}{i} 
   \fmfv{l=$l$,l.a=-90}{h12}\fmflabel{$j$}{j}
   \fmffreeze 
   \fmf{fillvx,left}{i,h16}
   \fmf{unfillwx,left}{h5,h12}
   \fmf{plain}{i,h5} \fmf{plain}{h12,j}
   \fmf{photon,left}{i,h16}
   \fmf{dashes,left}{h5,h12}
   \fmfv{d.shape=circle, d.size=0.6thick}{j}
   \fmfv{d.shape=circle, d.size=0.6thick}{h16}
\end{fmfgraph*}
$\quad\Big |\quad$

\begin{fmfgraph*}(30,25)\fmfkeep{wh3(2)}
   \fmfleft{i,d1}
   \fmfright{j,d2}
   \fmf{phantom}{i,h1}
   \fmfn{phantom}{h}{16}
   \fmf{phantom}{h16,j}
   \fmfv{l=$k$,l.a=-90}{h5}\fmflabel{$i$}{i} 
   \fmfv{l=$l$,l.a=-90}{h12}\fmflabel{$j$}{j}
   \fmffreeze 
   \fmf{fillwx,left}{i,j}
   \fmf{unfillvx,left}{h4,h13}
   \fmf{plain}{i,h5} \fmf{plain}{h12,j}
   \fmf{dashes,left}{i,j}
   \fmf{photon,left}{h4,h13}
   \fmfv{d.shape=circle, d.size=0.6thick}{h4}
   \fmfv{d.shape=circle, d.size=0.6thick}{h5}
   \fmfv{d.shape=circle, d.size=0.6thick}{h12}
   \fmfv{d.shape=circle, d.size=0.6thick}{h13}
\end{fmfgraph*}
$\quad\Big |\quad$
\begin{fmfgraph*}(30,25)\fmfkeep{wh3(3)}
   \fmfleft{i,d1}
   \fmfright{j,d2}
   \fmf{phantom}{i,h1}
   \fmfn{phantom}{h}{16}
   \fmf{phantom}{h16,j}
   \fmfv{l=$k$,l.a=-90}{h5}\fmflabel{$i$}{i} 
   \fmfv{l=$l$,l.a=-90}{h12}\fmflabel{$j$}{j}
   \fmffreeze 
   \fmf{fillwx,left}{i,j}
   \fmf{unfillvx,left}{h4,h12}
   \fmf{plain}{i,h5} \fmf{plain}{h12,j}
   \fmf{dashes,left}{i,j}
   \fmf{photon,left}{h4,h12}
   \fmfv{d.shape=circle, d.size=0.6thick}{h4}
   \fmfv{d.shape=circle, d.size=0.6thick}{h5}
\end{fmfgraph*}
$\quad\Big |\quad$
\begin{fmfgraph*}(30,25)\fmfkeep{wh3(4)}
   \fmfleft{i,d1}
   \fmfright{j,d2}
   \fmf{phantom}{i,h1}
   \fmfn{phantom}{h}{16}
   \fmf{phantom}{h16,j}
   \fmfv{l=$k$,l.a=-90}{h5}\fmflabel{$i$}{i} 
   \fmfv{l=$l$,l.a=-90}{h12}\fmflabel{$j$}{j}
   \fmffreeze 
   \fmf{fillwx,left}{i,j}
   \fmf{unfillvx,left}{h5,h13}
   \fmf{plain}{i,h5} \fmf{plain}{h12,j}
   \fmf{dashes,left}{i,j}
   \fmf{photon,left}{h5,h13}
   \fmfv{d.shape=circle, d.size=0.6thick}{h12}
   \fmfv{d.shape=circle, d.size=0.6thick}{h13}
\end{fmfgraph*}
$\quad\Big |\quad$
\end{figure}

\begin {figure}[ht]
\begin{fmfgraph*}(30,25)\fmfkeep{wh4(1)}
   \fmfleft{i,d1}
   \fmfright{j,d2}
   \fmf{phantom}{i,h1}
   \fmfn{phantom}{h}{16}
   \fmf{phantom}{h16,j}
   \fmfv{l=$k$,l.a=-90}{h5}\fmflabel{$i$}{i} 
   \fmfv{l=$l$,l.a=-90}{h12}\fmflabel{$j$}{j}
   \fmffreeze 
   \fmf{fillwx,left}{h1,j}
   \fmf{unfillwx,left}{h5,h12}
   \fmf{plain}{i,h5} \fmf{plain}{h12,j}
   \fmf{dashes,left}{h1,j}
   \fmf{dashes,left}{h5,h12}
   \fmfv{d.shape=circle, d.size=0.6thick}{i}
   \fmfv{d.shape=circle, d.size=0.6thick}{h1}
\end{fmfgraph*}
$\quad\Big |\quad$
\begin{fmfgraph*}(30,25)\fmfkeep{wh4(2)}
   \fmfleft{i,d1}
   \fmfright{j,d2}
   \fmf{phantom}{i,h1}
   \fmfn{phantom}{h}{16}
   \fmf{phantom}{h16,j}
   \fmfv{l=$k$,l.a=-90}{h5}\fmflabel{$i$}{i} 
   \fmfv{l=$l$,l.a=-90}{h12}\fmflabel{$j$}{j}
   \fmffreeze 
   \fmf{fillwx,left}{i,h16}
   \fmf{unfillwx,left}{h5,h12}
   \fmf{plain}{i,h5} \fmf{plain}{h12,j}
   \fmf{dashes,left}{i,h16}
   \fmf{dashes,left}{h5,h12}
   \fmfv{d.shape=circle, d.size=0.6thick}{j}
   \fmfv{d.shape=circle, d.size=0.6thick}{h16}
\end{fmfgraph*}
$\quad\Big |\quad$
\begin{fmfgraph*}(30,25)\fmfkeep{wh4(3)}
   \fmfleft{i,d1}
   \fmfright{j,d2}
   \fmf{phantom}{i,h1}
   \fmfn{phantom}{h}{16}
   \fmf{phantom}{h16,j}
   \fmfv{l=$k$,l.a=-90}{h5}\fmflabel{$i$}{i} 
   \fmfv{l=$l$,l.a=-90}{h12}\fmflabel{$j$}{j}
   \fmffreeze 
   \fmf{fillwx,left}{i,j}
   \fmf{unfillwx,left}{h4,h12}
   \fmf{plain}{i,h5} \fmf{plain}{h12,j}
   \fmf{dashes,left}{i,j}
   \fmf{dashes,left}{h4,h12}
   \fmfv{d.shape=circle, d.size=0.6thick}{h4}
   \fmfv{d.shape=circle, d.size=0.6thick}{h5}
\end{fmfgraph*}
$\quad\Big |\quad$

\begin{fmfgraph*}(30,25)\fmfkeep{wh4(4)}
   \fmfleft{i,d1}
   \fmfright{j,d2}
   \fmf{phantom}{i,h1}
   \fmfn{phantom}{h}{16}
   \fmf{phantom}{h16,j}
   \fmfv{l=$k$,l.a=-90}{h5}\fmflabel{$i$}{i} 
   \fmfv{l=$l$,l.a=-90}{h12}\fmflabel{$j$}{j}
   \fmffreeze 
   \fmf{fillwx,left}{i,j}
   \fmf{unfillwx,left}{h5,h13}
   \fmf{plain}{i,h5} \fmf{plain}{h12,j}
   \fmf{dashes,left}{i,j}
   \fmf{dashes,left}{h5,h13}
   \fmfv{d.shape=circle, d.size=0.6thick}{h12}
   \fmfv{d.shape=circle, d.size=0.6thick}{h13}
\end{fmfgraph*}
$\quad\Big |\quad$
\begin{fmfgraph*}(30,25)\fmfkeep{wh5}
   \fmfleft{i,d1}
   \fmfright{j,d2}
   \fmf{plain}{i,k1,k}
   \fmf{phantom}{k,m,l}
   \fmf{plain}{l,l1,j}
   \fmfv{l=$k$,l.a=-90}{k}\fmflabel{$i$}{i} 
   \fmfv{l=$l$,l.a=-90}{l}\fmflabel{$j$}{j}
   \fmffreeze
   \fmf{fillwx,left}{i,k}
   \fmf{fillwx,left}{l,j}
   \fmf{dashes,left}{i,k}
   \fmf{dashes,left}{l,j}
   \fmf{plain}{i,k1,k}\fmf{plain}{l,l1,j}
\end{fmfgraph*}
$\quad\Big |\quad$
\begin{fmfgraph*}(30,25)\fmfkeep{wh6}
   \fmfleft{i,d1}
   \fmfright{j,d2}
   \fmf{phantom}{i,h1}
   \fmfn{phantom}{h}{16}
   \fmf{phantom}{h16,j}
   \fmfv{l=$k$,l.a=-90}{h7}\fmflabel{$i$}{i} 
   \fmfv{l=$l$,l.a=-90}{h13}\fmflabel{$j$}{j}
   \fmfv{l=$r$,l.a=-90}{h3}
   \fmffreeze 
   \fmf{fillwx,left}{i,h3}
   \fmf{fillwx,left}{h4,j}
   \fmf{unfillwx,left}{h7,h13}
   \fmf{plain}{i,h7} \fmf{plain}{h13,j}
   \fmf{dashes,left}{h7,h13}
   \fmf{dashes,left}{h4,j}
   \fmf{dashes,left}{i,h3}
   \fmfv{d.shape=circle, d.size=0.6thick}{h3}
   \fmfv{d.shape=circle, d.size=0.6thick}{h4}
\end{fmfgraph*}
$\quad\Big |\quad$

\begin{fmfgraph*}(30,25)\fmfkeep{wh7}
   \fmfleft{i,d1}
   \fmfright{j,d2}
   \fmf{phantom}{i,h1}
   \fmfn{phantom}{h}{16}
   \fmf{phantom}{h16,j}
   \fmfv{l=$k$,l.a=-90}{h7}\fmflabel{$i$}{i} 
   \fmfv{l=$l$,l.a=-90}{h13}\fmflabel{$j$}{j}
   \fmfv{l=$r$,l.a=-90}{h3}
   \fmffreeze 
   \fmf{fillvx,left}{i,h3}
   \fmf{fillvx,left}{h4,j}
   \fmf{unfillwx,left}{h7,h13}
   \fmf{plain}{i,h7} \fmf{plain}{h13,j}
   \fmf{dashes,left}{h7,h13}
   \fmf{photon,left}{h4,j}
   \fmf{photon,left}{i,h3}
   \fmfv{d.shape=circle, d.filled=empty, d.size=3pt}{h3}
   \fmfv{d.shape=circle, d.filled=empty, d.size=3pt}{h4}
\end{fmfgraph*}
$\quad\Big |\quad$
\begin{fmfgraph*}(30,25)\fmfkeep{wh8}
   \fmfleft{i,d1}
   \fmfright{j,d2}
   \fmf{phantom}{i,h1}
   \fmfn{phantom}{h}{16}
   \fmf{phantom}{h16,j}
   \fmfv{l=$k$,l.a=-90}{h7}\fmflabel{$i$}{i} 
   \fmfv{l=$l$,l.a=-90}{h13}\fmflabel{$j$}{j}
   \fmfv{l=$r$,l.a=-90}{h3}
   \fmffreeze 
   \fmf{fillwx,left}{h4,h7}
   \fmf{fillwx,left}{i,j}
   \fmf{unfillwx,left}{h3,h13}
   \fmf{plain}{i,h7} \fmf{plain}{h13,j}
   \fmf{dashes,left}{i,j}
   \fmf{dashes,left}{h3,h13}
   \fmf{dashes,left}{h4,h7}
   \fmfv{d.shape=circle, d.size=0.6thick}{h3}
   \fmfv{d.shape=circle, d.size=0.6thick}{h4}
\end{fmfgraph*}
$\quad\Big |\quad$
\begin{fmfgraph*}(30,25)\fmfkeep{wh9}
   \fmfleft{i,d1}
   \fmfright{j,d2}
   \fmf{phantom}{i,h1}
   \fmfn{phantom}{h}{16}
   \fmf{phantom}{h16,j}
   \fmfv{l=$k$,l.a=-90}{h7}\fmflabel{$i$}{i} 
   \fmfv{l=$l$,l.a=-90}{h13}\fmflabel{$j$}{j}
   \fmfv{l=$r$,l.a=-90}{h3}
   \fmffreeze 
   \fmf{fillvx,left}{h4,h7}
   \fmf{fillwx,left}{i,j}
   \fmf{unfillvx,left}{h3,h13}
   \fmf{plain}{i,h7} \fmf{plain}{h13,j}
   \fmf{dashes,left}{i,j}
   \fmf{photon,left}{h3,h13}
   \fmf{photon,left}{h4,h7}
   \fmfv{d.shape=circle, d.filled=empty, d.size=3pt}{h3}
   \fmfv{d.shape=circle, d.filled=empty, d.size=3pt}{h4}
\end{fmfgraph*}
$\quad\Big |\quad$

\begin{fmfgraph*}(30,25)\fmfkeep{wh10}
   \fmfleft{i,d1}
   \fmfright{j,d2}
   \fmf{phantom}{i,h1}
   \fmfn{phantom}{h}{16}
   \fmf{phantom}{h16,j}
   \fmfv{l=$k$,l.a=-90}{h5}\fmflabel{$i$}{i} 
   \fmfv{l=$l$,l.a=-90}{h10}\fmflabel{$j$}{j}
   \fmfv{l=$s$,l.a=-90}{h13}
   \fmffreeze 
   \fmf{fillwx,left}{h14,j}
   \fmf{fillwx,left}{i,h13}
   \fmf{unfillwx,left}{h5,h10}
   \fmf{plain}{i,h5} \fmf{plain}{h10,j}
   \fmf{dashes,left}{i,h13}
   \fmf{dashes,left}{h5,h10}
   \fmf{dashes,left}{h14,j}
   \fmfv{d.shape=circle, d.size=0.6thick}{h13}
   \fmfv{d.shape=circle, d.size=0.6thick}{h14}
\end{fmfgraph*}
$\quad\Big |\quad$
\begin{fmfgraph*}(30,25)\fmfkeep{wh11}
   \fmfleft{i,d1}
   \fmfright{j,d2}
   \fmf{phantom}{i,h1}
   \fmfn{phantom}{h}{16}
   \fmf{phantom}{h16,j}
   \fmfv{l=$k$,l.a=-90}{h5}\fmflabel{$i$}{i} 
   \fmfv{l=$l$,l.a=-90}{h10}\fmflabel{$j$}{j}
   \fmfv{l=$s$,l.a=-90}{h13}
   \fmffreeze 
   \fmf{fillvx,left}{h14,j}
   \fmf{fillvx,left}{i,h13}
   \fmf{unfillwx,left}{h5,h10}
   \fmf{plain}{i,h5} \fmf{plain}{h10,j}
   \fmf{photon,left}{i,h13}
   \fmf{dashes,left}{h5,h10}
   \fmf{photon,left}{h14,j}
   \fmfv{d.shape=circle, d.filled=empty, d.size=3pt}{h13}
   \fmfv{d.shape=circle, d.filled=empty, d.size=3pt}{h14}
\end{fmfgraph*}
$\quad\Big |\quad$
\begin{fmfgraph*}(30,25)\fmfkeep{wh12}
   \fmfleft{i,d1}
   \fmfright{j,d2}
   \fmf{phantom}{i,h1}
   \fmfn{phantom}{h}{16}
   \fmf{phantom}{h16,j}
   \fmfv{l=$k$,l.a=-45}{h5}\fmflabel{$i$}{i} 
   \fmfv{l=$l$,l.a=-90}{h10}\fmflabel{$j$}{j}
   \fmfv{l=$s$,l.a=-90}{h13}
   \fmffreeze 
   \fmf{fillwx,left}{h10,h13}
   \fmf{fillwx,left}{i,j}
   \fmf{unfillwx,left}{h5,h14}
   \fmf{plain}{i,h5} \fmf{plain}{h10,j}
   \fmf{dashes,left}{i,j}
   \fmf{dashes,left}{h5,h14}
   \fmf{dashes,left}{h10,h13}
   \fmfv{d.shape=circle, d.size=0.6thick}{h13}
   \fmfv{d.shape=circle, d.size=0.6thick}{h14}
\end{fmfgraph*}
$\quad\Big |\quad$

\begin{fmfgraph*}(30,25)\fmfkeep{wh13}
   \fmfleft{i,d1}
   \fmfright{j,d2}
   \fmf{phantom}{i,h1}
   \fmfn{phantom}{h}{16}
   \fmf{phantom}{h16,j}
   \fmfv{l=$k$,l.a=-45}{h5}\fmflabel{$i$}{i} 
   \fmfv{l=$l$,l.a=-90}{h10}\fmflabel{$j$}{j}
   \fmfv{l=$s$,l.a=-90}{h13}
   \fmffreeze 
   \fmf{fillvx,left}{h10,h13}
   \fmf{fillwx,left}{i,j}
   \fmf{unfillvx,left}{h5,h14}
   \fmf{plain}{i,h5} \fmf{plain}{h10,j}
   \fmf{dashes,left}{i,j}
   \fmf{photon,left}{h5,h14}
   \fmf{photon,left}{h10,h13}
   \fmfv{d.shape=circle, d.filled=empty, d.size=3pt}{h13}
   \fmfv{d.shape=circle, d.filled=empty, d.size=3pt}{h14}
\end{fmfgraph*}
$\quad\Big |\quad$
\begin{fmfgraph*}(30,25)\fmfkeep{wh14}
   \fmfleft{i,d1}
   \fmfright{j,d2}
   \fmf{phantom}{i,r,k,h,l,s,j}
   \fmfv{l=$k$,l.a=-90}{k}\fmflabel{$i$}{i} 
   \fmfv{l=$l$,l.a=-90}{l}\fmflabel{$j$}{j}
   \fmfv{l=$r$,l.a=-90}{r} \fmfv{l=$s$,l.a=-90}{s}
   \fmffreeze 
   \fmf{fillwx,left}{i,j}
   \fmf{unfillwx,left}{k,l}
   \fmf{fillvx,left}{r,s}
   \fmf{unfillwx,left}{k,l}
   \fmf{plain}{i,r,k}\fmf{plain}{l,s,j}
   \fmf{dashes,left}{k,l}
   \fmf{photon,left}{r,s}
   \fmf{dashes,left}{i,j}
\end{fmfgraph*}
$\quad\Big |\quad$
\begin{fmfgraph*}(30,25)\fmfkeep{wh15}
   \fmfleft{i,d1}
   \fmfright{j,d2}
   \fmf{phantom}{i,h1}
   \fmfn{phantom}{h}{16}
   \fmf{phantom}{h16,j}
   \fmfv{l=$k$,l.a=-90}{h6}\fmflabel{$i$}{i} 
   \fmfv{l=$l$,l.a=-120}{h11}\fmflabel{$j$}{j}
   \fmfv{l=$r$,l.a=-90}{h3} \fmfv{l=$s$,l.a=-90}{h14}
   \fmffreeze 
   \fmf{fillwx,left}{i,j}
   \fmf{unfillwx,left}{h3,h14}
   \fmf{fillwx,left}{h4,h13}
   \fmf{unfillwx,left}{h6,h11}
   \fmf{dashes,left}{h3,h14}
   \fmf{dashes,left}{h4,h13}
   \fmf{dashes,left}{i,j}
   \fmf{dashes,left}{h6,h11}
   \fmf{plain}{i,h1,h2,h3,h4,h5,h6}\fmf{plain}{h11,j}
   \fmfv{d.shape=circle, d.size=0.6thick}{h3}
   \fmfv{d.shape=circle, d.size=0.6thick}{h4}
   \fmfv{d.shape=circle, d.size=0.6thick}{h13}
   \fmfv{d.shape=circle, d.size=0.6thick}{h14}
\end{fmfgraph*}
$\quad\Big |\quad$
\end{figure}

\begin {figure}[ht]
\begin{fmfgraph*}(30,25)\fmfkeep{wh16}
   \fmfleft{i,d1}
   \fmfright{j,d2}
   \fmf{phantom}{i,h1}
   \fmfn{phantom}{h}{16}
   \fmf{phantom}{h16,j}
   \fmfv{l=$k$,l.a=-90}{h7}\fmflabel{$i$}{i} 
   \fmfv{l=$l$,l.a=-120}{h10}\fmflabel{$j$}{j}
   \fmffreeze 
   \fmf{fillwx,left}{i,h13}
   \fmf{unfillwx,left}{h3,h10}
   \fmf{fillwx,right}{h4,j}
   \fmf{unfillwx,right}{h7,h14}
   \fmf{plain}{i,h7} \fmf{plain}{h10,j}
   \fmf{dashes,left}{i,h13}
   \fmf{dashes,left}{h3,h10}
   \fmf{dashes,right}{h4,j}
   \fmf{dashes,right}{h7,h14}
   \fmfv{d.shape=circle, d.size=0.6thick}{h3}
   \fmfv{d.shape=circle, d.size=0.6thick}{h4}
   \fmfv{d.shape=circle, d.size=0.6thick}{h13}
   \fmfv{d.shape=circle, d.size=0.6thick}{h14}
\end{fmfgraph*}
$\quad\Big |\quad$
\begin{fmfgraph*}(30,25)\fmfkeep{wh17}
   \fmfleft{i,d1}
   \fmfright{j,d2}
   \fmf{phantom}{i,h1}
   \fmfn{phantom}{h}{16}
   \fmf{phantom}{h16,j}
   \fmfv{l=$k$,l.a=-90}{h4}\fmflabel{$i$}{i} 
   \fmfv{l=$l$,l.a=-90}{h9}\fmflabel{$j$}{j}
   \fmffreeze 
   \fmf{fillwx,left}{i,h14}
   \fmf{unfillwx,left}{h4,h12}
   \fmf{fillwx,right}{h9,j}
   \fmf{unfillwx,right}{h11,h15}
   \fmf{plain}{i,h4} \fmf{plain}{h9,j}
   \fmf{dashes,left}{i,h14}
   \fmf{dashes,left}{h4,h12}
   \fmf{dashes,right}{h9,j}
   \fmf{dashes,right}{h11,h15}
   \fmfv{d.shape=circle, d.size=0.6thick}{h11}
   \fmfv{d.shape=circle, d.size=0.6thick}{h12}
   \fmfv{d.shape=circle, d.size=0.6thick}{h14}
   \fmfv{d.shape=circle, d.size=0.6thick}{h15}
\end{fmfgraph*}
$\quad\Big |\quad$
\begin{fmfgraph*}(30,25)\fmfkeep{wh18}
   \fmfleft{i,d1}
   \fmfright{j,d2}
   \fmf{phantom}{i,h1}
   \fmfn{phantom}{h}{16}
   \fmf{phantom}{h16,j}
   \fmfv{l=$k$,l.a=-90}{h4}\fmflabel{$i$}{i} 
   \fmfv{l=$l$,l.a=-90}{h9}\fmflabel{$j$}{j}
   \fmffreeze 
   \fmf{fillvx,left}{i,h14}
   \fmf{unfillwx,left}{h4,h12}
   \fmf{fillwx,right}{h9,j}
   \fmf{unfillvx,right}{h11,h15}
   \fmf{plain}{i,h4} \fmf{plain}{h9,j}
   \fmf{photon,left}{i,h14}
   \fmf{dashes,left}{h4,h12}
   \fmf{dashes,right}{h9,j}
   \fmf{photon,right}{h11,h15}
   \fmfv{d.shape=circle, d.size=0.6thick}{h11}
   \fmfv{d.shape=circle, d.size=0.6thick}{h12}
   \fmfv{d.shape=circle, d.filled=empty, d.size=3pt}{h14}
   \fmfv{d.shape=circle, d.filled=empty, d.size=3pt}{h15}
\end{fmfgraph*}
$\quad\Big |\quad$

\begin{fmfgraph*}(30,25)\fmfkeep{wh19}
   \fmfleft{i,d1}
   \fmfright{j,d2}
   \fmf{phantom}{i,h1}
   \fmfn{phantom}{h}{16}
   \fmf{phantom}{h16,j}
   \fmfv{l=$k$,l.a=-90}{h4}\fmflabel{$i$}{i} 
   \fmfv{l=$l$,l.a=-90}{h9}\fmflabel{$j$}{j}
   \fmffreeze 
   \fmf{fillwx,left}{i,h14}
   \fmf{unfillvx,left}{h4,h12}
   \fmf{fillwx,right}{h9,j}
   \fmf{unfillvx,right}{h11,h15}
   \fmf{plain}{i,h4} \fmf{plain}{h9,j}
   \fmf{dashes,left}{i,h14}
   \fmf{photon,left}{h4,h12}
   \fmf{dashes,right}{h9,j}
   \fmf{photon,right}{h11,h15}
   \fmfv{d.shape=circle, d.filled=empty, d.size=3pt}{h11}
   \fmfv{d.shape=circle, d.filled=empty, d.size=3pt}{h12}
   \fmfv{d.shape=circle, d.size=0.6thick}{h14}
   \fmfv{d.shape=circle, d.size=0.6thick}{h15}
\end{fmfgraph*}
$\quad\Big |\quad$
\begin{fmfgraph*}(30,25)\fmfkeep{wh20}
   \fmfleft{i,d1}
   \fmfright{j,d2}
   \fmf{phantom}{i,h1}
   \fmfn{phantom}{h}{16}
   \fmf{phantom}{h16,j}
   \fmfv{l=$k$,l.a=-45}{h8}\fmflabel{$i$}{i} 
   \fmfv{l=$l$,l.a=-90}{h13}\fmflabel{$j$}{j}
   \fmffreeze 
   \fmf{fillwx,left}{h3,j}
   \fmf{unfillwx,left}{h5,h13}
   \fmf{fillwx,right}{i,h8}
   \fmf{unfillwx,right}{h2,h6}
   \fmf{plain}{i,h8} \fmf{plain}{h13,j}
   \fmf{dashes,left}{h3,j}
   \fmf{dashes,left}{h5,h13}
   \fmf{dashes,right}{i,h8}
   \fmf{dashes,right}{h2,h6}
   \fmfv{d.shape=circle, d.size=0.6thick}{h2}
   \fmfv{d.shape=circle, d.size=0.6thick}{h3}
   \fmfv{d.shape=circle, d.size=0.6thick}{h5}
   \fmfv{d.shape=circle, d.size=0.6thick}{h6}
\end{fmfgraph*}
$\quad\Big |\quad$
\begin{fmfgraph*}(30,25)\fmfkeep{wh21}
   \fmfleft{i,d1}
   \fmfright{j,d2}
   \fmf{phantom}{i,h1}
   \fmfn{phantom}{h}{16}
   \fmf{phantom}{h16,j}
   \fmfv{l=$k$,l.a=-45}{h8}\fmflabel{$i$}{i} 
   \fmfv{l=$l$,l.a=-90}{h13}\fmflabel{$j$}{j}
   \fmffreeze 
   \fmf{fillvx,left}{h3,j}
   \fmf{unfillwx,left}{h5,h13}
   \fmf{fillwx,right}{i,h8}
   \fmf{unfillvx,right}{h2,h6}
   \fmf{plain}{i,h8} \fmf{plain}{h13,j}
   \fmf{photon,left}{h3,j}
   \fmf{dashes,left}{h5,h13}
   \fmf{dashes,right}{i,h8}
   \fmf{photon,right}{h2,h6}
   \fmfv{d.shape=circle, d.filled=empty, d.size=3pt}{h2}
   \fmfv{d.shape=circle, d.filled=empty, d.size=3pt}{h3}
   \fmfv{d.shape=circle, d.size=0.6thick}{h5}
   \fmfv{d.shape=circle, d.size=0.6thick}{h6}
\end{fmfgraph*}
$\quad\Big |\quad$

\begin{fmfgraph*}(30,25)\fmfkeep{wh22}
   \fmfleft{i,d1}
   \fmfright{j,d2}
   \fmf{phantom}{i,h1}
   \fmfn{phantom}{h}{16}
   \fmf{phantom}{h16,j}
   \fmfv{l=$k$,l.a=-45}{h8}\fmflabel{$i$}{i} 
   \fmfv{l=$l$,l.a=-90}{h13}\fmflabel{$j$}{j}
   \fmffreeze 
   \fmf{fillwx,left}{h3,j}
   \fmf{unfillvx,left}{h5,h13}
   \fmf{fillwx,right}{i,h8}
   \fmf{unfillvx,right}{h2,h6}
   \fmf{plain}{i,h8} \fmf{plain}{h13,j}
   \fmf{dashes,left}{h3,j}
   \fmf{photon,left}{h5,h13}
   \fmf{dashes,right}{i,h8}
   \fmf{photon,right}{h2,h6}
   \fmfv{d.shape=circle, d.size=0.6thick}{h2}
   \fmfv{d.shape=circle, d.size=0.6thick}{h3}
   \fmfv{d.shape=circle, d.filled=empty, d.size=3pt}{h5}
   \fmfv{d.shape=circle, d.filled=empty, d.size=3pt}{h6}
\end{fmfgraph*}\bigskip
\caption{\label{fig:whxrecur}
Recursion for the $whx$ matrix.}
\end{figure}


\vfill\eject

\parindent -5mm
{\bf References}
 
Abrahams,~J.P., van der Berg,~M., van
Batenburg,~E. \& Pleij,~C.W.A.
(1990). 
Prediction of RNA secondary structure, including
pseudoknotting, by computer simulation.
{\it Nucl. Acids Res.} {\bf 18}, 3035--44.

Ahlquist,~P., Dasgupta,~R. \& Kaesberg,~P.
(1981). 
Near identity of $3^\prime$ RNA secondary structure
in bromoviruses and cucumber mosaic virus.
{\it Cell} {\bf 23}, 183-9.

Bjorken,~J.D.  \& Drell,~S.D.
(1965). 
{\it Relativistic Quantum Fields,}
McGraw-Hill, New York, NY.

Bonhoeffer,~S., McCaskill,~J.S, Stadler,~P.F. \& Schuster,~P.
(1993). 
Statistics of RNA secondary structure.
{\it Eur. Biophys. J. (EHU)} {\bf 22}, 13--24.

Brown,~M. 
(1996). 
RNA pseudoknot modeling using intersections of stochastic
context free grammars with applications to database search.
{\it Pacific Symposium on Biocomputing 1996}.

Cary,~R.B. \& Stormo,~G.D.
(1995).
Graph-theoretic approach to RNA modeling using comparative data. 
{\it ISBM-95}, Eds.: C. Rawlings \& others. AAAI Press. 75--80.

Chomsky,~D. 
(1959). 
On certain formal properties of grammars.
{\it Information and Control} {\bf 2}, 137--76.

Dumas,~P., Moras,~D., Florentz,~C., Gieg\'e,~R.,
Verlaan,~P., van Belkum,~A. \& Pleij,~C.W.A.
(1987).
{3-D} graphics modeling of the {tRNA}-like $3^\prime$ end of turnip
yellow mosaic virus {RNA}: structural and functional implications.
{\it J. Biomol. Sruct. Dyn.} {\bf 4}, 707-28.

Durbin,~R., Eddy,~S.R., Krogh,~A. \& Mitchison,~G.J.
(1998).
{\it Biological Sequence Analysis: Probabilistic Models 
of Proteins and Nucleic Acids,}
Cambridge University Press, Cambridge UK.

Eddy,~S.R. \& Durbin,~R.
(1994). 
RNA sequence analysis using covariance models.
{\it Nucl. Acids Res.} {\bf 22}, 2079--88.

Edmonds,~J. 
(1965). 
Maximum matching and polyhedron with $0,1$-vertices.
{\it J. Res. Nat. Bur. Stand.} {\bf 69B}, 125-30.

Florentz,~C., Briand,~J.P., Romboy,~P., Hirth,~L., Ebel,~J.P. \& Giege,~R.
(1982). 
The tRNA-like structure of turnip yellow mosaic virus RNA:
structural organization of the last $159$ nucleotides from the $3^\prime$
OH terminus.
{\it EMBO J.} {\bf 1}, 269-76.

Freier,~S., Kierzek,~R., Jaeger,~J.A., Sugimoto,~N., Caruthers,~M.H.,
Neilson,~T. \& Turner,~D.H.
(1986). 
Improved free-energy parameters for predictions of
RNA duplex stability.
{\it Proc. Natl. Acad. Sci. USA,} {\bf 83}, 9373--7.

Gabow, H.N.
(1976).
An efficient implementation of Edmonds' algorithm
for maximum matching on graphs.
{\it J. Asc. Com. Mach.} {\bf 23}, 221-34.

Gluick,~T.C. \& Draper,~D.E.
(1994). 
Thermodynamics of folding a pseudoknotted mRNA fragment.
{\it J. Mol. Biol.} {\bf 241}, 246-262.

Guilley,~H., Jonard,~G., Kukla,~B. \& Richards,~K.E.
(1979). 
Sequence of $1000$ nucleotides at the $3^\prime$ end
of tobacco mosaic virus RNA.
{\it Nucl. Acids Res.} {\bf 6}, 1287-308.

Gultyaev~A.P., van Batenburg~F.H.  \& Pleij~C.W.A.
(1995). 
The computer simulation of {RNA} folding pathways
 using a genetic algorithm.	
{\it J. Mol. Biol.} {\bf 250}, 37-51.

Huynen,~M., Gutell,~R., \& Konings,~D.
(1997). 
Assessing the reliability of RNA folding using statistical mechanics.
{\it J. Mol. Biol.} {\bf 267}, 1104-12.

Kolk,~M.H., van der Graff,~M., Wijmenga,~S.S., Pleij,~C.W.A.,
Heus,~H.A. \& Hilbers, C.\-W.
(1998). 
NMR structure of a classical pseudoknot: interplay of single- and 
double-stranded RNA.
{\it Science} {\bf 280}, 434-8.

Lefebvre,~F.
(1996). 
A grammar-based unification of several alignments and folding
algorithms.
{\it ISBM-96}, Eds.: C. Rawlings \& others. AAAI Press. 143--54.

Matthews, D.H., Andre, T.C., Kim, J., Turner, D.H. \& Zuker, M.
(1998).
An updated recursive algorithm for RNA secondary structure
prediction with improved free energy parameters. 
{\it Molecular Modeling of Nucleic Acids.} 
Eds.: N.~B.~Leontis \& J.~SantaLucia Jr.
American Chemical Society. 

McCaskill~J.S
(1990).
The equilibrium partition function and base pair bindings
probabilities for RNA secondary structure.
{\it Biopolymers (A5Z)} {\bf 29}, 1105--19.

Notredame,~C., O'Brien,~E.A., \& Higgins,~D.G.
(1997).
RAGA: RNA sequence alignment by genetic algorithm.
{\it  Nucl. Acids Res.} {\bf 25}, 4570--80.

Nussinov,~R., Pieczenik,~G., Griggs,~J.R., \& Kleitman,~D.J.
(1978).
Algorithms for loop matchings.
{\it SIAM J. Appl. Math.} {\bf 35}, 68--82.

Sakakibara,~Y., Brown,~M., Hughey,~R., Mian,~I.S., Sj\"olander,~K.,
Underwood,~R.C. \& Hussler,~D.
(1994). 
Stochastic context-free grammars for tRNA modeling.
{\it Nucl. Acids Res.} {\bf 22}, 5112--20.


Sankoff,~D. 
(1985). 
Simultaneous solution of the RNA folding alignment and 
protosequence problems.
{\it SIAM J. Appl. Math.} {\bf 45}, 810--25.

Schuster,~P., Fontana,~W., Stadler,~P.F. \& Hofacker,~I.L.
(1994). 
From sequences to shapes and back: a case study
in RNA secondary structure.
{\it Proc. R. Soc. Lond. B. Biol. Sci.} {\bf 255}, 279--84.
http://www.\-itc.uni\-vie.ac.at/~ivo/RNA

Schuster,~P., Fontana,~W., Stadler,~P.F. \& Renner,~A.
(1997). 
{RNA} structures and folding: from conventional to new issues 
in structure predictions.
{\it Curr. Opin. Struct. Biol.} {\bf 7}, 229--35.

Serra,~M.J. \& Turner,~D.H.
(1995). 
Predicting the thermodynamic properties of RNA.
{\it Meth. Enzymol.} {\bf 259}, 242--61.

Steinberg,~S., Misch,~A. \& Sprinzl,~M.
(1993).
Compilation of {RNA} sequences and sequences of {tRNA} genes.
{\it Nucl. Acids Res.} {\bf 21}, 3011--15.   

ten Dam~E., Pleij,~K. \& Draper,~D.
(1992).
Structural and functional aspects of {RNA} pseudoknots.
{\it Biochemistry} {\bf 31}, 11665-11676.

Tuerk, C., MacDougal, S. \& Gold, L. 
(1992). 
RNA pseudoknots that inhibit human immunodeficiency
virus type 1 reverse transcriptase. 
{\it Proc. Natl. Acad. Sci. USA,} {\bf 89}, 6988--92.

van Batenburg,~F.H.D., Gultyaev,~A.P. \& Pleij,~C.W.A.
(1995).
An {APL}-programmed genetic algorithm for the
prediction of {RNA} secondary structure.
{\it J. Theor. Biol.} {\bf 174}, 269--80.

Van Belkum,~A., Abrahams,~J.P., Pleij,~C.W.A. \& Bosch,~L.
(1985).
Five pseudoknots are present at the $204$ nucleotides
long $3^\prime$ non coding region of tobacco mosaic virus {RNA}.
{\it Nucl. Acids Res.} {\bf 13}, 7673--86.   

Van Belkum,~A., Bingkun,~J., Pleij,~C.W.A. \& Bosch,~L.
(1987). 
Structural similarities among valine-accepting tRNA-like
structures in tymoviral RNAs and elongator tRNAs.
{\it Biochemistry} {\bf 26}, 1144-51.

Walter,~A., Turner,~D., Kim,~J., Lyttle,~M.,
M\"uller,~P., Matthews,~D. \& Zuker,~M.
(1994). 
Coaxial stacking of helixes enhances binding of
oligoribonucleotides and improves predictions of
RNA folding.
{\it Proc. Natl. Acad. Sci. USA,} {\bf 91}, 9218--22.

Woese~C.R. \& Pace~N.R.
(1993).
Probing {RNA} structure, function, and history by comparative analysis.
{\it The RNA World}. 
Eds.: R.~F. Gesteland \& J.~F. Atkins.
Cold Spring Harbor Laboratory Press.
New York NY. 91-117.

Wyatt,~J.R., Puglisi,~J.D. \& Tinoco,~I.Jr.
(1990). 
RNA pseudoknots: stability and loop size requirements.
{\it J. Mol. Biol.} {\bf 214}, 455-70.

Zuker,~M. \& Stiegler,~P. 
(1981).
Optimal computer folding of large RNA sequences using thermodynamics
and auxiliary information.
{\it Nucl. Acids Res.} {\bf 9}, 133--48.

Zuker,~M. \& Sankoff,~D.
(1984).
RNA secondary structure and their prediction.
{\it Bull. Math. Biol.} {\bf 46}, 591--621.

Zuker,~M. 
(1989). 
Computer prediction of {RNA} structure.
{\it Meth. Enzymol.} {\bf 180}, 262-88.

Zuker,~M. 
(1989). 
On finding all suboptimal foldings of an RNA molecule.
{\it Science} {\bf 244}, 48--52.

Zuker,~M. 
(1995). 
``Well-determined'' regions in RNA secondary structure
prediction: analysis of small subunit ribosomal RNA.
{\it Nucl. Acids Res.} {\bf 23}, 2791--8.

\hfill\eject

\end{fmffile}
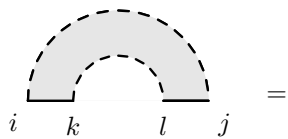
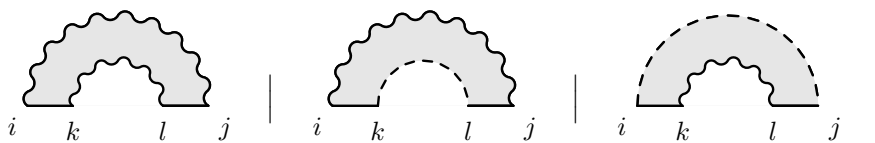
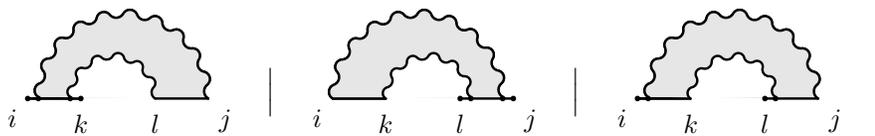
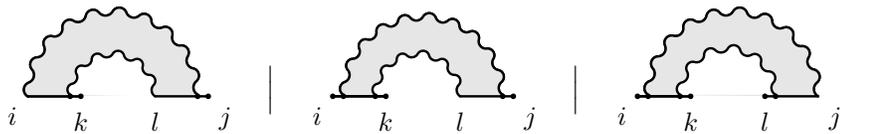
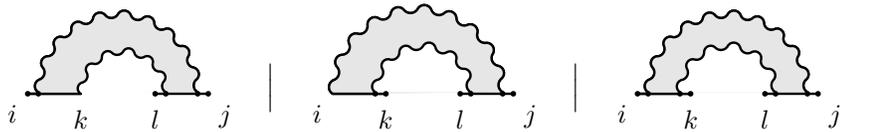
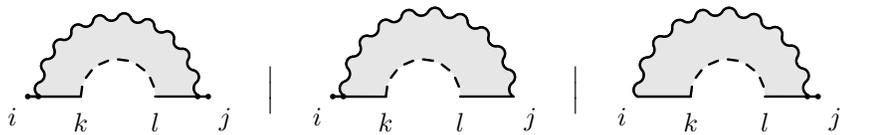
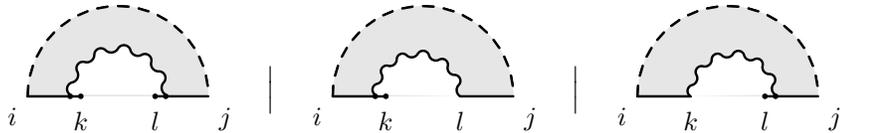
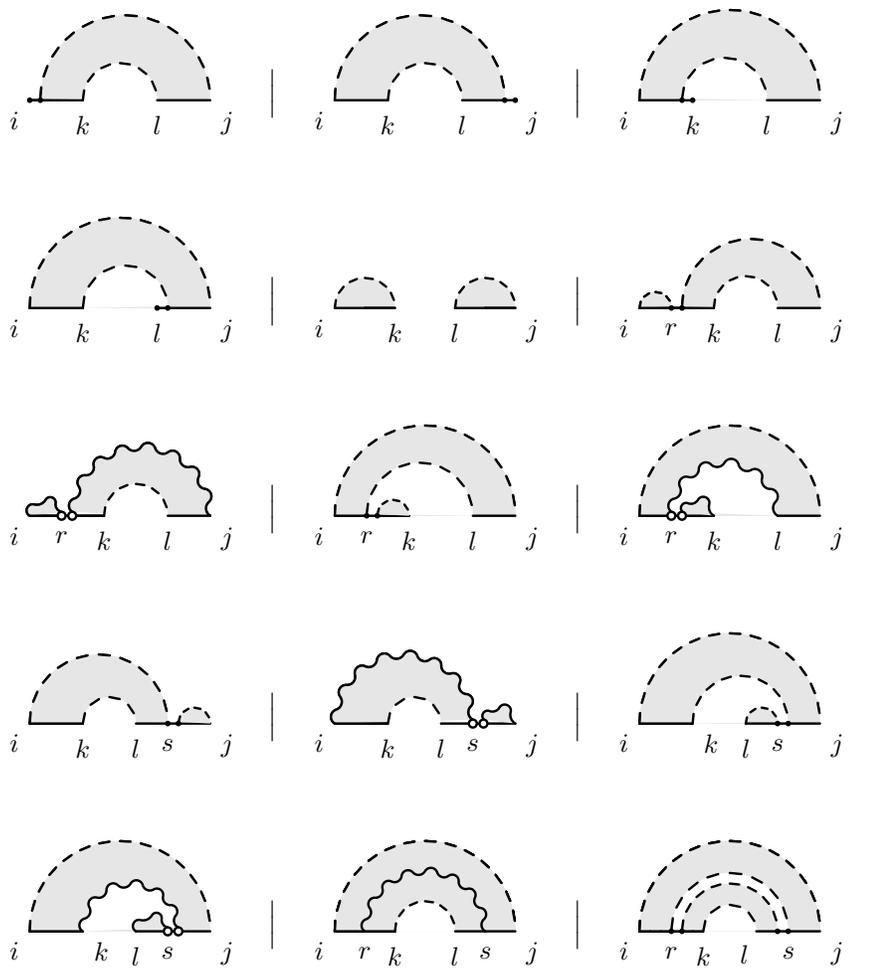
\end{document}